\documentclass[twocolumn]{aastex631}

\usepackage[utf8]{inputenc}
\usepackage{natbib}
\usepackage{amssymb, amsmath}
\usepackage{bm}
\usepackage{hyperref}
\usepackage{graphicx}

\newcommand{\mrm}[1]{\mathrm{#1}}

\newcommand{\lya}{Ly$\alpha$}
\newcommand{\oii}{[O\,{\sc ii}]}
\newcommand{\oiill}{\oii $\lambda \lambda\ 3726,3729$}
\newcommand{\ew}{W_\lambda}
\newcommand{\ewlya}{W_\lambda ({\mrm{Ly\alpha}})}
\newcommand{\fesc}{$f_{esc}^{\mrm{Ly\alpha}}$}
\newcommand{\snr}{$S/N$}

\newcommand{\arcsecs}{^{\prime\prime}}
\newcommand{\msun}{\mrm{M_\odot}}
\newcommand{\zsun}{\mrm{Z_\odot}}
\newcommand{\ang}{\mrm{\AA}}
\newcommand{\mic}{\mrm{\mu m}}
\newcommand{\flux}{\mrm{erg\ s^{-1}\ cm^{-2}}}

\newcommand{\hd}{HETDEX}
\newcommand{\hdr}{HDR2}
\newcommand{\hdra}{HDR 2.1.2}
\newcommand{\hdrb}{HDR 2.1.3}
\newcommand{\hst}{\mbox{$HST$}}
\newcommand{\spitzer}{\mbox{$Spitzer$}}

\newcommand{\ninspect}{842}
\newcommand{\nlyalines}{94}
\newcommand{\nsamp}{72}
\newcommand{\gn}{GOODS-N}
\newcommand{\lae}{LAE}
\newcommand{\laes}{LAEs}
\newcommand{\plae}{P(\lya)}

\newcommand{\zrange}[2]{$#1 < z < #2$}
\newcommand{\lrange}[2]{$#1 \ \ang < \lambda < #2 \ \ang $}

\newcommand{\code}[1]{\textsc{#1}}
\newcommand{\elixer}{ELiXer}
\newcommand{\bagpipes}{\code{Bagpipes}}
\newcommand{\emcee}{\code{emcee}}
\newcommand{\srcex}{\code{SourceExtractor}}
\newcommand{\python}{\code{python}}
\newcommand{\multinest}{\code{MultiNest}}

\begin{document}

\title{Stellar Populations of Ly$\alpha$ Emitting Galaxies in the \hd\ Survey I:\\ An Analysis of \laes\ in the \gn\ Field}

\author[0000-0002-3912-9368]{Adam P. McCarron}
\affiliation{Department of Astronomy, The University of Texas at Austin, 2515 Speedway, Austin, TX 78712, USA}
\email{apm.astro@utexas.edu}

\author[0000-0001-8519-1130]{Steven L. Finkelstein}
\affiliation{Department of Astronomy, The University of Texas at Austin, 2515 Speedway, Austin, TX 78712, USA}

\author[0000-0003-2332-5505]{Oscar A. Chavez Ortiz}
\affiliation{Department of Astronomy, The University of Texas at Austin, 2515 Speedway, Austin, TX 78712, USA}

\author[0000-0002-8925-9769]{Dustin Davis}
\affiliation{Department of Astronomy, The University of Texas at Austin, 2515 Speedway, Austin, TX 78712, USA}

\author[0000-0002-2307-0146]{Erin Mentuch Cooper}
\affiliation{Department of Astronomy, The University of Texas at Austin, 2515 Speedway, Austin, TX 78712, USA}
\affiliation{McDonald Observatory, University of Texas at Austin, 2515 Speedway, Austin, TX 78712, USA}

\author[0000-0003-1187-4240]{Intae Jung}
\affil{Department of Physics, The Catholic University of America, Washington, DC 20064, USA }
\affil{Astrophysics Science Division, Goddard Space Flight Center, Greenbelt, MD 20771, USA}
\affil{Center for Research and Exploration in Space Science and Technology, NASA/GSFC, Greenbelt, MD 20771}
 
\author[0000-0002-7707-9437]{Delaney R. White}
\affiliation{Department of Astronomy, The University of Texas at Austin, 2515 Speedway, Austin, TX 78712, USA}

\author[0000-0002-9393-6507]{Gene C. K. Leung}
\affiliation{Department of Astronomy, The University of Texas at Austin, 2515 Speedway, Austin, TX 78712, USA}

\author[0000-0002-8433-8185]{Karl Gebhardt}
\affiliation{Department of Astronomy, The University of Texas at Austin, 2515 Speedway, Austin, TX 78712, USA}

\author[0000-0002-6788-6315]{Viviana Acquaviva}
\affiliation{Physics Department, NYC College of Technology, 300 Jay Street, Brooklyn, NY 11201, USA}
\affiliation{Center for Computational Astrophysics, Flatiron Institute, New York, NY 10010, USA}

\author[0000-0003-4381-5245]{William P. Bowman}
\affil{Department of Astronomy \& Astrophysics,
The Pennsylvania State University, University Park, PA 16802, USA}
\affil{Institute for Gravitation and the Cosmos, The Pennsylvania
State University, University Park, PA 16802, USA}

\author[0000-0002-1328-0211]{Robin Ciardullo}
\affil{Department of Astronomy \& Astrophysics,
The Pennsylvania State University, University Park, PA 16802, USA}
\affil{Institute for Gravitation and the Cosmos, The Pennsylvania
State University, University Park, PA 16802, USA}

\author[0000-0003-1530-8713]{Eric Gawiser}
\affiliation{Department of Physics and Astronomy, Rutgers, The State University, Piscataway, NJ 08854, USA}

\author{Caryl Gronwall}
\affil{Department of Astronomy \& Astrophysics,
The Pennsylvania State University, University Park, PA 16802, USA}
\affil{Institute for Gravitation and the Cosmos, The Pennsylvania
State University, University Park, PA 16802, USA}

\author[0000-0001-6717-7685]{Gary J. Hill}
\affiliation{McDonald Observatory, University of Texas at Austin, 2515 Speedway, Austin, TX 78712, USA}
\affiliation{Department of Astronomy, The University of Texas at Austin, 2515 Speedway, Austin, TX 78712, USA}

\author{Wolfram Kollatschny}
\affiliation{Institut f\"ur Astrophysik, Universit\"at G\"ottingen, Friedrich-Hund Platz 1, D-37077 G\"ottingen, Germany}

\author[0000-0003-1838-8528]{Martin Landriau}
\affiliation{Lawrence Berkeley National Laboratory, 1 Cyclotron Road, Berkeley, CA 94720, USA}

\author[0000-0001-5561-2010]{Chenxu Liu}
\affiliation{Department of Astronomy, The University of Texas at Austin, 2515 Speedway, Austin, TX 78712, USA}

\author[0000-0003-4237-2470]{Daniel N. Mock}
\affiliation{Department of Physics, Florida State University, Tallahassee, Florida 32306}

\author[0000-0003-1198-831X]{Ariel G. S\'anchez}
\affiliation{Max-Planck-Institut f\"ur extraterrestrische Physik,
Postfach 1312, Giessenbachstr., 85748 Garching, Germany}

\begin{abstract}
    We present the results of a stellar-population analysis of \nsamp\ \lya\ emitting galaxies (\laes) in \gn\ at \zrange{1.9}{3.5} spectroscopically identified by the Hobby-Eberly Telescope Dark Energy Experiment (HETDEX\null). We provide a method for connecting emission-line detections from the blind spectroscopic survey to imaging counterparts, a crucial tool needed as \hd\ builds a massive database of $\sim$ 1 million \lya\ detections. Using photometric data spanning as many as 11 filters covering $0.4 < \lambda \ (\mrm{\mu m}) < 4.5$ from the \textit{Hubble} and {\it Spitzer} Space Telescopes, we study the objects' global properties and explore which properties impact the strength of \lya\ emission.  We measure a median stellar mass of $0.8^{+2.9}_{-0.5} \times 10^{9} \ \msun$ and conclude that the physical properties of \hd\ spectroscopically-selected \laes\ are comparable to \laes\ selected by previous deep narrow band studies. We find that stellar mass and star formation rate correlate strongly with the \lya\ equivalent width. We then use a known sample of $z>7$ \laes\ to perform a proto-study of predicting \lya\ emission from galaxies in the Epoch of Reionization, finding agreement at the $1\sigma$ level between prediction and observation for the majority of strong emitters.
\end{abstract}

\section{Introduction}
\label{sec:intro}

Lyman-alpha emitting galaxies (hereafter \laes) have fascinated astronomers for decades, from when \citet{partridge67} first predicted that primitive galaxies in formation could emit a detectable \lya\ line, through their discovery by \citet{cowie98} and \citet{rhoads00}. These objects exhibit strong emission of the \lya\ photon corresponding to the $n \! = \! 2$ to $n \! = \! 1$ resonant transition in hydrogen atoms. These photons are expected to face high optical depths from neutral hydrogen to escape the galaxies in which they are generated, and dust grains along their paths can absorb them. To date, despite enormous effort (see \citealt{ouchi20} for review), the community has not formed a strong consensus on exactly how \lya\ radiation escapes its host galaxy, and no reliable model exists to predict the \lya\ luminosity or equivalent width, $\ewlya$, of a galaxy given its global physical properties, such as stellar mass, metallicity, age, star formation rate, and dust extinction. 

Part of the problem arises from discrepant conclusions drawn from studying LAEs identified using different selection techniques. Locally ($z \ll 1$), the ultraviolet (UV) flux measured in wide or narrow-band filters often defines LAE samples, biasing studies to brighter, higher mass systems than those found spectroscopically \citep{hayes14}. Observations in the nearby universe paint \laes\ as low mass galaxies with young stellar ages as determined from spectral energy distribution (SED) fitting, and many studies concur on trends showing an increase in \lya\ luminosity with decreasing dust and metals \citep{hayes15}. Nonetheless, many galaxies show stronger \lya\ emission than models would predict based on dust extinction (e.g., \citealt{martin15}, \citealt{atek14}, \citealt{scarlata09}, \citealt{fink09}), and a satisfactory explanation of this \lya\ enhancement does not currently exist. 

With narrow-band selected LAEs at higher redshift, discrepant results still persist. \citet{fink09} found \laes\ at $z \sim 4.5$ represent a diverse population in terms of stellar age, mass, and dust extinction. \citet{keely15} modeled the SEDs of IRAC-detected LAEs at $z \sim 5$ and found a third have old stellar populations, contrasting with the young populations found in the local universe, and \citet{guaita11} observed similar heterogeneous populations in a narrow-band selected sample at $z \simeq 2.1$. Moreover, \citet{gawiser07} found NB-selected LAEs at $z=3.1$ to generally be low mass, dust-free objects, but their model allowed for both young and more evolved stellar populations, and \citet{acquaviva12} found LAEs at $z=3.1$ to be older than those at $z=2.1$. \citet{kornei10} compiled a UV continuum selected sample of $z\sim3$ galaxies, finding those with strong \lya\ emission had older stellar populations with lower star formation rates and less dust. Recently, \citet{santos20} used SED fitting of nearly 4000 LAEs in the COSMOS field at $2<z<6$  to find that LAEs were younger and/or more dust-poor than other UV-selected objects based on their UV slopes. 

Studies of LAE samples compiled using detection of the Ly$\alpha$ emission line itself in the high-redshift universe confound consensus as well. \citet{hagen16} used the Hobby Eberly Telescope Dark Energy Experiment (HETDEX) pilot survey (\citealt{adams11}, \citealt{blanc11}) to compare properties of LAEs at $z \sim 2$ with optical emission line-selected galaxies (oELGs) and found no significant differences between the populations. Remarkably, even the UV-slope did not differ in the two samples, implying either that diffuse dust in the interstellar medium (ISM) did not modulate \lya\ emission or that oELGs strongly emit \lya. Recently, spectroscopic surveys have also yielded confusing results about LAEs at $z>2$. Using data from the VANDELS survey, \citet{marchi19} suggested LAEs have low mass and low dust extinction, but found no correlation with star formation rate. From the VIMOS Ultra-Deep Survey, \citet{hathi16} concurred with LAEs having lower mass and lower dust extinction, but they found that the objects have lower SFRs than non-LAEs. Approaching the problem from the other direction, \citet{oyarzun17} found from studying the spectra of stellar mass selected galaxies at $3<z<4.6$ that a negative correlation existed between \lya\ equivalent width and both stellar mass and star formation rate. A review of the field's current knowledge of high-redshift \lya\ emission can be found in \citet{ouchi20}.

A deeper understanding of what makes \laes\ unique from other star forming galaxies (SFGs) tantalizes astronomers because of the profound implications for leveraging \laes\ as sensitive probes of reionization at $z \gtrsim 6$.  Whether the Universe re-ionized rapidly at late times (e.g. \citealt{robertson15}) or gradually  beginning very early in its history (e.g. \citealt{fink19}) can determine if massive, rare galaxies or low-mass, ubiquitous objects emitted the needed ionizing photons. Answering such a fundamental cosmological question hinges on our ability to detect neutral hydrogen in the Universe's infancy. Crucially, the attenuation of \lya\ photons can probe the presence of neutral hydrogen in the intergalactic medium (IGM) \citep[e.g,][]{miralda-escude98, malhotra04, dijkstra14}, but the photons also undergo complicated resonant scattering within the galaxy, complicating our understanding of how much of the emission exits the ISM and circumgalactic medium (CGM) and enters the IGM in the first place. Recent attempts to use Ly$\alpha$ as a reionization probe have struggled to account for the intrinsic effects of host galaxy properties on the Ly$\alpha$ luminosity before the radiation encounters the IGM, leaving an unknown systematic uncertainty present in their results. The most detailed spectroscopic studies of post-reionization LAEs point to the covering fraction of optically thick neutral hydrogen (e.g. \citealt{reddy21}) as the key predictor of \lya\ escape, but such observations remain expensive and time intensive. Finding correlations between \lya\ emission and global properties such as mass and star formation activity, which photometry can reliably measure even at very high redshifts, could be a path forward to predicting galaxies' intrinsic \lya\ output.

Small LAE sample sizes ($<20$) were typical a decade ago, and although recently large samples with $>$1000 objects have been amassed using narrow-band surveys (e.g. \citealt{sobral18}, \citealt{ono2021}), spectroscopically confirmed samples remain small. This has statistically hindered the efficacy of studies of global property correlations with Ly$\alpha$ emission. The HETDEX project (\citealt{hill08}, \citealt{hill21}, \citealt{gebhardt21}) is in the process of discovering a transformative sample of LAEs, clearing the way for the community to obtain a better understanding of this intriguing population.  The un-targeted (targets not pre-selected), spectroscopically selected 
\hd\ \lae\ sample at \zrange{1.9}{3.5} provides a unique vantage point of galaxy evolution, as these galaxies probe the lower-mass end of the galaxy distribution, making them analagous to typical galaxies discovered in the epoch of reionization (e.g., \citealt{fink10}). 

As the first step toward realizing \hd's ability to unlock \laes\ as probes of reionization, we present an initial study detailing how to link detections from the survey to imaging counterparts, and we provide an SED fitting analysis of their stellar population properties. Our modest sample of \nsamp\ \laes\ in the \gn\ field will pave the way for future large samples from \hd\ to obtain the best understanding of \laes\ to date. In \S\ref{sec:method} we describe how we built our sample and selected imaging counterparts. In \S\ref{sec:analysis} we describe our SED fitting procedure. We present our results in \S\ref{sec:results}, comparing them to other studies, and we discuss our interpretations in \S\ref{sec:discussion}. Finally, we attempt to predict the \lya\ emission from a sample of epoch of reionization (EoR) galaxies in \S\ref{sec:predictions} and summarize this study in \S\ref{sec:summary}. In our analysis, we adopt a flat $\Lambda$CDM cosmology with $H_0 = 70 \ \mrm{km\ s^{-1}\ Mpc^{-1}}$ and $\Omega_{\mathrm{m}} = 0.30$.

\section{Methodology}
\label{sec:method}

In order to explore how \lya\ emission from galaxies depends on stellar population properties, we built a sample of \laes\ using emission line detections from the \hd\ survey, carefully identifying them as \lya\ or other contaminant features, such as \oiill, which is unresolved at \hd\ resolution. We then created a procedure for assigning the line detections to imaging counterparts in \hst\ data so that we could proceed with fitting their SEDs. 

\subsection{The HETDEX Survey}

With \hd\, 
the upgraded Hobby-Eberly Telescope (\citealt{rams94}, \citealt{hill21}) is observing an area of $540 \ \mathrm{deg}^2$ in the north Galactic cap and on the celestial equator using up to 78 pairs of integral-field spectrographs that span $350-550 \ \mathrm{nm}$ at $R \sim 800$. Each spectrograph pair is fed by an integral field unit (IFU) of 448 1.5$\arcsec$-diameter fibers which cover a 51$\arcsec$ $\times$ 51$\arcsec$ region on the sky with 1$/$3 fill factor \citep{kelz14, hill21}.  Each HETDEX observation consists of three 6-min dithered exposures to fill in the area between fibers, each with $>$30,000 individual fibers. The majority of these fibers just contain blank sky, but some subset contain continuum sources such as stars or emission lines from both nearby and distant galaxies.

\citet{gebhardt21} describe the data reduction and calibrations needed to convert the raw observations into a three dimensional spectroscopic data set as well as the methods used to detect emission lines contained in the millions of observed spectra. As a brief summary, HETDEX reductions involve three types of calibration frames:  biases (taken nightly), pixel flats (taken yearly using a laser-driven light source), and twilight sky flats (taken nightly and averaged monthly), which are used for bias subtraction, bad pixel masking, fiber profile tracing, wavelength calibration, scattered light removal, spectral extraction, fiber normalization, spectral masking, and sky subtraction. These frames, combined with sky background on science images, produce a wavelength calibrated, sky-subtracted spectrum for each fiber in the array.

Astrometric calibrations are achieved by measuring the centroid of each field star and comparing their positions on the IFUs to the stars' equatorial coordinates in the Sloan Digital Sky Survey \citep[SDSS;][]{york00, abazajian09} and \textit{Gaia} \citep{gaia18} catalogs.  This process typically results in global solutions which are good to $\sim 0.2\arcsec$ and no worse than $\sim 0.5\arcsec$, with the exact precision of a measurement dependent upon the number of IFUs in operation at the time of the observation.  


To find emission lines, the data pipeline searched every spatial and spectral resolution element in the internal HETDEX data release 2 (HDR2) to look for a peak in signal. Regions of enhanced signal were fit with a single Gaussian model with a constant continuum level, a model found adequate for potentially asymmetric line profiles by \cite{gebhardt21} because of the low resolution of the VIRUS spectrographs and low signal to noise (\snr) of typical sources. The exact location was determined by rastering on a grid and maximizing the line's signal-to-noise. An internal catalog of high-quality emission lines was generated by Mentuch Cooper et al. (in preparation),
and we drew our initial sample from the \hdr\ version of that catalog. The catalog reduced the raw detected line emission sources as described in \citet{gebhardt21} into a more robust sample by passing the observations through a quality assessment pipeline and limiting various fitted line parameters. Specifically, emission lines were required to have a quality of fit, $\chi^2 <1.2$ and a linewidth, $\sigma$, in the Gaussian model between 1.7\,\AA~and 8\,\AA.
The full \hd\ survey will eventually detect $\sim1$ million LAEs, providing an incredible opportunity to study such objects, but our analysis is focused on \laes\ discovered in 2018--2020 data from a \hd\ science verification field in \gn, a roughly $10^\prime \times 16^\prime$ field centered at (J2000) $\mrm{12^h 36^m 55^s}, 62^\circ 14^\mrm{m} 15^\mrm{s}$ (\citealt{giavalisco04}, \citealt{grogin11}, \citealt{koekemoer11}) because we required deep, multi-band imaging to study each galaxy's stellar populations.  

\subsection{Sample Selection}
\label{ssec:sampleselection}

We visually inspected \hd\ detections in \gn\ to obtain a clean sample of \laes. To get initial candidates, we applied various quality cuts to the curated catalog for data release \hdra\ (Mentuch Cooper at al., in preparation). We restricted emission line detections to those with signal-to-noise ratio $S/N>5.5$ to limit the fraction of spurious detections from noise fluctuations to less than 5\% (see \citealt{gebhardt21}) as well as $\chi^2 < 1.6$ for the  Gaussian model fit, which was a value tuned to remove the most obvious artifacts while retaining the largest sample for inspection. We required emission line full-width at half-maximum (FWHM) between 3.4 \AA\ and 24 \AA, where the lower bound removed exceedingly narrow peaks arising from unidentified cosmic rays and the upper bound removed emission generated by broad-line AGNs, which we considered contaminants in this study (see \S\ref{ssec:sedfitting}). We further only included observations with throughput $>0.07$ for reliable flux measurements minimally affected by cloud cover, and seeing below 2.8$\arcsecs$ to enable continuum counterpart identification. We did not remove ``repeat'' detections coincident spatially and spectrally resulting from the survey revisiting the field multiple times in order to ensure we found as many \lya\ detections as possible. We excluded data in \gn\ taken prior to 2018 as they included significant artifacts from early CCDs that had been replaced by 2018.

Finally, we did not initially remove any detections based on the Bayesian probability values used to help determine the identity of an emission line as \lya\ vs \oiill, such as \plae. These probabilities, which are calculated by the \hd\ team based on the work of \citet{leung17} and \citet{farrow21}, leverage the inherent differences between the emission line luminosity and equivalent width (EW), $\ew$, distribution functions of LAEs and \oii\ emitters to identify single emission line detections using information about the line flux and continuum emission, when available. During the process of visual inspection, we used the statistic to guide our identifications, and we make recommendations for using quality cuts based off this statistic at the end of this section. We do not believe that keeping this statistic visible to the classifier biased our results because we implemented an independent procedure (see \S \ref{ssec:cptident}) to distinguish LAEs from low-redshift counterparts that relied on SED fitting.

After applying quality cuts, we began with \ninspect\ detections (of which $\sim$500 were ``unique'' in the sense that there were no other emission line detections within 3$\arcsec$ spatially and 6\,\AA\  spectrally). To inspect each detection, we used the \hd\ Emission Line eXplorer tool \elixer\ (Davis et al., in preparation), 
which shows measured quantities for the emission lines such as \snr, line width, line fit $\chi^2$, the continuum estimate, the Bayesian probability for \lya\ emission described above, as well as useful visual information, such as cutouts of the 2D spectra for several fibers containing the feature, the Gaussian model fit to the feature, the full 1D spectrum, and any imaging and catalog data uploaded in the \hd\ pipeline.

\begin{figure*}
    \centering
    \includegraphics[width=0.75\textwidth]{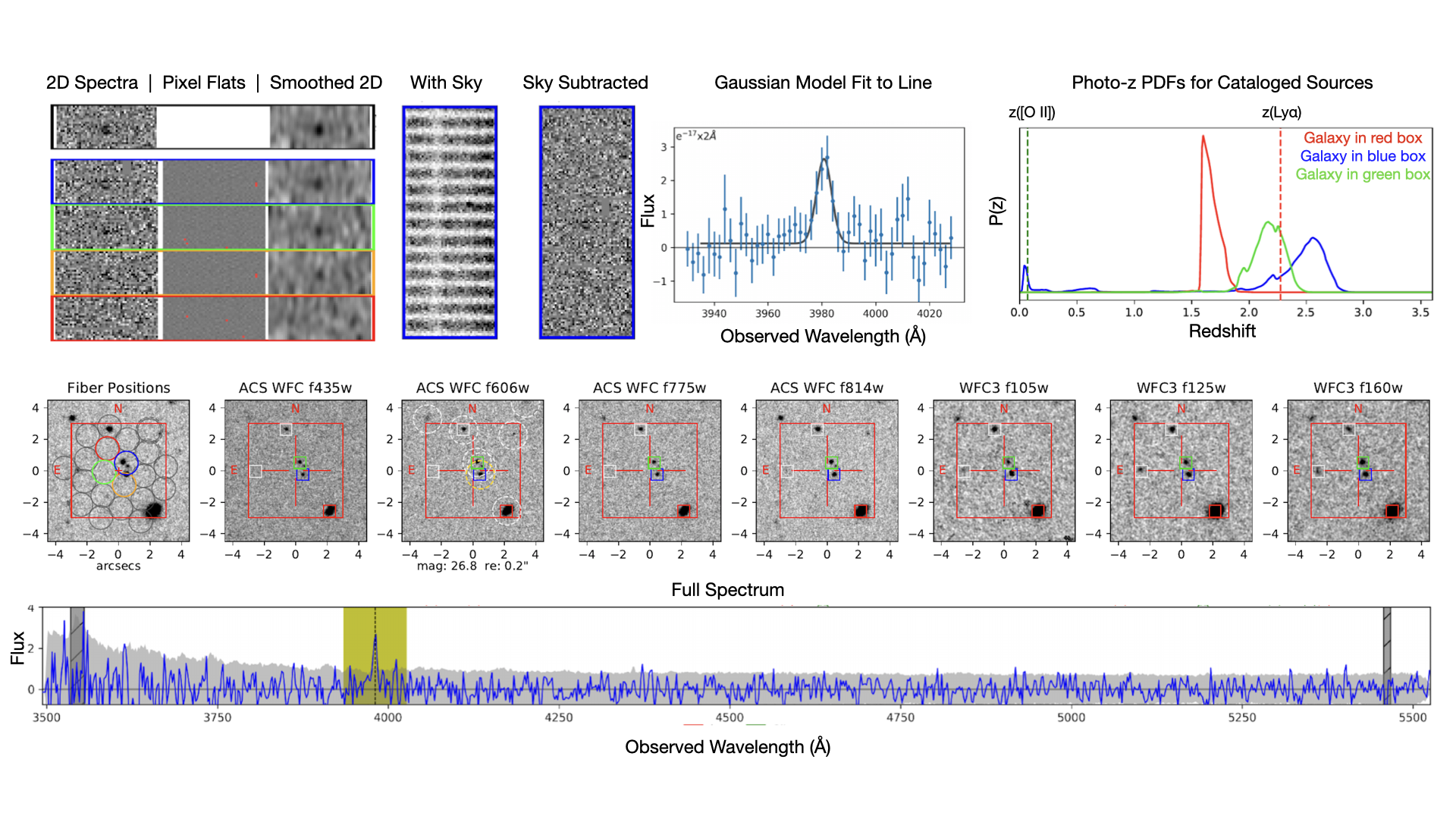}
    \caption{A section of an \elixer\ report for detection ID 2100325857, the line from an LAE in our sample. The report contains information about the detected line as well as imaging at the detection position. The 2D spectra from the four fibers contributing the highest \snr\ to the detection are in the top left corner; the stacked signal is shown on the top row, outlined in black. The pixel flats and smoothed 2D spectra are displayed in the right two columns. Plots to the right show the sky subtraction and the emission line model fit (black line). The middle row contains $10\arcsec$ imaging cutouts with the fiber positions shown in gray in the first image and the locations of cataloged sources marked with colorful boxes in all subsequent images. White boxes indicate sources too far to be considered, and white circles show the aperture for brightness measurement in F606W. The AB magnitude of the nearest likely source is reported. The top right corner shows the photo-$z$ probability distributions calculated by the CANDELS team (B. Andrews et al., in preparation) for sources of matching color in the imaging, and the \oii\ and \lya\ redshifts are shown as vertical dashed green and red lines. Finally, the bottom row shows the full 1D spectrum. Flux densities have units of $\mathrm{erg\ s^{-1}\ cm^{-2}\ 2\AA^{-1}}$}.
    \label{fig:elixer_lae}
\end{figure*}

We rated our confidence in a detection on a scale of 0-5 using a customized widget tool that allows interactive classification of detected sources based on the its Elixer Report (see Figure~\ref{fig:elixer_lae}). Additionally, other classifications include ``artifact,'' a false detection caused by a malfunction in the instrument or the reduction pipeline, low-redshift sources, and ``other'' for miscellaneous objects like meteors. To qualify for a classification of 4 or 5 (a high-confidence LAE by our definition), a detection had to meet the following criteria:
\begin{itemize}
    \item A clear emission line in at least one fiber in the un-smoothed 2D spectrum, or a probable emission line in at least two fibers. Since each point-spread-function (PSF) covers multiple fibers (due to the dithering pattern), we expected strong emission to be seen in more than one fiber, increasing the likelihood of a real detection.
    \item No obvious defects at the emission line location in the pixel flat or sky subtraction cutouts.  This eliminated hot pixels, sky model residuals, charge traps, and other artifacts from the sample.
    \item A Gaussian plus constant continuum model fit that adequately matched the data and did not have a FWHM far below the spectral resolution of $\sim 6$\,\AA\null.
    \item A line peak that exceeded the typical noise level in multiple pixels in the 1D spectrum.
    \item No source at the line's detection position brighter than roughly $m_{AB} = 24$ in the imaging cutouts, if available. The high equivalent widths of sources fainter than this threshold drastically decrease the likelihood of contamination by \oii\ emitters (see Figure 6 in \citealt{leung17}), though a few low equivalent width, luminous LAEs can be missed with this requirement. 
\end{itemize}

As the \oiill\ emission feature falls into the \lrange{3500}{5500} spectral range for $z<0.5$, the imaging proved crucial in choosing between high-redshift \laes\ and interloping \oii\ emitting galaxies. 

\begin{figure*}
    \centering
    \includegraphics[width=0.75\textwidth]{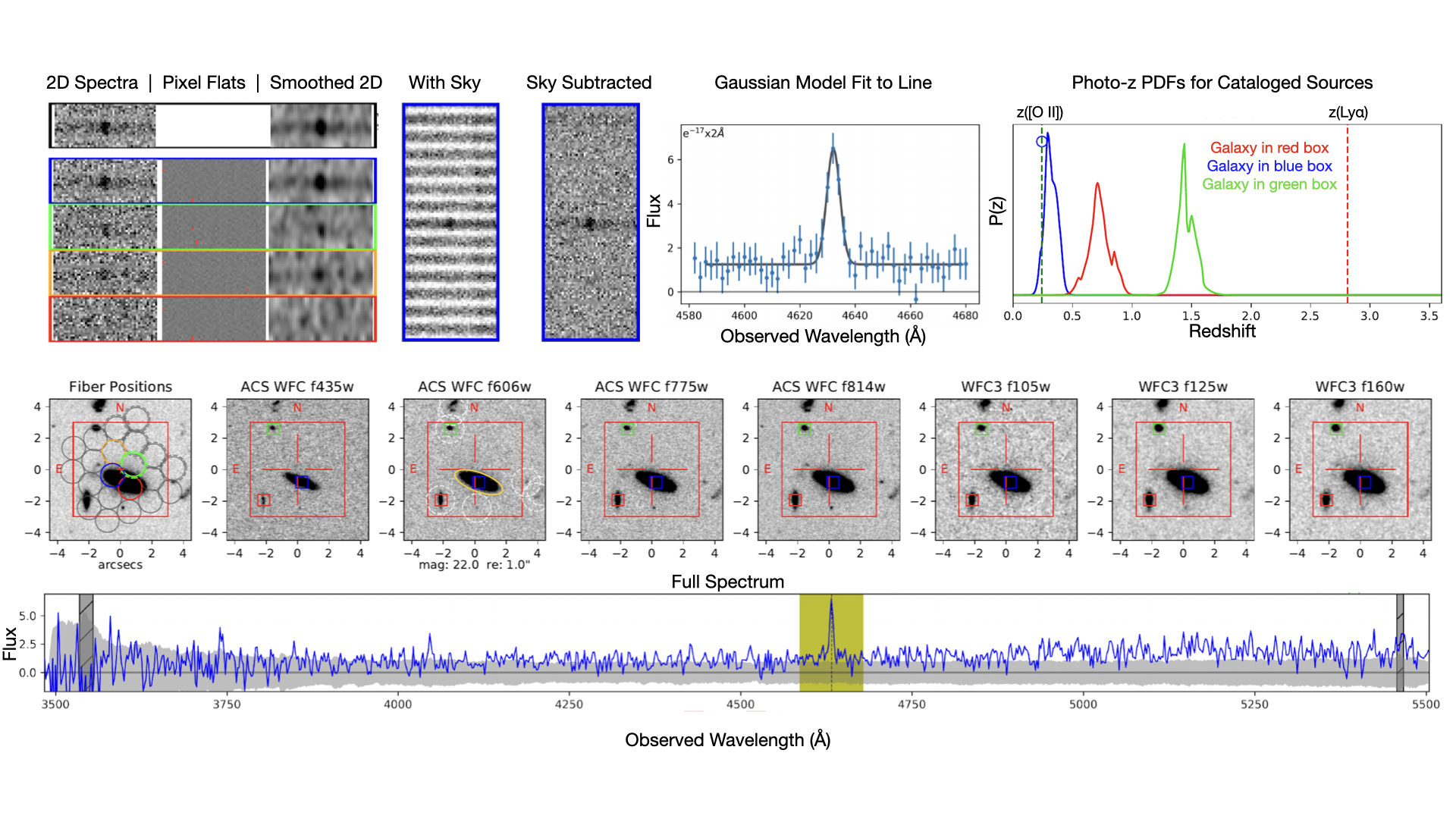}
    \caption{A section of an \elixer\ report, in the same format as Figure~\ref{fig:elixer_lae}, for detection ID 2100037191, corresponding to the \oiill\ feature in a galaxy at $z \approx 0.24$. The black trace in the 2D fiber spectrum (blue rectangle) indicates a clear detection of continuum emission, which is also evident in the 1D spectrum in at the bottom of the figure. The imaging shows a large, bright source ($m_{AB}=22.0$) centered on the detection position, and the source has a cataloged spectroscopic redshift consistent with \oii\, indicated as an open blue circle in the photometric redshift plot in the upper-right corner.}
    \label{fig:elixer_lowz}
\end{figure*}

Figure~\ref{fig:elixer_lae} shows an example \elixer\ report for a source classified as a high-redshift \lae\null. Note that for readability, tabulated numeric information such as \plae, line flux, line model $\chi^2$, and more was cropped out of this visualization, but was visible to the classifier. In Figure~\ref{fig:elixer_lae}, A clear emission feature is present as a black signal in three out of the four 2D un-smoothed fiber spectra, the sky subtraction looks clean, the model fit accurately represents the data, and the image stamps show a number of faint sources with photometric redshift estimates reasonably close to the \lya\ redshift (shown by the vertical red dashed line). Figure~\ref{fig:elixer_lowz} shows a clear example of a low-redshift object detected by its \oii\ emission line. As in Figure~\ref{fig:elixer_lae}, the line appears strong in multiple fibers, and the sky subtraction and model fit present no concerns. Characteristically of a brighter low-redshift galaxy, continuum emission is visible as a horizontal black trace in the fiber spectra, and a large, bright object appears in the \hst\ image stamps. In this case, the object is in fact a cataloged \oii\ emitter, but even without such information this would be a clear low-redshift classification. In both of these cases, no other emission lines are detected, or would be expected to be detectable, across the observed wavelength range.   

\begin{figure}[b]
    \centering
    \includegraphics[width=0.9\columnwidth]{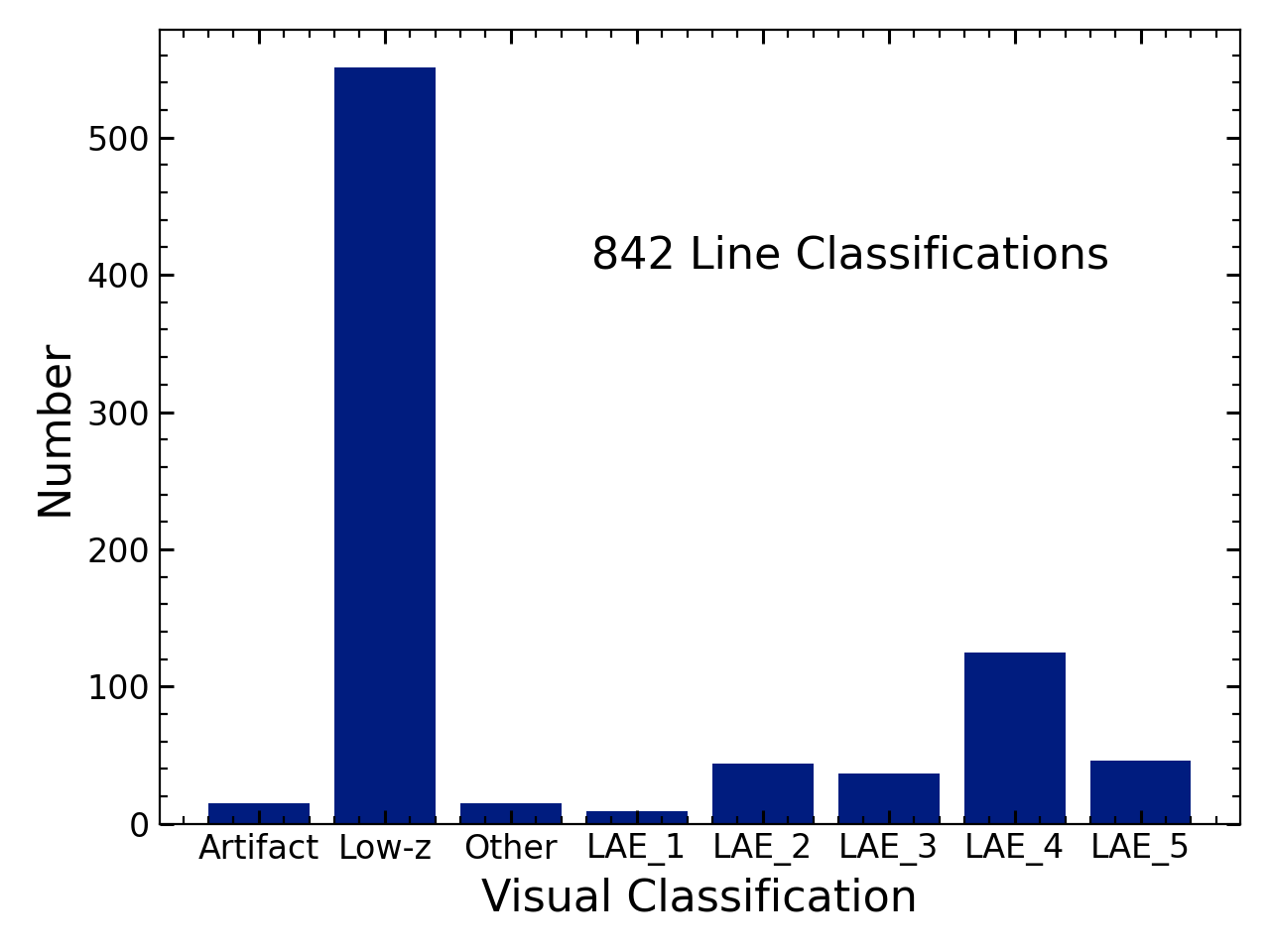}
    \caption{The distribution of visual classifications of candidate detections. Making cuts based on \plae\ to remove low-z sources can dramatically reduce the visual inspection workload.}
    \label{fig:class_dist}
\end{figure}

\begin{figure}[b]
    \centering
    \includegraphics[width=0.9\columnwidth]{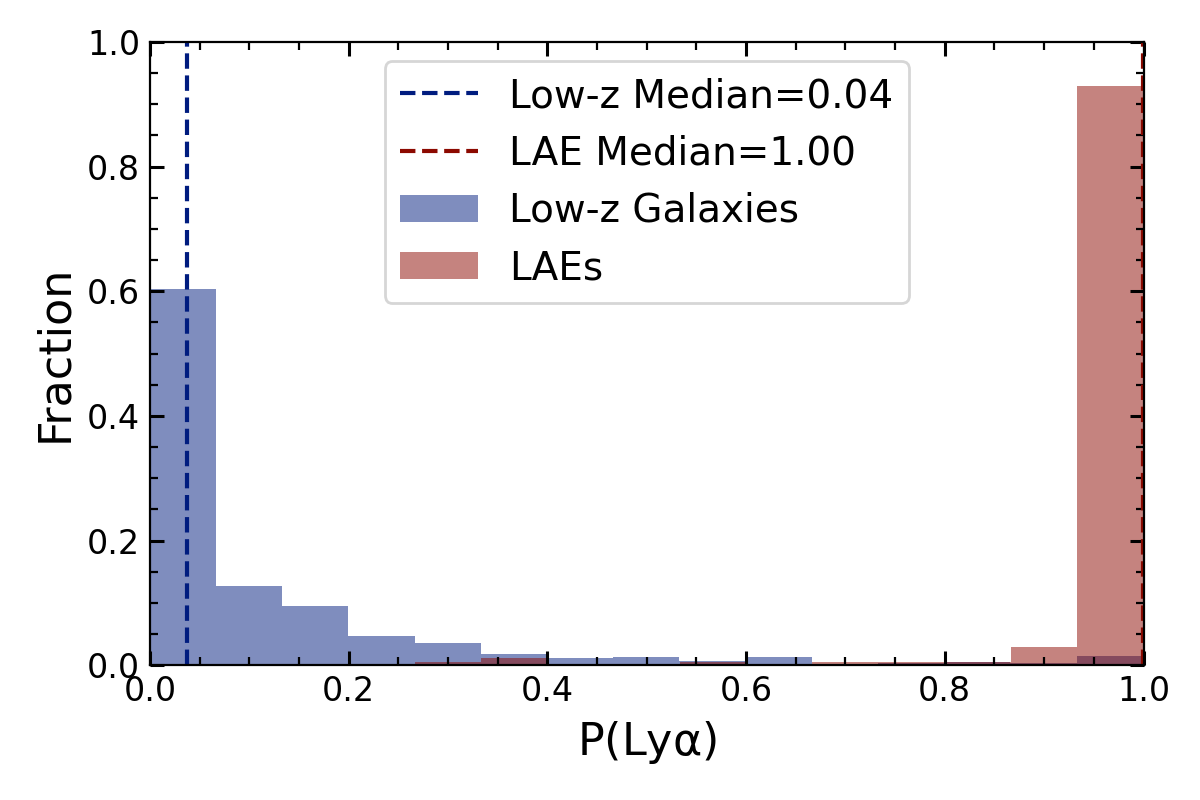}
    \caption{The distribution of \plae\ for detections visually classified as low-redshift galaxies (light blue) and high-confidence (scores of 4 and 5) LAEs (light red). Adopting a minimum threshold for \plae\ can remove a large fraction of low-redshift interlopers without eliminating very many LAEs.}
    \label{fig:plya_comp}
\end{figure}

After classifying each detection, we obtained $\sim 200$ detections categorized as high confidence \laes\ (scores of 4-5) and almost three times as many classified as low-$z$ sources (Figure~\ref{fig:class_dist}). Note that we did not include detections with scores of 3 or below for initial study as we want the cleanest sample possible. To assess the \hd\ collaboration's built-in Bayesian classification probability, \plae, we plotted that statistic for all of our detections classified as either low-$z$ galaxies or \laes. Figure~\ref{fig:plya_comp} shows that true LAE detections rarely score low in the \plae\ statistic, but a few low-$z$ sources can score in the intermediate range. For this reason, we suggest future studies can dramatically reduce the amount of visual inspections needed by adopting a cutoff of \plae $\gtrsim 0.6$ for \lae\ candidates.

To finalize our sample, we removed detections of the same source (since the \gn\ field was observed multiple times between 2018--2020), by selecting the highest \snr\ measurement of all detections grouped within 2$\arcsec$ and one spectral resolution element (6~\AA\null). Our final emission line sample consisted of \nlyalines\ high-confidence \lya\ detections (with classification scores of 4-5).

\subsection{Counterpart Identification}
\label{ssec:cptident}

In order to study the stellar populations of the \laes\ in our sample, we developed a method to match the un-targeted spectroscopic detections to counterparts in \hst\ imaging of the \gn\ field. 

The  overall astrometric precision of a \hd\ observation is $\sim$0.2\arcsec. However, due to the 1.5\arcsec\ diameters of the fibers, the typical seeing, and the 3-dither pattern, the position of an individual (faint) LAE is known to no better than $\sim$0.5\arcsec.   Since the HST images have a resolution that is $\sim$20 times higher than this, great care is needed to ensure an emission-line source is matched with the correct counterpart. 

We used the imaging obtained by the Great Observatories Origins Deep Survey \citep[GOODS][]{giavalisco04} with the optical ACS camera, and the Cosmic Assembly Near-infrared Deep Extragalactic Legacy Survey \citep[CANDELS][]{grogin11,koekemoer11} with the WFC3/IR infrared camera, using the internal CANDELS team's reduced mosaics for each filter.  This dataset consists of imaging in nine filters (F435W, F606W, F775W, F814W, F850LP with ACS, and F105W, F125W, F140W and F160W with WFC3/IR).  We made use of the photometric catalogs derived by \citet{fink21} which used \srcex\ \citep{sextractor} in two-image mode to create a F160W-selected catalog, coupled with using the Tractor \citep{lang16} to perform deblended photometry on the deep S-CANDELS \citep{ashby15} {\it Spitzer}/IRAC 3.6 and 4.5 $\mu$m imaging.  Further details on the cataloguing process are available in \citet{fink21}.  Similar to the widget used to classify detections as \lya\ , we created a visual inspection tool that provided information about the distance between the centroid of the HETDEX emission location and a given imaging source, the HETDEX emission-line strength when re-extracted centered at the imaging counterpart position, and the goodness of an SED fit assuming the \lya\ redshift, $z_{\mrm{Ly\alpha}}$.

Before selecting counterpart candidates, we optimized our search by developing a deep photometric catalog using a stacked image across all \hst \ filters in \gn. Each pixel value in this image and its error was computed using an inverse variance weighted average across $N=9$ filters with pixel value $p_i$ and rms error $\sigma_i$ given by Equation~\ref{eqn:weightedsum}. 
\begin{equation}
    \label{eqn:weightedsum}
    \bar{p} = \frac{\sum_i^N p_i \sigma_i^{-2}}{\sum_i^N \sigma_i^{-2}} \ , \ 
    \sigma_{\bar{p}} = \left( \sum_i^N \sigma_i^{-2} \right)^{-1/2}
\end{equation}
Since LAEs are often low-mass, faint systems, this stacked image improved our chances of identifying the continuum source corresponding to the detected emission line.

\begin{figure}
    \centering
    \includegraphics[width=0.9\columnwidth]{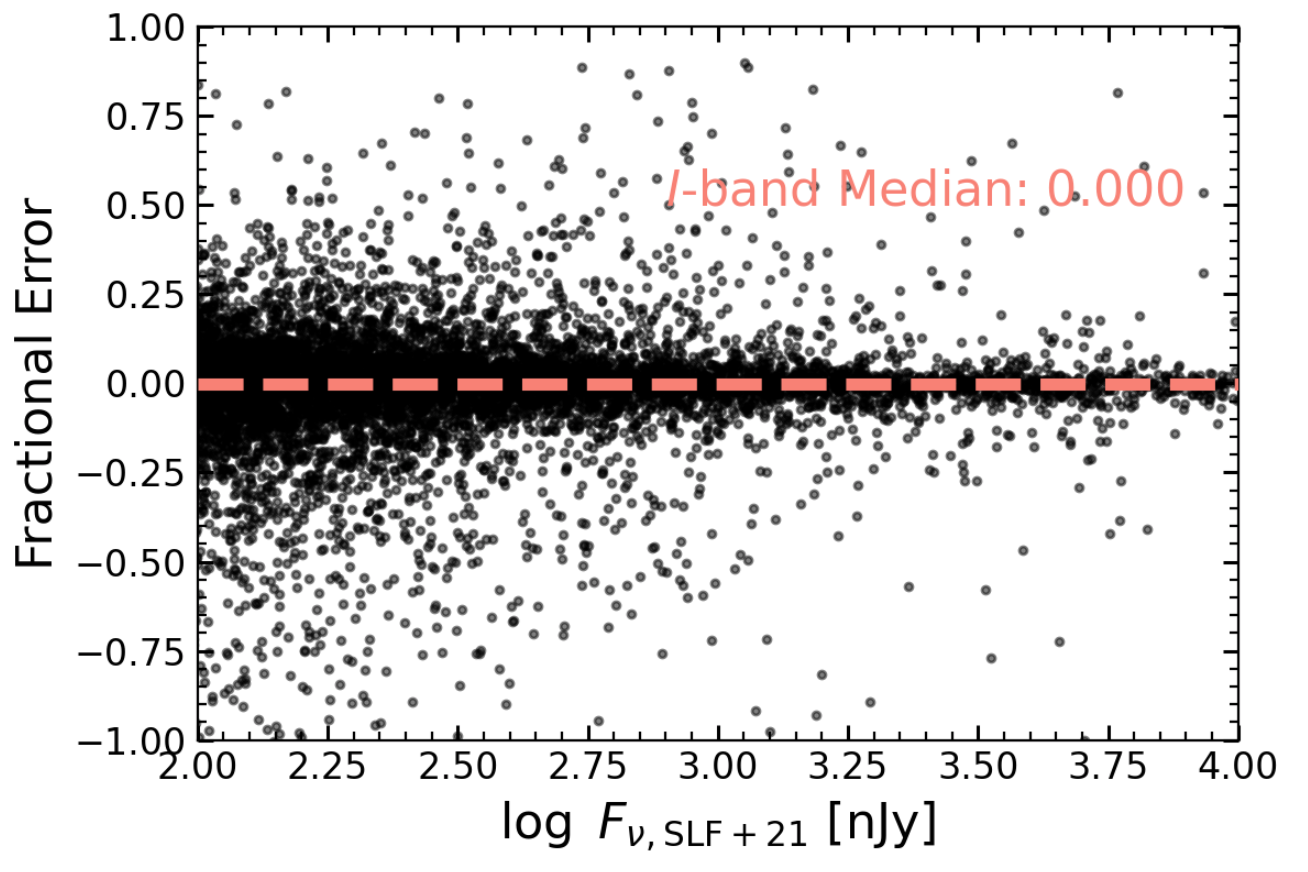}
    \caption{The fractional error between $I$-band (F775W) fluxes of sources measured in our derived stacked-detection catalog matched to sources in the original catalog from \citet{fink21} as a function of flux in the latter catalog. The median offset is indicated with the dashed pink line, and its value is given with text, showing good agreement between these two catalogs.}
    \label{fig:photflux}
\end{figure}

We then used \srcex\ \citep{sextractor} to detect the faintest possible sources in the stacked image, requiring a source to have 5 contiguous pixels with \snr$>1.6$. Following the procedures outlined in \citet{fink21} we used the same software in two-image mode to measure the flux in each filter and applied the appropriate aperture correction obtained from simulations. We performed extinction corrections using a Cardelli extinction law with $R_V=3.1$ for the Milky Way \citep{cardelli89}. We then compared the fluxes measured in this catalog to the F160W-selected catalog of \citet{fink21} and found the flux measurements to have no systematic offset and minimal scatter. Figure~\ref{fig:photflux} shows the fractional error of the stacked catalog photometry compared to the \citealt{fink21} photometry as a function of source brightness in the $I$-band. The median offset is zero with scatter of roughly 25\% for fluxes near 100 nJy, in agreement with the typical error bars for such sources, providing confidence in the fidelity of the stacked catalog. In all subsequent analysis, we defaulted to using measurements from the \citet{fink21} catalog for sources detected in both, and we only used photometry from the stacked catalog for five LAEs in our sample unique to it.

After generating the catalog from stacked imaging, we identified all imaging sources within $3\arcsec$ of the \hd\ detection position as possible \lae\ counterparts. Since the typical image quality of the HETDEX observations used here has a point spread function (PSF) of $\sim 1.7\arcsec$, the $3\arcsec$ annulus served as a generous aperture around the \lya\ centroid to encompass all possible counterparts for the detected emission. 

\begin{figure*}
    \centering
    \includegraphics[width=.7\textwidth]{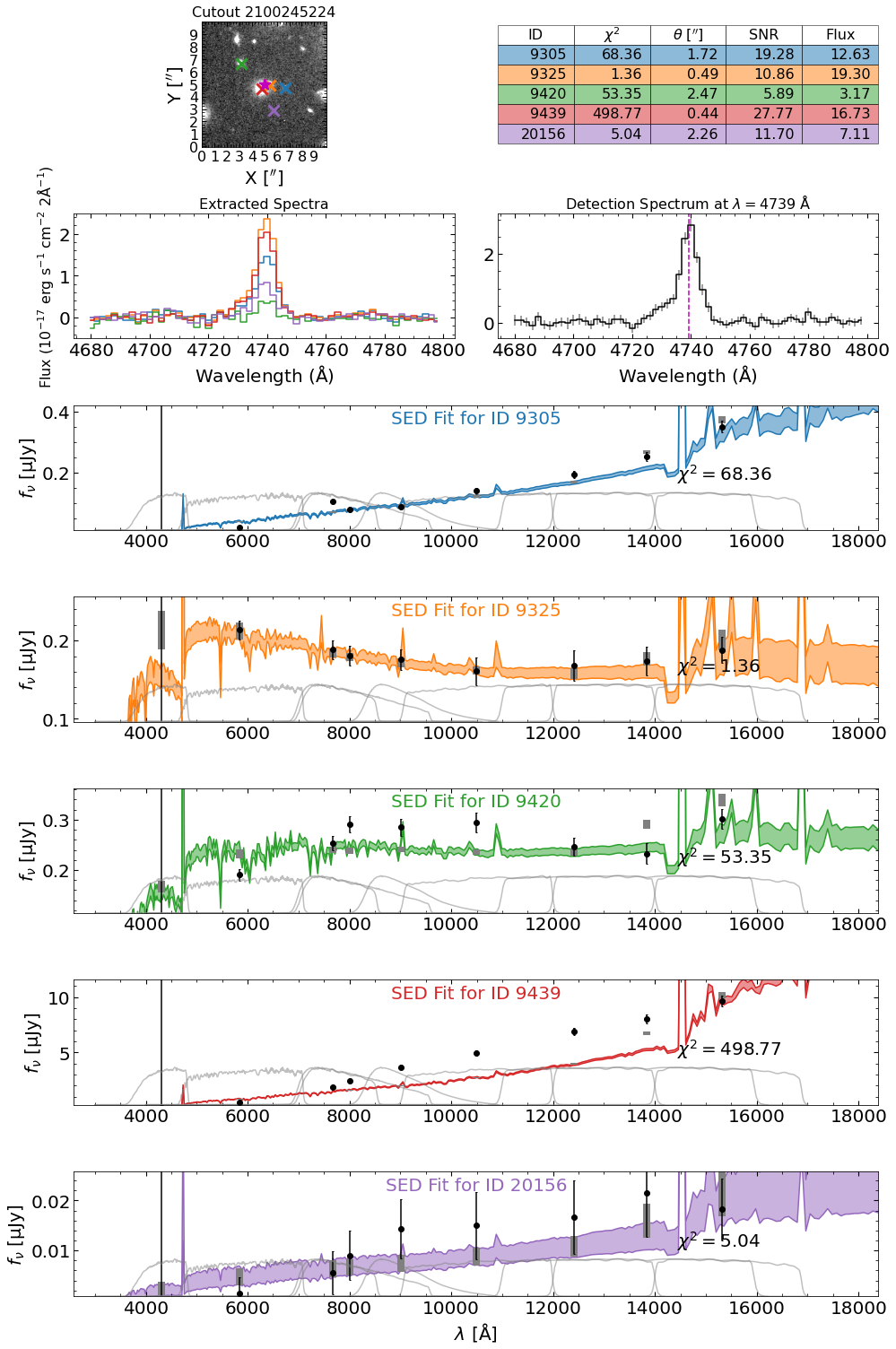}
    \caption{An example of the visualizations used to select imaging counterparts for \lya\ detections. The figure shows a 10$\arcsec$ image cutout with the detection positioned marked with a magenta star and sources marked with ``X''s of various colors, 1D spectra extracted at each source position, the original detection 1D spectrum, and SED fits (with redshift fixed assuming \lya\ emission) of all sources with significant measured fluxes. The table in the top-right contains the $\chi^2$ value of each SED fit, the separation between each source and the emission line detection position (labeled $\theta$), and the significance (labeled SNR) and the line flux in $\flux$ of the emission line extracted at the source position. The colors are consistent across all plots and tables, so each source corresponds to a unique color. The SED plots also contain normalized filter response curves as gray lines. In this case, while the red and orange sources have similarly small distances from the detection position and similar line fluxes, the SED fit $\chi^2$ strongly favors the orange source to be an LAE at $z=2.90$.}
    \label{fig:cpt_inspect}
\end{figure*}

We selected imaging counterparts based on the neighboring sources' angular distances from the detection, significance of emission extracted at the source positions, and goodness of SED fits performed by fixing the redshift assuming a \lya\ detection. First, we measured the on-sky angular separation from the detection position to the position of each possible source in the photometric catalog (labeled $\theta$ in Figure~\ref{fig:cpt_inspect}). Then, for each source, we used the \hd\ API script, \href{https://github.com/HETDEX/hetdex_api}{\tt get\_spectrum.py}\footnote{https://github.com/HETDEX/hetdex\_api}, to perform an aperture-weighted optimized spectral extraction (following \citealt{horne86}) at the source position to obtain a 1D spectrum. We created a Markov chain Monte Carlo (MCMC) line-fitting code using \emcee\ \citep{emcee} to fit a model to the feature to estimate its flux and significance. Our model consisted of two components: a linear trend with slope $m$ and intercept $b$, which captured any underlying continuum, and a Gaussian with total flux $F$ and standard deviation $\sigma$ to fit the line profile.

\begin{equation}
    \label{eqn:linefit}
    f_\lambda = m(\lambda - \lambda_0) + b + \frac{F}{\sqrt{2\pi} \sigma}
    \exp \left[- \frac{(\lambda - \lambda_0)^2}{2 \sigma^2} \right]
\end{equation}

In the model, $\lambda_0$, the wavelength of the emission line, was allowed to vary by $\pm$ one pixel (2\,\AA) from the detection wavelength reported by \hd\null. For each fit, we measured an effective \snr\ ratio (labeled SNR in Figure~\ref{fig:cpt_inspect}) by comparing the median value of the line flux to the standard deviation of the line flux for the last 20\% of the MCMC sampling chain, which had converged at that stage of sampling. To limit computation time for future counterpart identification steps, we ruled out any counterpart candidates that did not have any indication ($S/N > 1$) of an  emission feature at the pixel corresponding to the detected wavelength. Finally, for those sources with significant emission, we performed SED fitting with \bagpipes\ (see Section~\ref{ssec:sedfitting} for a full description of this procedure), fixing the redshift as $z_{{\rm Ly}\alpha}$. Our simple SED model for counterpart identification included free parameters for stellar mass, metallicity, dust extinction, and SFH, and we adopted the \citet{calzetti94} dust attenuation law, the \cite{chabrier03} initial mass function, and a delayed-$\tau$ SFH\null. At this stage, we did not include any IRAC fluxes in our fits since those fluxes depend sensitively on deblending, which is unreliable when sources are crowded. Furthermore, between $1.9 < z < 3.5$, there are no strong spectral features at the rest-frame wavelengths probed by IRAC, and  redshift-sensitive features such as the 4000\,\AA\ break are adequately covered by \hst.  We then visually inspected the separations, spectral extractions, and SED fits of all candidate counterparts to choose the one most likely to be the detected LAE. 

Figure~\ref{fig:cpt_inspect} shows an example of our approach. Separate sources are marked with an ``X'' and the color of the mark corresponds to the color of the table row, spectrum, and SED in the subsequent plots. In this case, the red and and orange sources within $0\farcs 5$ of the detection position (magenta star) show similar extracted emission line flux at the detection wavelength. Crucially, the SED fit for the red object poorly matches the data when fixing the redshift as $z_{Ly\alpha}$, but the orange object has a fit in excellent agreement with its observed SED based on the $\chi^2$ statistic. Therefore, in this example case we selected the orange object as the detected LAE at $z=2.90$. We followed the same process to identify counterparts for the other \lya\ lines in our sample. 

\begin{figure}
    \centering
    \includegraphics[width=0.8\columnwidth]{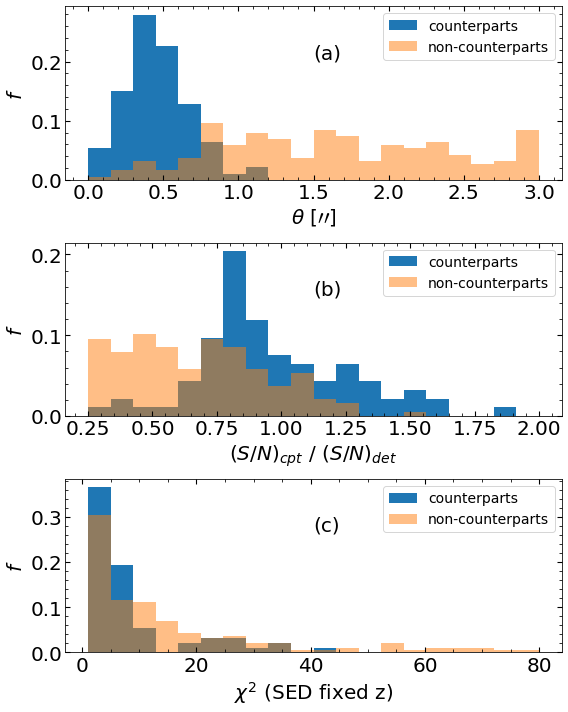}
    \caption{The distributions, expressed as the fraction of objects in a given bin, of (a) separation, (b) position-extracted emission line signal-to-noise (relative to that of the detection), and (c) SED $\chi^2$ assuming $z_{\mrm{Ly\alpha}}$, for objects identified as the detection imaging counterparts and those that happened to be spatially coincident. Histograms are normalized to the population size. The top panel indicated that finding a counterpart with an imaging separation larger than 1$\arcsec$ from the detection position is exceedingly rare.} 
    \label{fig:cpt_stats}
\end{figure}

\begin{figure}
    \centering
    \includegraphics[width=0.9\columnwidth]{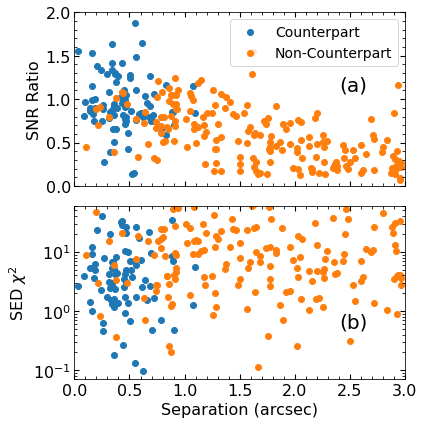}
    \caption{(a) The 2D distribution (in \snr\ and separation space) for objects identified as the imaging counterparts for emission line detections and those that happened to be spatially coincident. (b) The same plot in SED $\chi^2$ and separation space. These two figures show substantial overlap in these parameter spaces for true LAEs and neighboring sources, motivating the benefits of detailed visual inspection shown in Figure~\ref{fig:cpt_inspect}.}
    \label{fig:cpt_sep_snr}
\end{figure}

By studying the distribution of our counterparts in the parameter space of separation, signal-to-noise of emission, and SED $\chi^2$, we found no obvious way to select counterparts reliably based on these numbers alone, but we did find favorable regions. Figure~\ref{fig:cpt_stats}a shows the distribution of separation from the detection positions for sources we identified as \laes\ and sources that just happened to be nearby. Clearly, it was exceedingly unlikely that the true counterpart lay farther than 1$\arcsec$ away on-sky. For this reason, we could very reasonably shrink our selection criteria from all sources within 3$\arcsec$ to roughly 1$\arcsec$ without significant loss of \laes. In terms of emission line \snr\ (compared to the measured value of the detection itself), we found that, while typically the identified counterparts had stronger emission, the \hd\ PSF caused the extracted flux to not depend sensitively enough on position to clearly identify the counterpart for sources separated by less than 1$\arcsec$. This is clearest in Figure~\ref{fig:cpt_sep_snr}a, which shows that true counterparts and close neighbors show overlap in the \snr, separation plane. Note that the different on-sky centroids for emission line extraction between the counterparts and the original detection allow for the values of the \snr\ ratios in Figure \ref{fig:cpt_stats}b to be greater than unity. Finally, we note that, while most of the \laes\ in our sample had $\chi^2$ values in good agreement with the $z_\mrm{Ly\alpha}$ hypothesis, many neighboring galaxies also had low $\chi^2$, as shown in Figure~\ref{fig:cpt_stats}c. We attribute the low $\chi^2$ values for non-counterparts to our inclusion of such faint objects, which have large flux errors and are thus easily fit by a wide range of models.

After visually vetting all detections in our sample of \nlyalines\ \lya\ lines, we found 6 instances of detected emission with no continuum-detected counterpart. Since we could not study the properties of an LAE without photometry, we removed these objects from the final analysis. Furthermore, we removed 16 objects from the sample due to the following quality concerns. We eliminated the LAE corresponding to \hd\ detection ID 2100245124 (RA,DEC=189.346621$^\circ$,62.260662$^\circ$) from our sample as it was the only counterpart with an X-ray detection in the catalog of \citet{xue16}, indicating the galaxy hosted an AGN\null. Since our SED fitting code did not have an AGN template, we could not reliably report the physical properties of this object. We also eliminated the detection for ID 2100171783, as the counterpart inspection revealed the \lya\ emission line came from two probable LAEs separated by less than $0\farcs 5$ meaning we could not assign flux accurately to each source. Finally, we only analyzed objects detected in the $H$-band (F160W) of the \hst\ imaging as well as at least two bluer bands in order to span the rest-frame 4000\,\AA\ break at the sample redshift range. This serves as a crucial feature for constraining galaxy masses and ages with SED fitting (e.g., see \citealt{shapley03}). These choices limited our final sample size to \nsamp\ LAEs in \gn\ spanning \zrange{1.98}{3.48}. For 5 of these objects, photometry was not present in the catalog of \citet{fink21}, so we used photometry from the stacked catalog described in \S \ref{ssec:cptident}. Appendix \ref{sec:appendixb}, Figure~\ref{fig:all_lines} shows all the \hd\ emission lines for the LAEs in the final sample, and Figure~\ref{fig:hst_stamps} 
shows \hst\ imaging in F160W ($H$-band) for all objects.

\section{Analysis}
\label{sec:analysis}

After connecting \hd\ emission line detections with \hst\ imaging counterparts, we leveraged SED fitting to measure the galaxies' stellar population properties. From the SED fits and emission line detections, we also inferred the UV-slope  and \lya\ equivalent width.

\subsection{SED Fitting with \bagpipes} 
\label{ssec:sedfitting}

We fit all LAEs in our final sample with \bagpipes\ \citep{carnall18}, a flexible \python\ code that rapidly generates galaxy model spectra through stellar population synthesis using the 2016 version of the \citet{bruzual03} stellar spectral libraries. It explores the high-dimensional, multi-modal, and degenerate (e.g., age-dust-metallicity) model parameter space using the \multinest\ algorithm \citep{multinest}.

Our sample in \gn\ had photometry across nine \hst\ filters ranging from 0.4 to 1.6 $\mic$ as well as two \spitzer/IRAC channels centered at 3.6 $\mic$ and 4.5 $\mic$. Translating to the rest-frames of the objects in our sample at \zrange{1.9}{3.5}, these filters probed the UV, optical, and near-infrared (NIR) energy output of our objects. 

The filter coverage of our sample of \laes\  motivated our choice of SED modeling parameters. Table~\ref{table:sedparams} shows the names and units of the free parameters in our model, as well as the prior probability distributions assumed in our Bayesian framework. We adopted a delayed-$\tau$ SFH, defined as:
\begin{equation}
    \label{eqn:sfh}
    \mrm{SFR}(t) \propto
    \begin{cases} 
        (t-t_0) e^{-(t-t_0) / \tau} & t > t_0 \\
        0 & t < t_0
    \end{cases}
\end{equation}

This flexible SFH allows for star formation to be either rising, peaking, or falling, as opposed to the common exponentially declining model that only allows for falling SFRs over time. For example, \citet{lee10} found that SED fitting that adopted rising SFHs matched the stellar masses and SFRs from semi-analytic models for galaxies at \zrange{3}{6} better than exponentially declining models, while \citet{papovich11} found similar results favoring rising SFHs for real galaxies at $z =$ 4--7. We fit the $e$-folding scale of the SFH, $\tau$, the age of the Universe at the onset of star formation, $t_0$, the stellar mass formed, $M_\mrm{form}$, the global metallicity, $Z$, the dust extinction in the $V$-band, $A_V$, and the ionization parameter, $\log U$, defined as the log of the ratio of the number densities of ionizing photons and hydrogen atoms. Though we fit the total stellar mass formed by a galaxy, $M_\mrm{form}$,  we report its stellar mass at the redshift of observation excluding remnants, and we denote that stellar mass $M_\star$. We note that some of the parameters (namely $Z$ and $U$) are not expected to be well-constrained by our photometric data.  Nonetheless we allow them to vary within our imposed priors such that the uncertainties in the other parameters include the uncertainties in these parameters.  We adopted the \citet{calzetti94} dust attenuation law for star-forming galaxies and the \citet{chabrier03} initial mass function.

\begin{table}
    \centering
    \begin{tabular}{ |c|c|c|c| } 
        \hline
        Parameter & Prior & Bounds & Units \\ 
        \hline\hline
        $t_0$ & Uniform & 0, $T(z)$ & Gyr \\ 
        \hline
        $\tau$ & Uniform & 0.3, 10. & Gyr \\ 
        \hline
        $M_\mrm{form}$ & Log Uniform & $10^6$, $10^{12}$ & $\msun$ \\ 
        \hline
        $Z$ & Log Uniform & $10^{-5}$, 2 & $\zsun$ \\ 
        \hline
        $A_V$ & Uniform & 0, 2 & mag \\ 
        \hline
        $\log U$ & Uniform & -4, -2 & - \\
        \hline
    \end{tabular}
    \caption{Free parameters and their prior probability distributions for SED fitting. In our galaxy models, the redshift, $z$, was fixed based on the observed wavelength of \lya\ from \hd. $T(z)$ refers to the age of the Universe at redshift $z$. Note that we fit the cumulative stellar mass formed, $M_\mrm{form}$, from which the stellar mass (excluding remnants) at the object redshift was computed within the \bagpipes\ \citep{carnall18} code.}
    \label{table:sedparams}
\end{table}

All 11 filters were not necessarily included for every galaxy SED fit in our sample. For example, due to the large PSF of the IRAC imager, modeling sources in crowded fields of view and deblending the flux contribution of each source is crucial to accurately measuring the NIR fluxes of our LAEs. Although the catalog we used performed deblended photometric modeling with the IRAC PSF, this process can fail in crowded regions.  We thus visually inspected all IRAC residual maps for objects in our sample and removed the IRAC fluxes from our SED fitting if there were obvious problems in the deblending procedure. For the 5 objects not present in the catalog of \citet{fink21}, we did not have IRAC measurements. Furthermore, because the purpose of our analysis was to study the SED-derived properties of our \laes\ in relation to their \lya\ emission, we did not want \bagpipes's modeling of \lya\ emission or the IGM attenuation to bias our results. For this reason, we masked out all filters whose bandpass extended blue-ward of the observed \lya\ line; thus the $B$-band (F435W) and sometimes the $V$-band (F606W) was excluded, depending on redshift. 

\begin{figure*}
    \centering
    \includegraphics[width=0.9\textwidth]{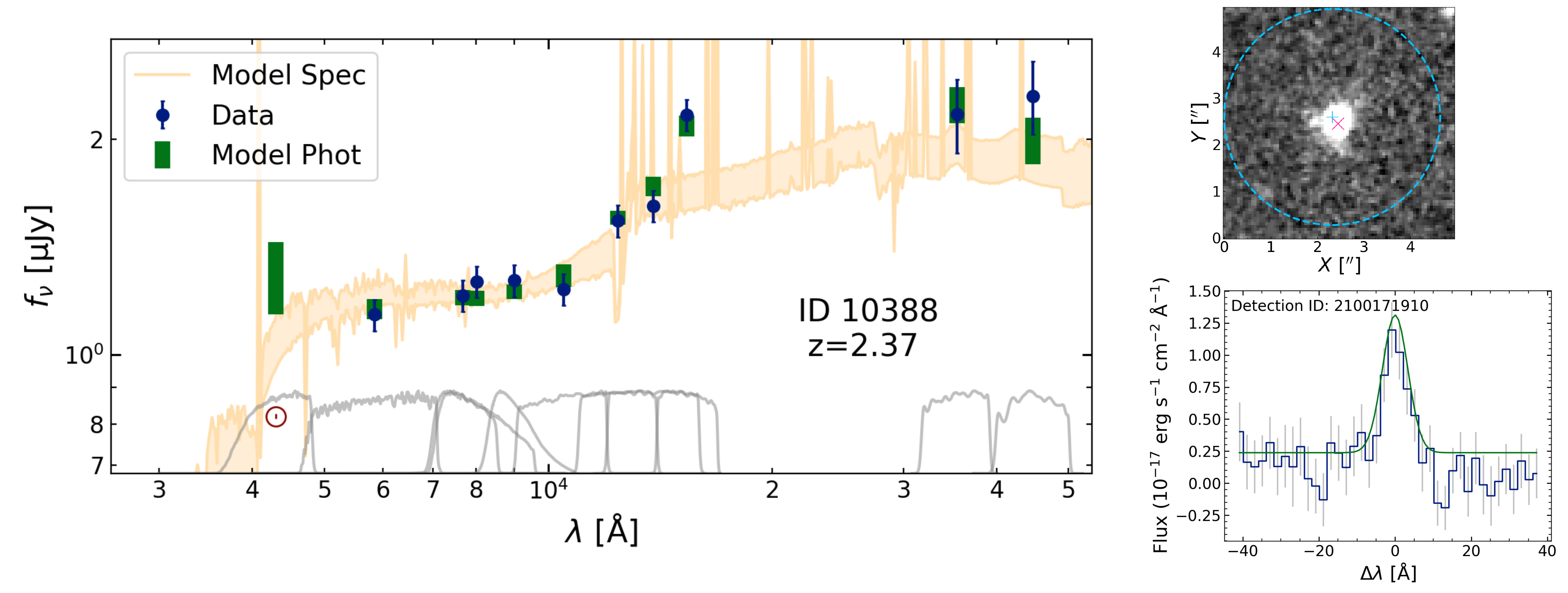}
    \caption{(\textit{Left}): An example fit to the SED of an LAE in the sample. Data are shown by blue circles, and the 68\% spread for the posterior model photometry and spectrum are shown by the green rectangles and orange shaded lines, respectively. A maroon open circle indicates the measured flux for a filter masked during SED fitting. In this case, the $B$-band was masked out since it includes the \lya\ emission line. Our imaging data constrain the $4000 \ \ang$ break and rest-optical colors. (\textit{Top-right}): $5\arcsec$ square image cutout for the source in the F160W \hst\ filter. The pink cross indicates the source position, and the blue plus sign and dashed blue circle indicate the detection position and the FWHM of the HETDEX fiber PSF. (\textit{Bottom-right}): 1D extracted spectrum for this source, centered on an $80 \ \ang$ window around the \lya\ emission line. The solid green line indicates the \hd\ Gaussian model fit to the data. }
    \label{fig:sedfit}
\end{figure*}

\begin{figure}
    \centering
    \includegraphics[width=\columnwidth]{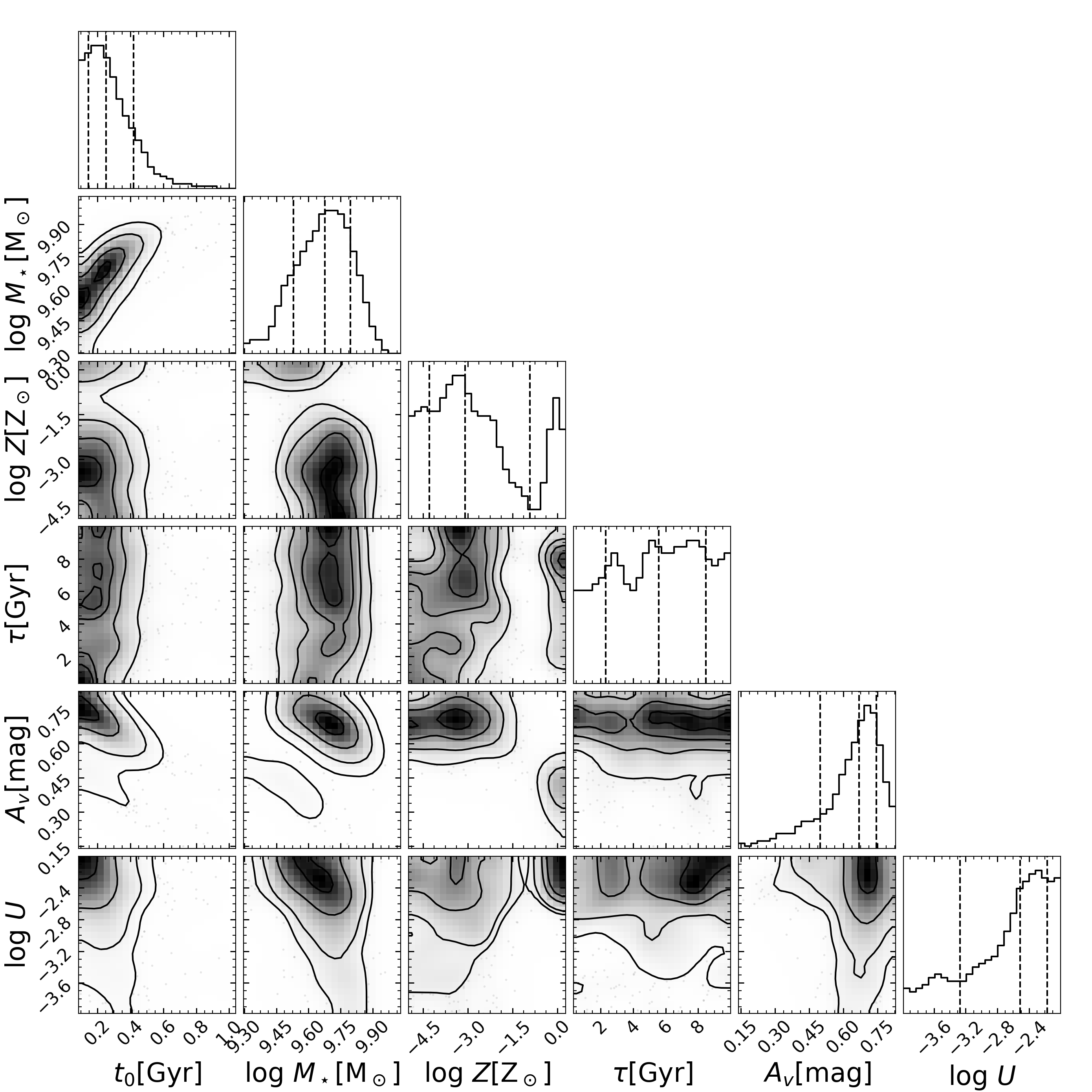}
    \caption{A ``corner'' plot of the fit in Figure~\ref{fig:sedfit} for object ID 10388. The 1D histograms are shown on the diagonal for the posterior distribution of each free parameter in our model (see Table~\ref{table:sedparams}). The 2D histograms show the correlations of all parameters with one another, where  contour lines are drawn for each $\sigma$ level. With our broadband photometry, we constrained ages, masses, and dust extinctions well.}
    \label{fig:corner}
\end{figure}

Figure~\ref{fig:sedfit} shows an example \bagpipes\ SED fit for an \lae\ in our sample. We plotted the $1\sigma$ spread on the model photometry as rectangles as well as the $1\sigma$ spread on the underlying model spectrum computed by evaluating the 16th and 84th percentiles of the posterior models. In this example, the fit did an excellent job matching salient features like the rest-frame 4000\,\AA\ break and nebular emission in the rest-frame optical region. We estimated galaxy properties using the posterior distributions for all free parameters explored by \bagpipes. Figure~\ref{fig:corner} shows an example ``corner'' plot (produced via \citealt{corner}), where all free parameters are plotted against each other for easy assessment of constraints and correlations. Stellar mass, time since the onset of star formation, and dust extinction were constrained well, while metallicity, $\tau$,  and ionization parameter were not well-constrained by our broadband photometry data. Figure~\ref{fig:all_sed_2} in Appendix \ref{sec:appendixb} shows the SED fits for all \nsamp\ LAEs in our final sample.

\subsection{Measuring $\ewlya$ and $\beta$}

Emission line strengths can be represented by the parameter equivalent width (EW or $\ew$), which represents the width of a rectangle drawn to the same height as the continuum needed for the rectangular area to match the area under the emission line. To estimate the equivalent width of \hd\ \lya\ detections, we used the measured line flux and error from the internal  \hdrb\ catalog computed by optimally extracting flux from all fibers within a 3.5 \arcsec~ radius circular aperture (roughly 15-20 individual fiber spectra)  contributing to the emission line detection (following \citealt{horne86}), weighted by the PSF of a point-source. We approximated the continuum flux density using the \bagpipes\ sampled model spectra from the SED fit. We took the continuum flux density to be the median value of all 500 sampled spectra averaged between 1250 and 1300\,\AA\  in the given object's rest-frame, and we computed the $1\sigma$ error using half the spread between the 16th and 84th percentiles of those values. This method allowed us to take advantage of complex computations performed by \bagpipes\ to get a statistically representative estimate of the continuum flux density instead of using a coarse approximation based off the flux in one of our photometric bands. We evaluated the \lya\ flux and the continuum flux density in the observer-frame and translated to the galaxy rest-frame by dividing by a factor of $(1+z)$ using the detected wavelength of \lya. 

\begin{equation}
    \label{eqn:eqwidth}
    \ewlya =  \frac{F_{\mrm{Ly\alpha}}}{f_{\lambda}} (1+z)^{-1}
\end{equation}

We measured  $\beta$, the UV continuum slope (un-corrected for dust), using the model spectra for galaxies in our sample following the method described in \citet{fink12a}. We masked the stellar and interstellar absorption features in the rest-frame UV using the windows provided by \citet{calzetti94}, and we fit a linear model to the spectrum in log space ($\log f_\lambda = \beta \log\lambda + C$) using \textit{polyfit} from the \python\ package \code{Numpy} \citep{numpy}. We determined $1\sigma$ uncertainties on $\beta$ for each object by measuring the distribution of values fitted to 500 spectral models sampled from the posterior by \bagpipes.

\section{Results} 
\label{sec:results}

We measured various physical properties of objects in our \lae\ sample using the posterior distributions returned by \bagpipes' exploration of the parameter space. We took the 16th and 84th percentiles of the posterior distributions to represent the error bars on physical properties. Examples of such measurements are shown in Figure~\ref{fig:corner} for a representative LAE in our sample. 

\subsection{SED-Derived Properties}
\label{ssec:sedprops}

\begin{figure*}
    \centering
    \includegraphics[width=\textwidth]{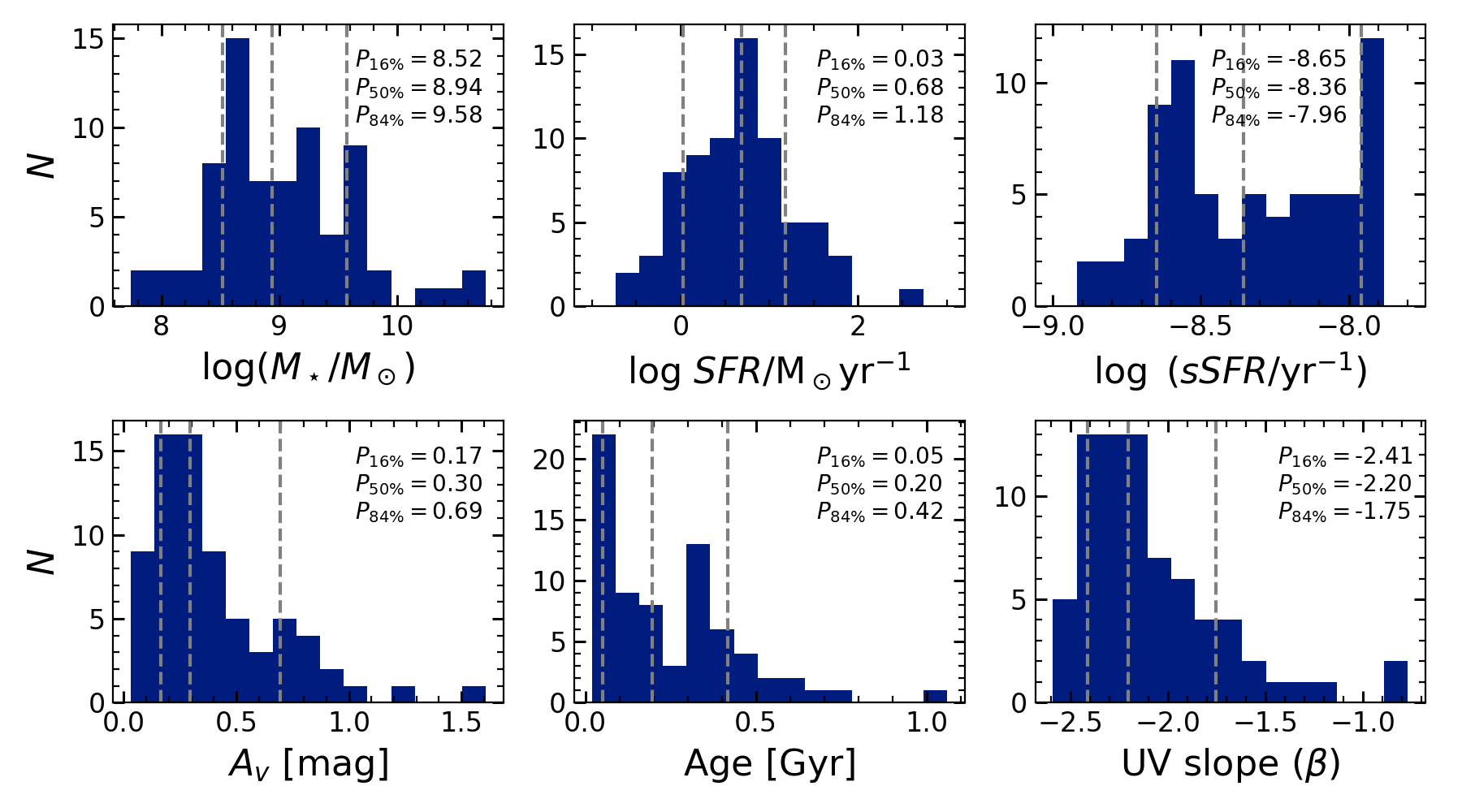}
    \caption{The distributions of posterior median values for (clockwise from top left)  stellar mass, SFR, sSFR, dust extinction (in $V$-band mag), mass-weighted age, and UV-slope for all objects in the sample. The 16$^{th}$, 50$^{th}$, and 84$^{th}$ percentiles are indicated by vertical dashed grey lines, and their values are indicated with text in the same units as the x-axis labels. The \laes\ in our sample exhibit average properties similar to other \lae\ samples compiled at comparable redshifts using narrow band selection.}
    \label{fig:distributions}
\end{figure*}

Figure~\ref{fig:distributions} shows the 1D distributions of posterior median values of stellar mass ($M_\star$), star formation rate (SFR), specific star formation rate (sSFR), dust extinction ($A_V$), mass-weighted age, and UV-slope ($\beta$) for all objects in our final LAE sample. We found the median stellar mass of our \hd\ \laes\ to be $0.8^{+2.9}_{-0.5} \times 10^{9} \ \msun$. This stellar mass value lies near the median masses of \laes\ selected in narrowband imaging surveys covering redshifts similar to this study (e.g. \citealt{guaita11}, \citealt{gawiser07}, \citealt{vargas14}, \citealt{kusakabe18}, \citealt{santos20}) and well below typical masses of Lyman-break selected objects (e.g. \citealt{shapley03}, \citealt{papovich01}, \citealt{trainor19}), which often have minimum masses an order of magnitude larger due to the depth of the broadband imaging used in their selection. 

We used our SED fitting procedure to obtain the attenuation in the $V$-band of starlight due to dust for galaxies in the sample, and we obtained a median value of $A_V = 0.3^{+0.4}_{-0.1}$ mag. The presence of dust has been measured in many other samples of \laes\, with values of $A_V$ or $E(B-V)$ often falling within a factor of two of this study (e.g. \citealt{guaita11}, \citealt{fink09}, \citealt{hathi16}, \citealt{kusakabe18}, \citealt{matthee21}).

Similar to dust reddening, our \lae\ sample has similar ages and star formation rates to \lae\ samples in the literature compiled using narrow band or continuum selection methods. Our SED-derived mass-weighted ages, typically spanning 0.05-0.5 Gyr, broadly agree with the narrow band samples of \citet{acquaviva11}, \citet{fink09}, \citet{gawiser07}, and \cite{vargas14}. Our median SFR, $4.8^{+10.4}_{-3.8} \mathrm{M_\odot / yr}$, falls near values reported by \citet{gawiser07}, \citet{hathi16}, and \citet{kusakabe18}, but falls 1 dex above the median SFR for \laes\ found in the MUSE HUDF Survey \citep{feltre20}. This discrepancy does not surprise us since the MUSE HUDF LAE sample had a median mass roughly 0.5 dex lower than this study, and their sample spanned \zrange{2.9}{4.6}, probing an era of lower star formation activity in the Universe than the one studied here (see \citealt{madau14}). 

Our model included stellar metallicity and ionization parameter as free parameters, but our broadband photometric data could not constrain those values precisely (see Figure~\ref{fig:corner}), since reliable estimates typically require sensitive emission line diagnostics (e.g. \citealt{reddy21}), which were coarsely probed at best by our filter set. For this reason, we do not present or discuss our galaxies' metallicities or ISM ionization conditions, but we note that by letting these parameters vary, our posterior constraints on all other parameters include the uncertainties in these quantities.

\subsection{$\ewlya$ Distribution}
The equivalent width distribution of LAEs has been modeled by various authors as exponential with the form given by Equation~\ref{eqn:ewdist} (e.g. \citealt{gronwall07}, \citealt{guaita10}, \citealt{wold14}, \citealt{jung18}).

\begin{equation}
    \label{eqn:ewdist}
    \frac{\mrm{d}N}{\mrm{d}\ew} \propto e^{-\ew / W_0}
\end{equation}

\begin{figure}
    \centering
    \includegraphics[width=0.9\columnwidth]{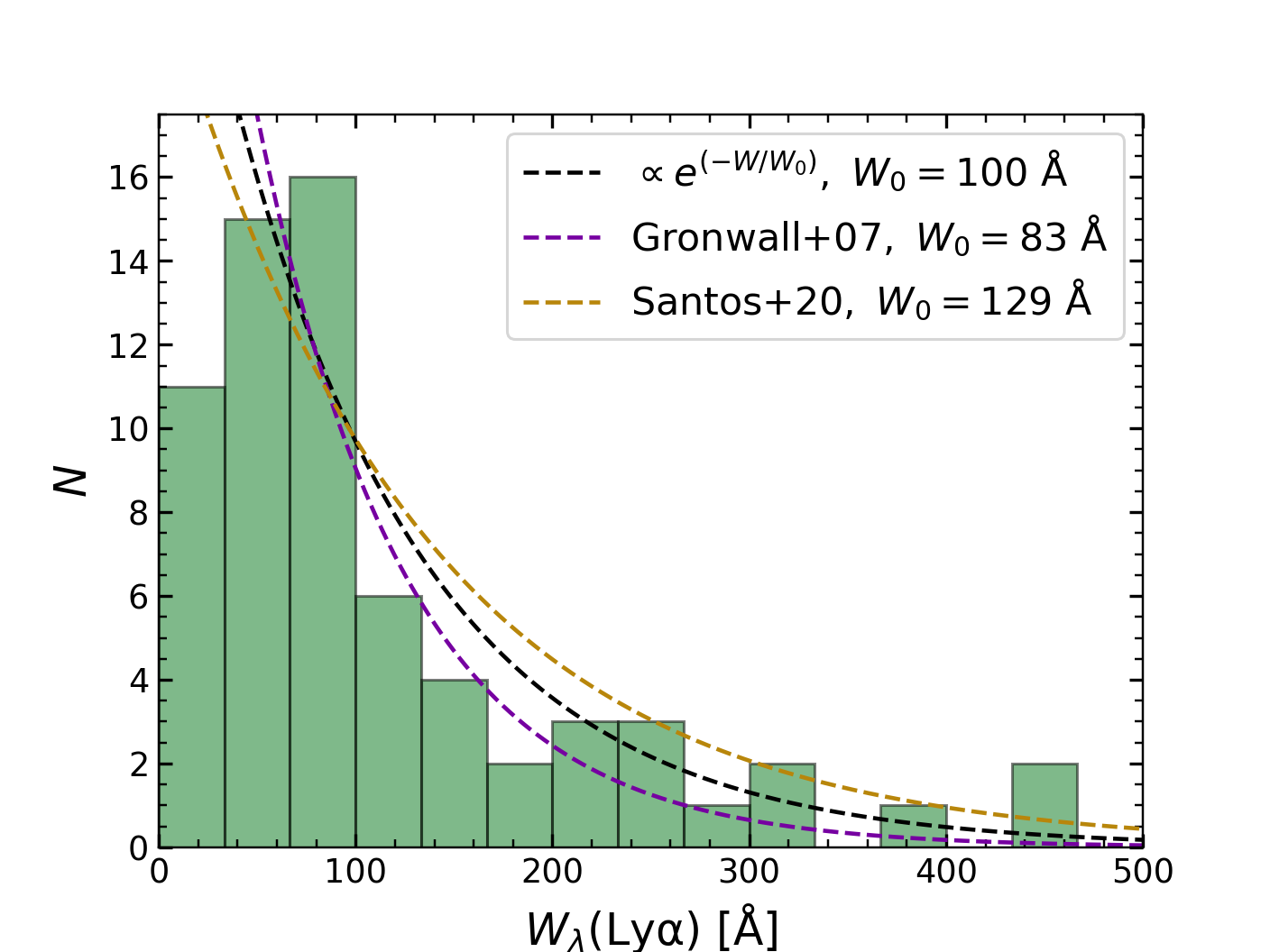}
    \caption{The equivalent width distribution of LAEs in the sample. An exponential distribution with $W_0=100\ \ang$ is drawn in red for comparison, as well as models fit by \citet{gronwall07} and \citet{santos20}. Our data favor models with larger values of $W_0$ to best match the number of high-EW sources.}
    \label{fig:ew_dist}
\end{figure}

We show our sample's rest-frame $\ewlya$ distribution in Figure~\ref{fig:ew_dist} with an $e$-folding scale $W_0=100 \ \ang$ drawn for comparison. We cannot measure the underlying distribution for LAEs from our sample since we have not measured the completeness as a function of equivalent width (which is complex due to our method of sample creation, and not crucial for our study of stellar population properties). Various other studies have precisely measured the Lyman-alpha equivalent width distribution, such as \citet{gronwall07}, who found an $e$-folding scale of $76^{+11}_{-8} \ \ang$ for a deep, narrow-band (NB) selected \lae\ sample at $z=3.1$, \citet{guaita10}, who measured $W_0=50 \pm 7 \ \ang$ for a NB sample at $z=2.1$, and recently \cite{santos20}, who measured $W_0=129 \pm 11 \ \ang$ for the full SC4K sample at \zrange{2}{6}. We plot some of these measured distributions in Figure~\ref{fig:ew_dist} for comparison. It is apparent that our sample becomes increasingly incomplete at EW $\lesssim$ 50 \AA, due to a combination of the HETDEX flux limit, the emission-line identification process, and our counterpart selection process.

\subsection{Correlations between $\ewlya$ and Galaxy Properties}
\label{ssec:corrs}

\begin{figure*}
    \centering
    \includegraphics[width=\textwidth]{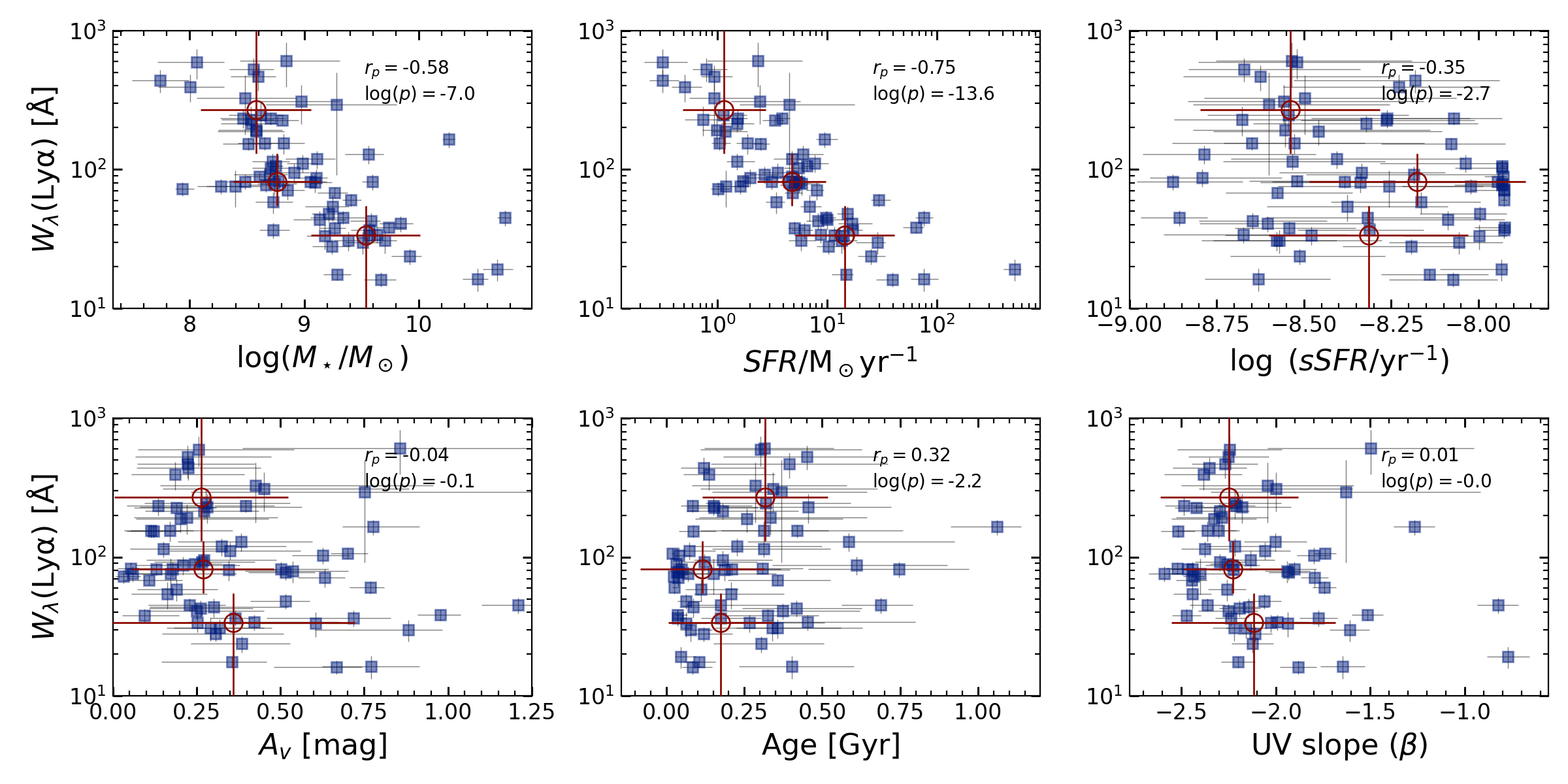}
    \caption{The relationship between $\ewlya$ and (clockwise from top left) stellar mass, SFR, sSFR, dust extinction (A$_V$), mass-weighted age, and UV-slope ($\beta$) for all objects in the sample. Grey lines indicate the $\pm 1\sigma$ error bars on physical properties and equivalent width. Open red circles show medians and standard deviations for properties in equivalent width bins having equal numbers of objects. Pearson correlation coefficients, $r_p$, and $p$-values are indicated for each plot. Stellar mass and SFR exhibit the strongest correlations with $\ewlya$, while age and sSFR correlate moderately. Surprisingly, no strong correlation exists with dust extinction.}
    \label{fig:correlations}
\end{figure*}

We combined our SED-derived galaxy properties with the $\ewlya$ measurements described above in order to assess correlations between \lya\ emission and global galaxy properties. We used $\ewlya$ as a proxy for the fraction of photons emitted as \lya\ as opposed to $L_{\mathrm{Ly\alpha}}$, for example, because the equivalent width more closely probes the physics governing \lya\ escape, whereas the flux also includes physics related to the \lya\ production rate. Figure~\ref{fig:correlations} shows $M_\star$, specific star formation rate (sSFR), star formation rate (SFR), dust extinction ($A_V$), mass-weighted stellar population age, and UV-slope ($\beta$) plotted against each galaxy's $\ewlya$ measurement. In the figure, error bars denote the 16th to 84th percentile range, and we indicate Pearson's linear correlation coefficient, $r_p$, and its significance ($p-$value) with text. 

Stellar mass and star formation rate both correlate strongly with $\ewlya$, with low mass, low SFR systems achieving larger $\ewlya$ than higher mass systems. The correlation with mass has been established in the literature from studies of a wide variety of galaxies such as LBGs, oELGs, and LAEs. It was noticed early by \citealt{ando06} and measured recently by many works such as \citealt{du18}, \citealt{marchi19}, \citealt{oyarzun17}, and  \citealt{shimakawa17}. Specifically, \citet{weiss21} found a negative correlation between the \lya\ escape fraction, \fesc, and stellar mass using data from the \hd\ survey. Additionally, \citet{khostovan21} found an intrinsic, negative correlation between H$\alpha$ equivalent width and galaxy stellar mass from a NB survey at $z \sim 5$. While the lack of low-EW, low-mass systems can be driven by selection incompleteness, we should be complete to high-mass, high-EW systems, yet these are seemingly rare.

Notably, our results show no significant anti-correlation between $\ewlya$ and dust extinction ($A_V$), whereas numerous other studies of Lyman-alpha emission measured a clear relationship that indicates dust hinders the ability of the \lya\ photon to escape the galaxy. For example, \citet{shapley03}, \citet{guaita11}, \citet{du18}, \citet{hathi16}, \citet{huang21}, \citet{marchi19}, \citet{matthee16}, \citet{reddy21}, \citet{trainor19}, and \citet{weiss21}, all showed that dustier galaxies exhibit weaker Lyman-alpha emission measured as $\ewlya$ or have smaller \fesc.  However, the lack of a significant anti-correlation may be due to our limited sample size and small dynamic range in dust attenuation. Moreover, objects with significant amounts of dust that suppress their \lya\ fluxes would not become members of our science sample in the first place.  The majority of our sample has $A_V <$ 0.3.  We do observe multiple galaxies with $A_V >$ 0.5, and interestingly these do not all have low $\ewlya$, implying that \lya\ can escape even from modestly dusty galaxies, which could indicate enhanced escape due to outflows (e.g., \citealt{steidel10}, \citealt{erb12}) or a multi-phase ISM (e.g. \citealt{fink09}, \citealt{neufeld91}).

Our Pearson correlation coefficient suggests a moderate correlation between $\ewlya$ and galaxy stellar mass-weighted age ($r_p=0.32$) in the sense that older galaxies exhibit larger $\ewlya$. \citet{marchi19} found a similar result, obtaining a Spearman rank correlation coefficient of 0.40. This contrasts with \citet{pentericci09} and \citet{pentericci10} who found no strong dependence of \lya\ equivalent width on age for \laes\ and LBGs, as well as \citet{reddy21} who found a weak negative correlation between the two measurements for star-forming galaxies in the same redshift range probed by this study.

Finally, a moderate negative correlation exists between sSFR and $\ewlya$, though the large error bars for our measurements of sSFR weaken the reliability of the correlation. For comparison, \citet{hathi16} found no significant correlation between the two properties for a sample including \lya\ in absorption and emission.

\begin{figure}
    \centering
    \includegraphics[width=0.9\columnwidth]{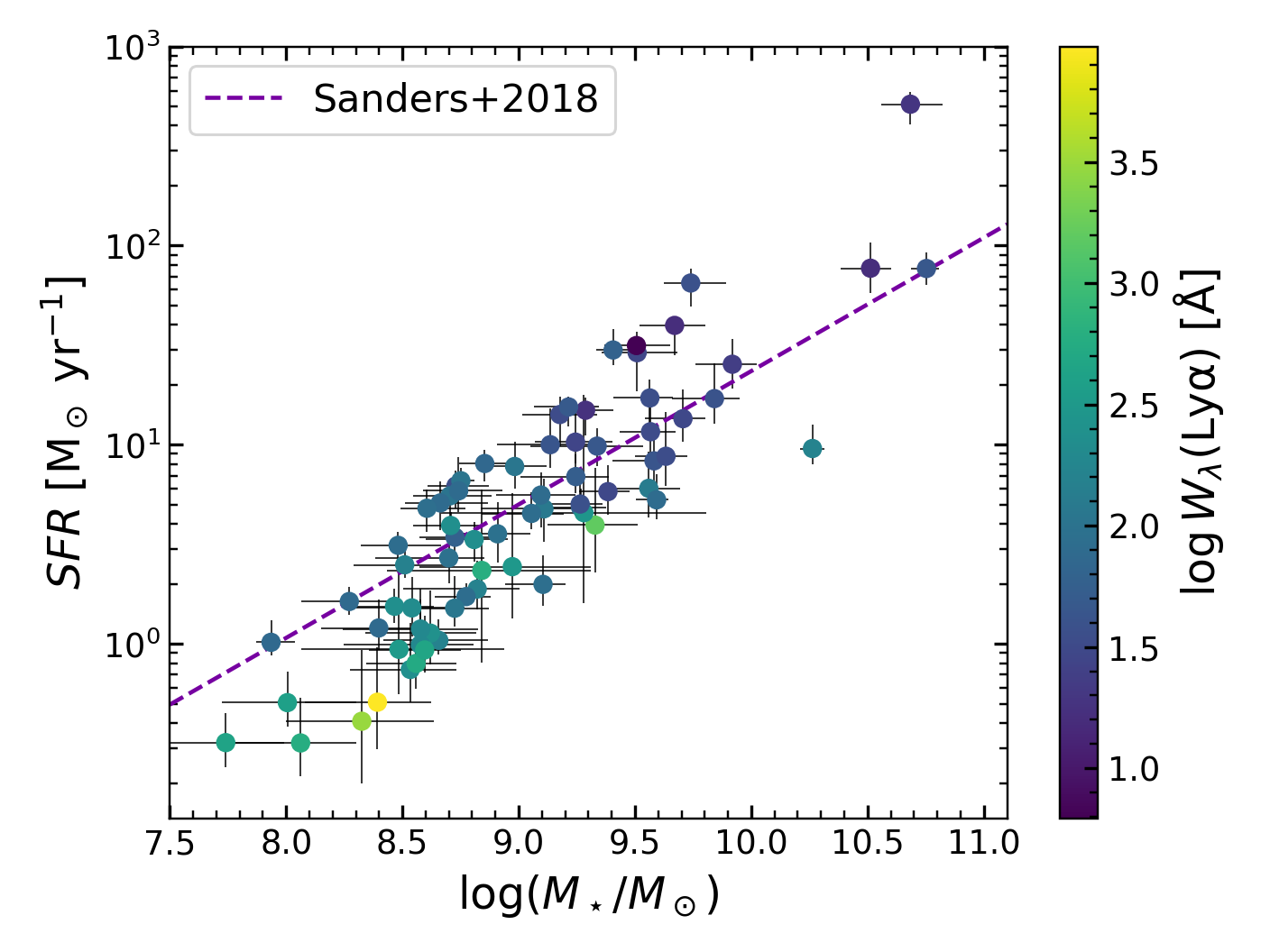}
    \caption{The stellar mass - star formation rate correlation for LAEs in our sample. The trend fit by \citet{sanders18} for $z \sim 2.3 $ star-forming galaxies is drawn in dashed purple for comparison. LAEs in our sample largely fall on the SFMS, though the lowest mass sources ($M_\star < 10^9 \ \msun$) tend to fall below the relation.}
    \label{fig:sfms}
\end{figure}

We also plot SFR against $M_\star$ for all objects in our sample in Figure~\ref{fig:sfms} to see how our galaxies compare to other objects at similar redshift in relation to the star-forming main sequence (SFMS). We include the best-fit line found by \citet{sanders18} for star forming galaxies in the MOSDEF survey at $z \sim 2.3$. Note that masses derived for that study used the \cite{chabrier03} IMF and \cite{calzetti00} dust curve but stellar population synthesis models from \cite{conroy09}. We also use a colorbar to show the value of $\ewlya$ for each galaxy. The position of LAEs on the SFMS remains somewhat controversial. Studies such as \citet{vargas14}, \cite{keely15}, \citet{hagen16}, and \cite{santos20} found LAEs to lie above the relation, while other studies have interpreted them as lying directly on the low-mass end of the relation (e.g. \citealt{kusakabe18}). Figure~\ref{fig:sfms} shows that the LAEs in our sample lie largely on the SFMS, though a significant fration lie below the relation of \cite{sanders18} for $M_\star < 10^9 \ \msun$.

In Appendix \ref{sec:appendixa}, we explore the model-dependence of our measured galaxy properties, since the parameters derived from SED fitting can be systematically different using different models (see \citealt{conroy13}). We conclude that our results, including the median physical properties and the correlations with $\ewlya$ are not driven by our specific choice of model.

\section{Discussion}
\label{sec:discussion}

\subsection{Are HETDEX \laes\ Special?}

The question, ``What is a HETDEX LAE?'' holds particular importance for astronomers studying galaxy science with this survey. A vast sample of HETDEX LAEs is upcoming, and samples of such objects selected by emission line detection from a blind spectroscopic survey remain rare in the literature (with the exception of the HETDEX Pilot Survey (\citealt{adams11}, \citealt{blanc11}), which probed a smaller area to a brighter flux limit, and MUSE surveys, which probe much smaller areas to fainter flux limits with only a small overlap in redshift with \hd). Characterizing any idiosyncrasies in the HETDEX LAE population will put these objects in context relative to the numerous LAEs found by previous studies, and it will aid the interpretation of future blind spectroscopic surveys for these objects in the EoR. 

As described above, in our f$_{Ly\alpha}$ $\gtrsim 6 \times 10^{-17}\ \flux$ flux-limited sample \citep{gebhardt21},
the median galaxy mass of $0.8^{+2.9}_{-0.5} \times 10^{9} \ \msun$ lies very close to many LAE samples selected through narrow band imaging. For example, \cite{gawiser07} found a median mass of $1^{+0.6}_{-0.4} \times 10^9 \ \msun$ with a flux limit of $1.5 \times 10^{-17} \mrm{erg \ s^{-1} \ cm^{-2}}$ at $z=3.1$. \citet{guaita11} pushed to an even lower median mass of $\sim 4 \times 10^8 \ \msun$, roughly a factor of two less massive than this sample's median, with a flux limit of $2.0 \times 10^{-17} \ \flux$ at $z=2.1$. The MUSE HUDF went even deeper, finding sources at $z>3$ with \lya\ line fluxes as small as $\sim2 \times 10^{-18} \ \flux$ and obtaining a median sample mass of $\sim 2.5 \times 10^8 \ \msun$. The sample of \citet{santos20} was limited by medium-band line flux limits spanning $3.0-4.8 \times 10^{-17} \ \flux$ over \zrange{2}{6} \citep{sobral18} and measured a median \lae\ mass of $~2 \times 10^9 \ \msun$, consistent with this study. Of course, the mass range probed by HETDEX falls far below samples selected using the Lyman/Lyman-alpha break (for example, the lowest mass probed by \citet{papovich01} was $10^{10} \ \msun$ at \zrange{2.0}{3.5}). Thus, the HETDEX flux limit explores an \lae\ mass range comparable to NB surveys, yet slightly more massive than the deepest NB and spectroscopic surveys. At the expense of sensitivity, the HETDEX survey can find fairly low-mass LAEs over a large continuous redshift interval, reducing the effects of cosmic variance compared to NB observations. 

As mentioned in \S\ref{ssec:sedprops}, the LAEs in this sample do not stand out from NB samples at similar redshift in terms of age, star formation rate, and dust extinction. Thus, we can conclude that the HETDEX survey selects a typical LAE having properties consistent with the general NB-selected population, but it may have slightly higher stellar mass based on the line flux limit of the survey. 

Nonetheless, our sample may stand out in its relation to the SFR-$M_\star$ relation shown in Figure~\ref{fig:sfms}. Compared to the relation measured in \citet{sanders18}, LAEs in the sample with $M_\star \lesssim 10^9 \ \msun$ appear to lie below the trend. This contrasts markedly with the work of \cite{hagen16}, who compiled their sample using the HETDEX Pilot survey (\citealt{adams11}, \citealt{blanc11}) and found their LAEs to lie above the SFMS. Interestingly, the LAEs lying below the SFMS in Figure~\ref{fig:sfms} have very high $\ewlya$, which correlates with lower $M_\star$ and SFR in Figure~\ref{fig:correlations}. We are not surprised that the lowest mass systems in our sample have the highest values of $\ewlya$ given the negative correlation with $M_\star$ and the fact that low mass objects need large $\ewlya$ to be detected by \hd, but their position below the SFMS is peculiar. It could be related to the weak negative correlation we found between $\ewlya$ and sSFR, or could simply be an artifact of our small sample size. This motivates further study of the positions of LAEs on the SFMS with larger samples.

\subsection{Which Properties Drive \lya\ Emission?}

While the size of the sample analyzed in this study is small, we were still able to extract important information linking galaxy stellar-population properties to \lya\ emission strength. As the number of LAEs detected by HETDEX grows in fields with rich photometric data, such as the Spitzer-HETDEX Exploratory Large-Area Survey (SHELA) \citep{papovich16}, the number of LAEs with measured galaxy properties will grow by many orders of magnitude.  This will provide a trove of useful data for explaining why some galaxies shine brightly in \lya\ while others do not, as well as exploring the effects of galaxy environment on \lya\ emission. 

We found a significant, strong negative correlation between $\ewlya$ and stellar mass in our sample (see the top left panel of Figure~\ref{fig:correlations}). This trend is often theoretically attributed to low mass, star-forming galaxies having less neutral gas to resonantly scatter the \lya\ photon (as well as less dust) leading to a shorter total path length to exit the galaxy without absorption by dust (see \citealt{ando06}). In this sample, $\ewlya$ also negatively correlated (even more strongly) with SFR, and the fact that stellar mass and star formation rate correlate strongly with each other complicates the interpretation of this result. \citet{weiss21} addressed this issue by binning their sample of [O\,{\sc iii}]-emitting galaxies with \lya\ line flux measurements from \hd\ according to stellar mass and SFR. They found mass to better predict \fesc\ at fixed SFR than SFR did at fixed mass.

Fascinatingly, we did not find even a weak correlation between dust extinction and $\ewlya$. This seems surprising given that many authors have noted such a correlation and that the theoretical explanation is inarguable: resonantly scattered \lya\ photons can get absorbed readily in the presence of even a small amount of dust. A partial explanation for our sample's behavior with $A_V$ could be that it consists of systems exhibiting strong \lya\ emission, not absorption. For example, \citet{reddy21} studied systems with \lya\ in net absorption or emission and found a strong correlation between $\ewlya$ and $E(B-V)$. If our sample contained objects with negative $\ewlya$, perhaps those objects would reveal the correlation. Nevertheless, other studies of only emitters ($\ewlya > 0$) have also noted a trend with dust extinction, such as \citet{marchi19}, though a close examination of their Figure 7 shows that the negative correlation is largely driven by weak emitters with $\ewlya < 10\ \ang$. Our small dynamic range in $\ewlya$ may obfuscate a correlation with dust extinction. This interpretation may also be complicated by the \lya\ photon's ability to escape the galaxy even in the presence of large amounts of dust. Given a clumpy ISM geometry, clumps of gas and dust can act as mirrors to \lya\ photons, which ``bounce'' of the surfaces of these clumps through resonant scattering by neutral gas, while continuum photons pass through and thus experience extinction. \citet{gronke16} found that simulated \lya\ emission lines agreed well with observations for models with clumpy ISM geometries, and \citet{fink09} found that clumpy-ISM models better fit the SEDs of over half their NB-selected sample of LAEs at $z \sim 4.5$. \citet{vargas14} also found their sample of 20 NB-selected LAEs at z=2.1 favored clumpy-ISM models.   

Lastly, we found a moderate correlation between $\ewlya$ and galaxy mass-weighted age. The strength of \lya\ emission depends on both its production through recombination in HII regions as well as its escape through channels in the ISM with low neutral gas covering fractions, so the interplay between these processes determines $\ewlya$. As noted by \citet{marchi19}, who obtained a similar result, the trend with age could arise from older systems having experienced intense star formation in their past, where stellar winds and radiation cleared out neutral gas and dust, leaving channels for \lya\ escape. Through ongoing star formation or recent bursts, these objects can still produce \lya\ photons, and the ISM conditions favor their escape. For the youngest galaxies, even though the most massive, ionizing photon-producing stars are present, it is possible that a significant amount of dust and neutral gas has yet to be swept away, hindering the escape of \lya. 

\section{Predicting Lyman-alpha emission in the Epoch of Reionization}
\label{sec:predictions}

Using our knowledge of \lya\ emission from \hd\ galaxies situated in an ionized IGM, we can attempt to predict the intrinsic emission strength of LAEs at $z>7$, an era where starlight from galaxies was still actively re-ionizing the universe. 


\subsection{An LAE Sample in the Epoch of Reionization}
Our sample at \zrange{1.9}{3.5} provides a view of \lya\ emission unobscured by a significant IGM neutral fraction. By creating a predictive model that connects global galaxy properties to their intrinsic $\ewlya$ in this pristine era, we can apply it to LAEs in the EoR to derive their expected intrinsic $\ewlya$, then attributing any deficiency of \lya\ emission from objects in the EoR to an increasing neutral fraction. This does require the assumption that the production and escape of \lya\ photons does not evolve with redshift for fixed galaxy properties, which will require further testing. As a pilot attempt here, we took advantage of the sample of $z > 7$ \laes\ that \citet{jung20} found in \gn\ to test our ability to predict \lya\ emission from EoR galaxies. 

Using a deep, spectroscopic survey conducted with Keck/MOSFIRE, \citet{jung20} found 10 $>4\sigma$ \lya\ detections at $z > 7$ among 72 high-$z$ candidate galaxies. Such objects likely reside in ionized bubbles of the IGM, allowing the \lya\ photon to redshift away from the resonant-frequency therefore lowering the absorption cross-section with neutral hydrogen. These emitters thus serve as direct tests of our understanding of the galaxy properties that modulate \lya\ emission strength from the ISM/CGM. 

\begin{figure}
    \centering
    \includegraphics[width=0.9\columnwidth]{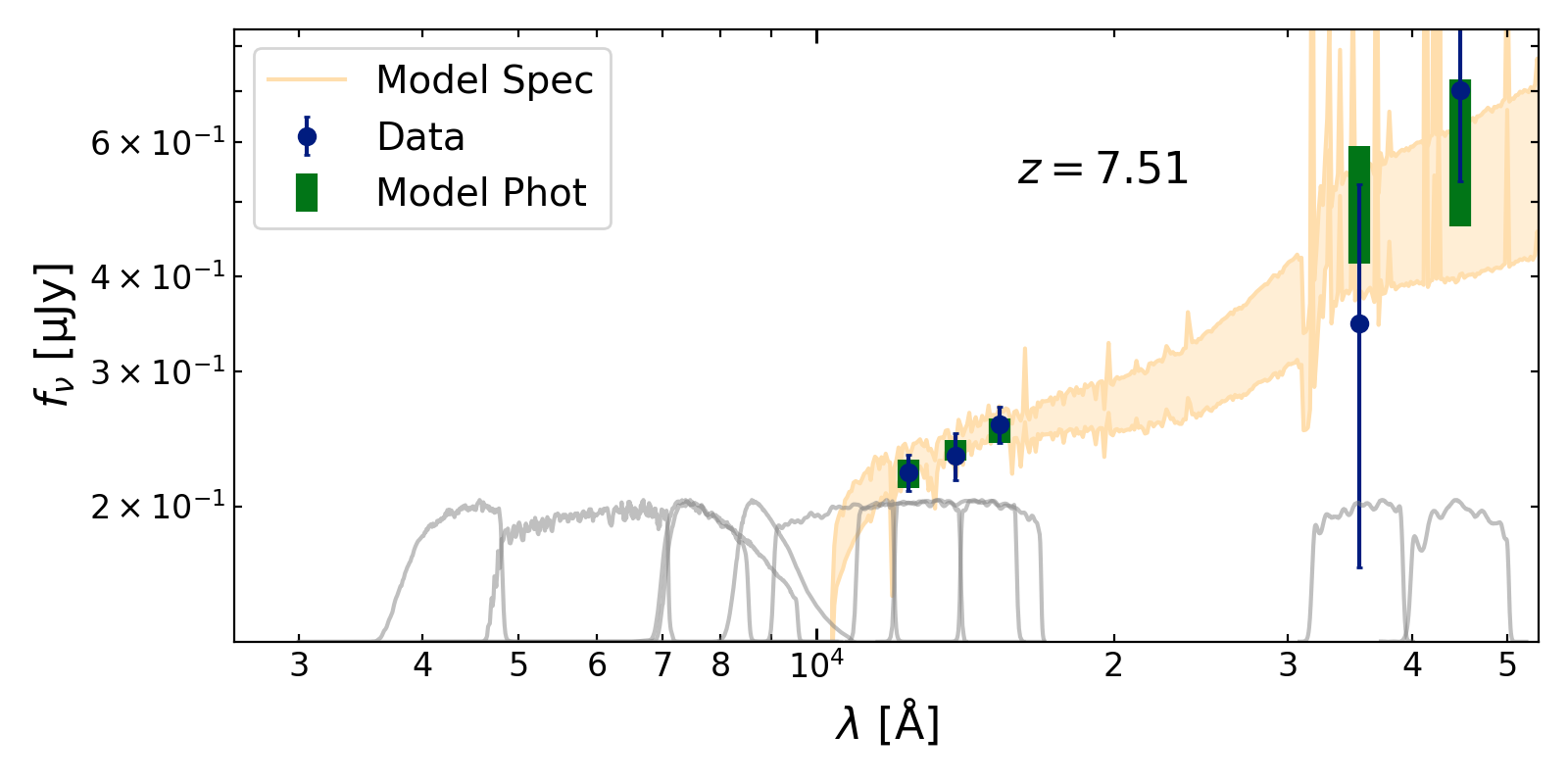}
    \caption{An example \bagpipes\ SED fit for LAE ID z7\_GND\_42912 at $z=7.51$  detected by \cite{jung20}. For scaling purposes, we do not show the upper-limits for non-detections in the \hst\ bands blueward of the \lya\ break. From our photometric data, we constrained the stellar population properties of ten LAEs in the EoR, allowing us to predict their intrinsic \lya\ emission using our \hd\ sample.}
    \label{fig:intae_sed}
\end{figure}

Because the photometric catalog for the \gn\ field contains the \laes\ discovered by \citet{jung20}, we performed the same SED analysis detailed in section~\ref{ssec:sedfitting} for those objects. We again masked all photometric bands including and blueward of \lya\ given the object's spectroscopic redshift. For most of the $z>7$ LAEs, this left 3 \hst\ filters as well as both \spitzer/IRAC channels. We again used \bagpipes\ to estimate the galaxy properties, adopting our fiducial model (delayed-$\tau$ SFH, \citealt{calzetti94} dust law). Figure~\ref{fig:intae_sed} shows an example fit for an object at $z=7.51$.

\subsection{A Predictive Model for $\ewlya$}

To predict the \lya\ equivalent widths of the $z>7$ sample, we chose several properties that strongly impact the emergent \lya\ emission from galaxies: stellar mass, dust extinction, and star formation rate. As discussed above, stellar mass may determine the amount of neutral  hydrogen gas (and thus dust) in the galaxy as well as the total path length needed to escape. In the presence of dust, \lya\ photons may terminate their resonant scattering process through absorption by a dust grain following re-emission at longer wavelengths, limiting likelihood of escape. Finally, the global star formation rate impacts the production of UV photons that can create \lya\ through recombination, and feedback from star formation may impact the structure of the ISM itself, creating ionized channels for escape. 

Using the posterior distributions sampled by \bagpipes, we matched each $z>7$ emitter to \laes\ in the \hd\ sample based on SED-derived properties. To do this, we calculated the ``separation" in the log mass, SFR, dust attenuation parameter space from the EoR \laes\ to each \lae\ in the \hd\ sample. For the separation calculation, we divided each parameter value by the full range of values in the sample to normalize the parameter space. For example, for log stellar mass, an object in the \hd\ sample with log mass halfway between the sample minimum and maximum would have a value of 0.5, so the difference between 0.5 and the EoR \lae\ log stellar mass scaled the same way would become input to the Euclidean distance formula. We then ranked the \hd\ \laes\ by separation in parameter space and constructed the prediction using the $N=3,5, \ \mrm{and}\ 7$ closest neighbors. We computed the posterior $\ewlya$ distribution by co-adding Gaussian distributions with mean and standard deviation set by the $\ewlya$ measurements and error bars in our sample. To give more importance to those LAEs that closely resembled the EoR galaxy, we weighted each Gaussian distribution by the inverse of its squared distance in parameter space from the EoR galaxy when co-adding to obtain the final prediction. The predicted $\ewlya$ distributions are normalized such that the integral over all equivalent widths equals unity.  

\begin{figure*}
    \centering
    \includegraphics[width=\textwidth]{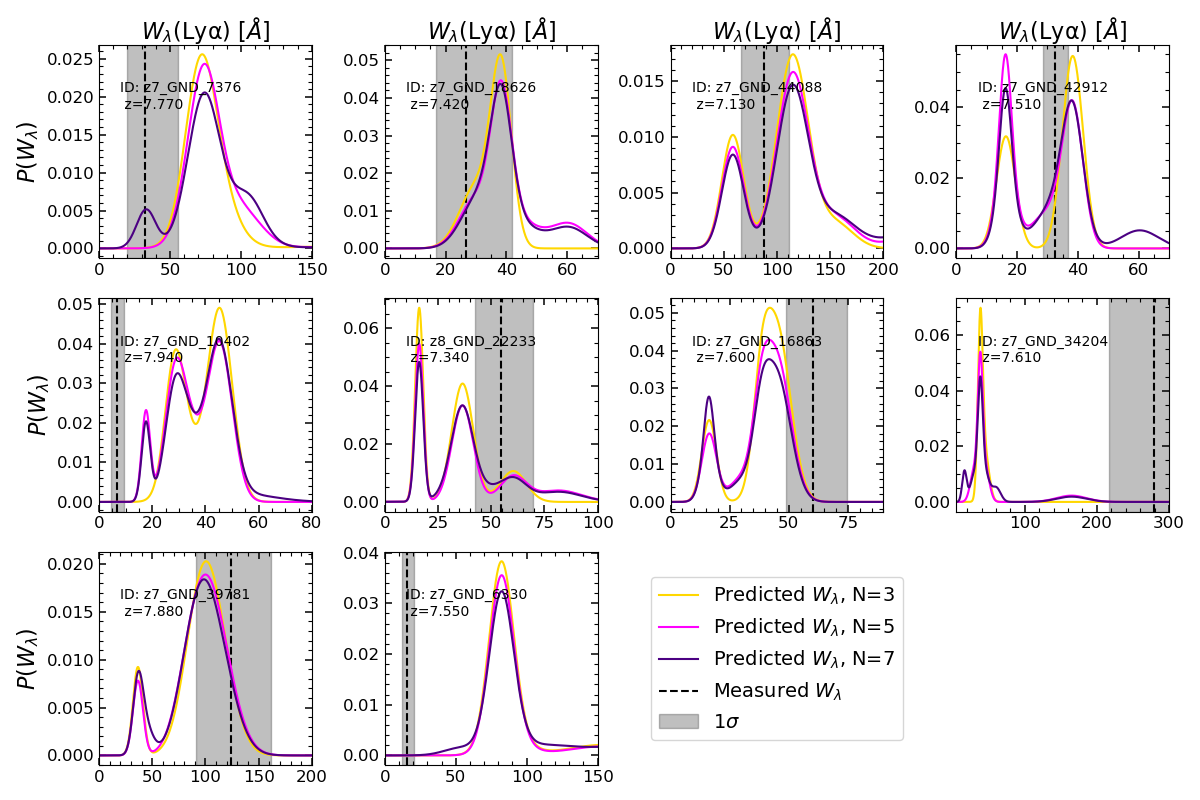}
    \caption{Probabilistic predictions of $\ewlya$ for ten \laes\ at $z>7$ having emission line $S/N>4$. The distributions were normalized by setting their integrals to unity. We chose stellar mass, SFR, and dust extinction as predictive properties for this calculation. Gold, magenta, and indigo lines show the probability distribution of our predictions using N=3,5, and 7 nearest neighbors, and the gray shaded region shows the 68\% confidence interval for the equivalent width measurements from \citet{jung20}. Object IDs and redshifts are indicated with text for each plot. We find good agreement between prediction and observation for the majority of strong emitters}
    \label{fig:ew_pred_sn4}
\end{figure*}

Figure~\ref{fig:ew_pred_sn4} shows our predicted $\ewlya$ distributions for \laes\ in the \citet{jung20} sample with Ly$\alpha$ $S/N>4$. We show predictions using three different values of $N$, the number of nearest neighbors in parameter space, to reveal any stochasticity in the prediction. The measured \lya\ equivalent widths from \citet{jung20} are indicated by vertical dashed lines with $1\sigma$ error intervals shaded grey. Importantly, we only expect our predictions to match the observed equivalent widths of EoR \laes\ if they exist in ionized bubbles. If the EoR \laes\ instead exist in regions of the IGM with significant neutral fractions, we expect to over-predict the \lya\ emission. On the other hand, an under-prediction of the \lya\ emission from an EoR object would imply our sample size is too small to account for the diversity in physical properties of the \lae\ population. 

\begin{figure}
    \centering
    \includegraphics[width=0.9\columnwidth]{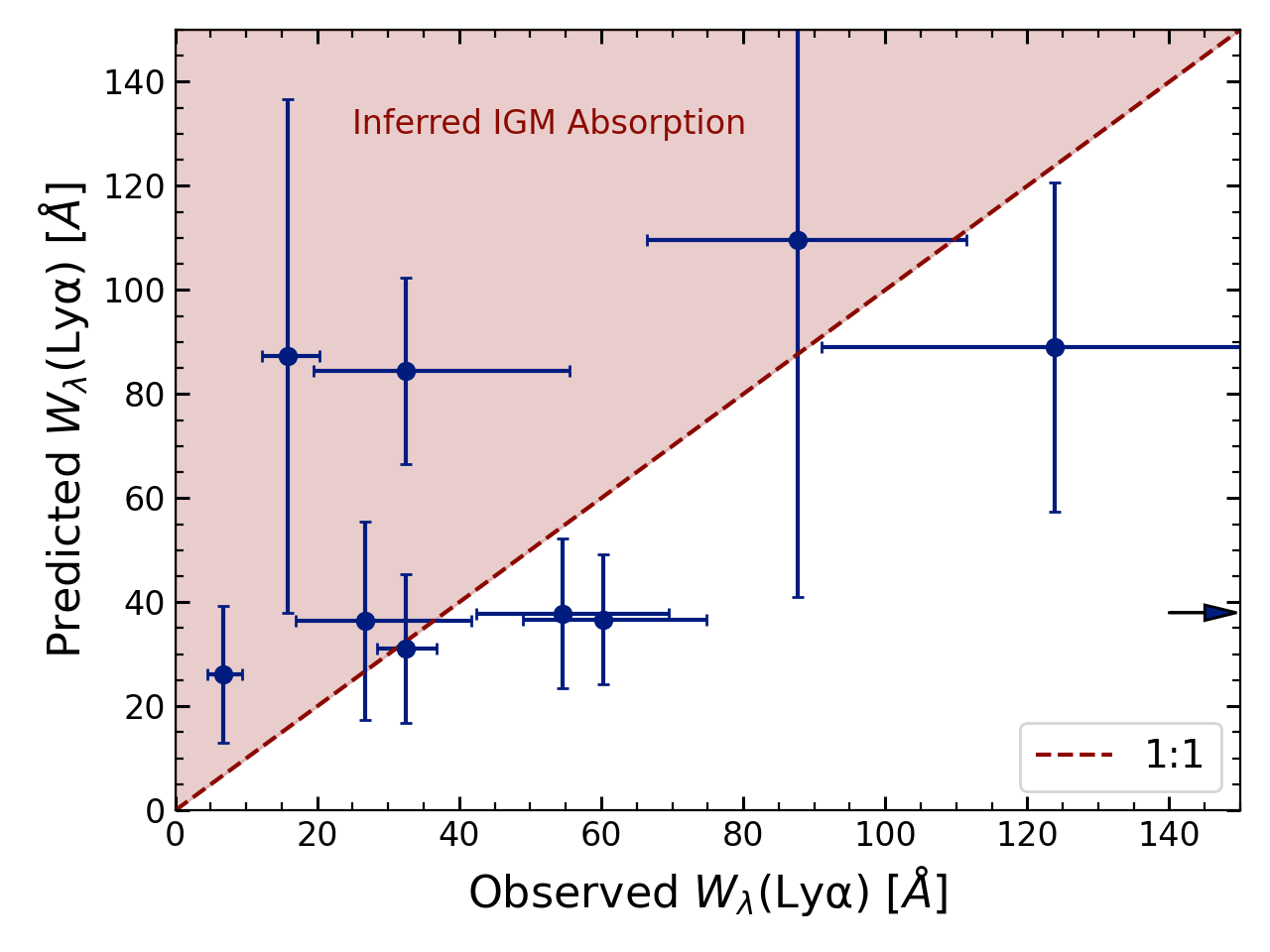}
    \caption{Predicted vs. observed $\ewlya$ for the \lae\ sample in Figure~\ref{fig:ew_pred_sn4}. We computed the first moment and square root of the second moment of each $N=5$ distribution in Figure~\ref{fig:ew_pred_sn4} for the predicted values and their error bars. A right-pointing arrow indicates the predicted value of object z7\_GND\_34204. A one-to-one dashed line is drawn to guide the eye, and points above this line (the region shaded red) could be the result of IGM absorption.}
    \label{fig:intae_pred_vs_obs}
\end{figure}

In Figure~\ref{fig:intae_pred_vs_obs}, we plot the predicted versus observed equivalents widths with a one-to-one line drawn to facilitate comparison. Each object's predicted value and error were calculated as the first moment and square root of the second moment of the $N=5$ curves in Figure~\ref{fig:ew_pred_sn4}, respectively. In five out of ten cases (ID z7\_GND\_18626, z7\_GND\_44088, z7\_GND\_42912, z7\_GND\_22233, and z7\_GND\_39781) the $1\sigma$ interval of our $\ewlya$ predictions overlapped with the $1\sigma$ interval of the observational measurement, indicating moderate agreement. For strong emitters (observed $\ewlya > 20\ \ang$), our prediction overlapped with observation five out of eight times. Furthermore, two strong emitters (z7\_GND\_42912 and z7\_GND\_16863), postulated by \citet{jung20} to inhabit ionized bubbles, had observed equivalent widths greater than or equal to the majority of our predicted $\ewlya$ distributions, as one might expect for sources with little IGM attenuation.

It is not surprising that our model failed to predict weak \lya\ emission accurately.  First, our model predicts \lya\ EWs in the absence of IGM absorption, thus an under prediction could imply significant absorption of \lya\ photons by neutral hydrogen in the IGM.  Second, as our sample by construction contains far more strong emitters than weak ones (see Figure~\ref{fig:ew_dist}), this could presently bias us towards an over-prediction of \lya\ emission strengtdrasticallyh.  We note that we  under-predicted the emission from ID z7\_GND\_34204 (indicated by an arrow in Figure \ref{fig:intae_pred_vs_obs}), which could be attributed to the dearth of objects in our sample with very high equivalent widths to match with that object's value, $\sim 280 \ \ang$. 

ID z7\_GND\_42912 offers a good example of how challenging predicting \lya\ emission can be. As $N$ increases, the peak of the predicted distribution shifts from agreeing well with the observation to under-predicting it. It is clear that our sample is presently too small to fully span the parameter space in both $\ewlya$ and physical properties. Future analyses with much larger samples made possible by \hd\ should be able to better capture the mean trends as well as variance in galaxy parameters that determine \lya\ emission strength.

Some of the predictions in Figure~\ref{fig:ew_pred_sn4} bode well for constraining the expected $\ewlya$ given a suite of galaxy properties measured from broadband SED fitting. With larger samples that suffer less from the inherent idiosyncratic behavior of \lya\ emission (for example, its dependence on the observer's line-of-sight), a rigorous, statistical understanding of the properties that drive that emission will arise, unlocking the potential of LAEs to probe cosmic reionization. We further note that, with larger samples, machine learning (ML) may prove an invaluable tool in making the nuanced connection between global galaxy properties and \lya\ emission strength, as the problem requires a regression analysis well suited for ML techniques.  

\section{Summary}
\label{sec:summary}

We used SED fitting to study the properties of a sample of LAEs from the \hd\ survey in \gn\ to better understand the phenomenology behind \lya\ emission and ultimately leverage these beacons of light in the distant Universe as probes of cosmic reionization.

To build the sample, we inspected \ninspect\ emission line detections to determine if the line was \lya\ or a feature from a low-redshift galaxy, such as \oii. We then created a procedure to synthesize information about angular separation from the emission line detection position, extracted emission line flux, and $\chi^2$ of SED fit assuming $z_\mathrm{Ly\alpha}$ to identify the continuum counterpart in our deep, mult-band \hst\ imaging in \gn. After removing detections with no counterparts, AGN contaminants, and sources with insufficient photometric data, we analyzed a sample of \nsamp\ LAEs using SED fitting performed by \bagpipes.

Our sample's properties were consistent with studies of LAEs from NB imaging surveys at similar redshifts. Our median sample mass was $0.8^{+2.9}_{-0.5} \times 10^{9} \ \msun$, and the galaxies' SFRs appeared to put them approximately on the star-forming main sequence, except for at $M_\star < 10^9\ \msun$. Using \lya\ emission line flux measurements from \hd, we also studied correlations between $\ewlya$ and galaxy properties. We found strong correlations between $\ewlya$ and stellar mass as well as SFR. We additionally found a moderate correlation where galaxies with older stellar populations had larger \lya\ equivalent widths. Interestingly, we did not find a significant impact of dust extinction on $\ewlya$, whereas many other studies have. Overall, this paints a picture of LAEs as low-mass systems with moderate star formation activity wherein \lya\ photons can escape even in the presence of dust. Also, the LAEs detected by \hd\ do not stand out significantly in terms of their stellar population properties from LAEs found using NB imaging with comparable flux limits.  

Finally, we used our LAE sample to try to predict the value of $\ewlya$ for ten LAEs at $z>7$ by matching the distinct samples in the parameter space of mass, SFR, and dust extinction. Our prediction matched the data at the $1\sigma$ level five out of ten times (5/8 for strong emitters); the three over-predictions could indicate significant absorption by a neutral hydrogen in the IGM. With large sample sizes in the near future and tools such as machine learning, we are optimistic about the ability of \hd\ LAEs to unlock the potential of \lya\ as a reliable reionization probe.

\section{Acknowledgements}

APM and SLF acknowledge support from the National Science Foundation, through grants AST-1908817 and AST-1614798.  I.J. acknowledges support from NASA under award number 80GSFC21M0002.

HETDEX is led by the University of Texas at Austin McDonald Observatory and Department of Astronomy with participation from the Ludwig-Maximilians-Universität München, Max-Planck-Institut für Extraterrestrische Physik (MPE), Leibniz-Institut für Astrophysik Potsdam (AIP), Texas A\&M University, Pennsylvania State University, Institut für Astrophysik Göttingen, The University of Oxford, Max-Planck-Institut für Astrophysik (MPA), The University of Tokyo and Missouri University of Science and Technology. In addition to Institutional support, HETDEX is funded by the National Science Foundation (grant AST-0926815), the State of Texas, the US Air Force (AFRL FA9451-04-2- 0355), and generous support from private individuals and foundations.

The observations were obtained with the Hobby-Eberly Telescope (HET), which is a joint project of the University of Texas at Austin, the Pennsylvania State University, Ludwig-Maximilians-Universität München, and Georg-August-Universität Göttingen. The HET is named in honor of its principal benefactors, William P. Hobby and Robert E. Eberly.

VIRUS is a joint project of the University of Texas at Austin, Leibniz-Institut f{\" u}r Astrophysik Potsdam (AIP), Texas A\&M University (TAMU), Max-Planck-Institut f{\" u}r Extraterrestrische Physik (MPE), Ludwig-Maximilians-Universit{\" a}t M{\" u}nchen, Pennsylvania State University, Institut f{\" u}r Astrophysik G{\" o}ttingen, University of Oxford, and the Max-Planck-Institut f{\" u}r Astrophysik (MPA). 

The authors acknowledge the Texas Advanced Computing Center (TACC) at The University of Texas at Austin for providing high performance computing, visualization, and storage resources that have contributed to the research results reported within this paper. URL: \url{http://www.tacc.utexas.edu}

The Institute for Gravitation and the Cosmos is supported by the Eberly College of Science and the Office of the Senior Vice President for Research at Pennsylvania State University.

\software{get\_spectrum.py \\ (https://github.com/HETDEX/hetdex\_api), emcee (Foreman-Mackey et al. 2013), Bagpipes (Carnall et al. 2018), Numpy (Harris et al. 2020)}

\bibliographystyle{aasjournal.bst}

\begin{thebibliography}{}
\expandafter\ifx\csname natexlab\endcsname\relax\def\natexlab#1{#1}\fi
\providecommand{\url}[1]{\href{#1}{#1}}
\providecommand{\dodoi}[1]{doi:~\href{http://doi.org/#1}{\nolinkurl{#1}}}
\providecommand{\doeprint}[1]{\href{http://ascl.net/#1}{\nolinkurl{http://ascl.net/#1}}}
\providecommand{\doarXiv}[1]{\href{https://arxiv.org/abs/#1}{\nolinkurl{https://arxiv.org/abs/#1}}}

\bibitem[{{Abazajian} {et~al.}(2009){Abazajian}, {Adelman-McCarthy},
  {Ag{\"u}eros}, {Allam}, {Allende Prieto}, {An}, {Anderson}, {Anderson},
  {Annis}, {Bahcall}, {Bailer-Jones}, {Barentine}, {Bassett}, {Becker},
  {Beers}, {Bell}, {Belokurov}, {Berlind}, {Berman}, {Bernardi}, {Bickerton},
  {Bizyaev}, {Blakeslee}, {Blanton}, {Bochanski}, {Boroski}, {Brewington},
  {Brinchmann}, {Brinkmann}, {Brunner}, {Budav{\'a}ri}, {Carey}, {Carliles},
  {Carr}, {Castander}, {Cinabro}, {Connolly}, {Csabai}, {Cunha}, {Czarapata},
  {Davenport}, {de Haas}, {Dilday}, {Doi}, {Eisenstein}, {Evans}, {Evans},
  {Fan}, {Friedman}, {Frieman}, {Fukugita}, {G{\"a}nsicke}, {Gates},
  {Gillespie}, {Gilmore}, {Gonzalez}, {Gonzalez}, {Grebel}, {Gunn},
  {Gy{\"o}ry}, {Hall}, {Harding}, {Harris}, {Harvanek}, {Hawley}, {Hayes},
  {Heckman}, {Hendry}, {Hennessy}, {Hindsley}, {Hoblitt}, {Hogan}, {Hogg},
  {Holtzman}, {Hyde}, {Ichikawa}, {Ichikawa}, {Im}, {Ivezi{\'c}}, {Jester},
  {Jiang}, {Johnson}, {Jorgensen}, {Juri{\'c}}, {Kent}, {Kessler}, {Kleinman},
  {Knapp}, {Konishi}, {Kron}, {Krzesinski}, {Kuropatkin}, {Lampeitl},
  {Lebedeva}, {Lee}, {Lee}, {French Leger}, {L{\'e}pine}, {Li}, {Lima}, {Lin},
  {Long}, {Loomis}, {Loveday}, {Lupton}, {Magnier}, {Malanushenko},
  {Malanushenko}, {Mandelbaum}, {Margon}, {Marriner}, {Mart{\'\i}nez-Delgado},
  {Matsubara}, {McGehee}, {McKay}, {Meiksin}, {Morrison}, {Mullally}, {Munn},
  {Murphy}, {Nash}, {Nebot}, {Neilsen}, {Newberg}, {Newman}, {Nichol},
  {Nicinski}, {Nieto-Santisteban}, {Nitta}, {Okamura}, {Oravetz}, {Ostriker},
  {Owen}, {Padmanabhan}, {Pan}, {Park}, {Pauls}, {Peoples}, {Percival}, {Pier},
  {Pope}, {Pourbaix}, {Price}, {Purger}, {Quinn}, {Raddick}, {Re Fiorentin},
  {Richards}, {Richmond}, {Riess}, {Rix}, {Rockosi}, {Sako}, {Schlegel},
  {Schneider}, {Scholz}, {Schreiber}, {Schwope}, {Seljak}, {Sesar}, {Sheldon},
  {Shimasaku}, {Sibley}, {Simmons}, {Sivarani}, {Allyn Smith}, {Smith},
  {Smol{\v{c}}i{\'c}}, {Snedden}, {Stebbins}, {Steinmetz}, {Stoughton},
  {Strauss}, {SubbaRao}, {Suto}, {Szalay}, {Szapudi}, {Szkody}, {Tanaka},
  {Tegmark}, {Teodoro}, {Thakar}, {Tremonti}, {Tucker}, {Uomoto}, {Vanden
  Berk}, {Vandenberg}, {Vidrih}, {Vogeley}, {Voges}, {Vogt}, {Wadadekar},
  {Watters}, {Weinberg}, {West}, {White}, {Wilhite}, {Wonders}, {Yanny},
  {Yocum}, {York}, {Zehavi}, {Zibetti}, \& {Zucker}}]{abazajian09}
{Abazajian}, K.~N., {Adelman-McCarthy}, J.~K., {Ag{\"u}eros}, M.~A., {et~al.}
  2009, \apjs, 182, 543, \dodoi{10.1088/0067-0049/182/2/543}

\bibitem[{{Acquaviva} {et~al.}(2011){Acquaviva}, {Gawiser}, \&
  {Guaita}}]{acquaviva11}
{Acquaviva}, V., {Gawiser}, E., \& {Guaita}, L. 2011, \apj, 737, 47,
  \dodoi{10.1088/0004-637X/737/2/47}

\bibitem[{{Acquaviva} {et~al.}(2012){Acquaviva}, {Vargas}, {Gawiser}, \&
  {Guaita}}]{acquaviva12}
{Acquaviva}, V., {Vargas}, C., {Gawiser}, E., \& {Guaita}, L. 2012, \apjl, 751,
  L26, \dodoi{10.1088/2041-8205/751/2/L26}

\bibitem[{{Adams} {et~al.}(2011){Adams}, {Blanc}, {Hill}, {Gebhardt}, {Drory},
  {Hao}, {Bender}, {Byun}, {Ciardullo}, {Cornell}, {Finkelstein}, {Fry},
  {Gawiser}, {Gronwall}, {Hopp}, {Jeong}, {Kelz}, {Kelzenberg}, {Komatsu},
  {MacQueen}, {Murphy}, {Odoms}, {Roth}, {Schneider}, {Tufts}, \&
  {Wilkinson}}]{adams11}
{Adams}, J.~J., {Blanc}, G.~A., {Hill}, G.~J., {et~al.} 2011, \apjs, 192, 5,
  \dodoi{10.1088/0067-0049/192/1/5}

\bibitem[{{Ando} {et~al.}(2006){Ando}, {Ohta}, {Iwata}, {Akiyama}, {Aoki}, \&
  {Tamura}}]{ando06}
{Ando}, M., {Ohta}, K., {Iwata}, I., {et~al.} 2006, \apjl, 645, L9,
  \dodoi{10.1086/505652}

\bibitem[{{Ashby} {et~al.}(2015){Ashby}, {Willner}, {Fazio}, {Dunlop}, {Egami},
  {Faber}, {Ferguson}, {Grogin}, {Hora}, {Huang}, {Koekemoer}, {Labb{\'e}}, \&
  {Wang}}]{ashby15}
{Ashby}, M.~L.~N., {Willner}, S.~P., {Fazio}, G.~G., {et~al.} 2015, \apjs, 218,
  33, \dodoi{10.1088/0067-0049/218/2/33}

\bibitem[{{Atek} {et~al.}(2014){Atek}, {Kunth}, {Schaerer}, {Mas-Hesse},
  {Hayes}, {{\"O}stlin}, \& {Kneib}}]{atek14}
{Atek}, H., {Kunth}, D., {Schaerer}, D., {et~al.} 2014, \aap, 561, A89,
  \dodoi{10.1051/0004-6361/201321519}

\bibitem[{{Bertin} \& {Arnouts}(1996)}]{sextractor}
{Bertin}, E., \& {Arnouts}, S. 1996, \aaps, 117, 393,
  \dodoi{10.1051/aas:1996164}

\bibitem[{{Blanc} {et~al.}(2011){Blanc}, {Adams}, {Gebhardt}, {Hill}, {Drory},
  {Hao}, {Bender}, {Ciardullo}, {Finkelstein}, {Fry}, {Gawiser}, {Gronwall},
  {Hopp}, {Jeong}, {Kelzenberg}, {Komatsu}, {MacQueen}, {Murphy}, {Roth},
  {Schneider}, \& {Tufts}}]{blanc11}
{Blanc}, G.~A., {Adams}, J.~J., {Gebhardt}, K., {et~al.} 2011, \apj, 736, 31,
  \dodoi{10.1088/0004-637X/736/1/31}

\bibitem[{{Bruzual} \& {Charlot}(2003)}]{bruzual03}
{Bruzual}, G., \& {Charlot}, S. 2003, \mnras, 344, 1000,
  \dodoi{10.1046/j.1365-8711.2003.06897.x}

\bibitem[{{Calzetti} {et~al.}(2000){Calzetti}, {Armus}, {Bohlin}, {Kinney},
  {Koornneef}, \& {Storchi-Bergmann}}]{calzetti00}
{Calzetti}, D., {Armus}, L., {Bohlin}, R.~C., {et~al.} 2000, \apj, 533, 682,
  \dodoi{10.1086/308692}

\bibitem[{{Calzetti} {et~al.}(1994){Calzetti}, {Kinney}, \&
  {Storchi-Bergmann}}]{calzetti94}
{Calzetti}, D., {Kinney}, A.~L., \& {Storchi-Bergmann}, T. 1994, \apj, 429,
  582, \dodoi{10.1086/174346}

\bibitem[{{Cardelli} {et~al.}(1989){Cardelli}, {Clayton}, \&
  {Mathis}}]{cardelli89}
{Cardelli}, J.~A., {Clayton}, G.~C., \& {Mathis}, J.~S. 1989, \apj, 345, 245,
  \dodoi{10.1086/167900}

\bibitem[{{Carnall} {et~al.}(2018){Carnall}, {McLure}, {Dunlop}, \&
  {Dav{\'e}}}]{carnall18}
{Carnall}, A.~C., {McLure}, R.~J., {Dunlop}, J.~S., \& {Dav{\'e}}, R. 2018,
  \mnras, 480, 4379, \dodoi{10.1093/mnras/sty2169}

\bibitem[{{Chabrier}(2003)}]{chabrier03}
{Chabrier}, G. 2003, \pasp, 115, 763, \dodoi{10.1086/376392}

\bibitem[{{Charlot} \& {Fall}(2000)}]{CharlotFall}
{Charlot}, S., \& {Fall}, S.~M. 2000, \apj, 539, 718, \dodoi{10.1086/309250}

\bibitem[{{Conroy}(2013)}]{conroy13}
{Conroy}, C. 2013, \araa, 51, 393, \dodoi{10.1146/annurev-astro-082812-141017}

\bibitem[{{Conroy} {et~al.}(2009){Conroy}, {Gunn}, \& {White}}]{conroy09}
{Conroy}, C., {Gunn}, J.~E., \& {White}, M. 2009, \apj, 699, 486,
  \dodoi{10.1088/0004-637X/699/1/486}

\bibitem[{{Cowie} \& {Hu}(1998)}]{cowie98}
{Cowie}, L.~L., \& {Hu}, E.~M. 1998, \aj, 115, 1319, \dodoi{10.1086/300309}

\bibitem[{{Dijkstra}(2014)}]{dijkstra14}
{Dijkstra}, M. 2014, \pasa, 31, e040, \dodoi{10.1017/pasa.2014.33}

\bibitem[{{Du} {et~al.}(2018){Du}, {Shapley}, {Reddy}, {Jones}, {Stark},
  {Steidel}, {Strom}, {Rudie}, {Erb}, {Ellis}, \& {Pettini}}]{du18}
{Du}, X., {Shapley}, A.~E., {Reddy}, N.~A., {et~al.} 2018, \apj, 860, 75,
  \dodoi{10.3847/1538-4357/aabfcf}

\bibitem[{{Erb} {et~al.}(2012){Erb}, {Quider}, {Henry}, \& {Martin}}]{erb12}
{Erb}, D.~K., {Quider}, A.~M., {Henry}, A.~L., \& {Martin}, C.~L. 2012, \apj,
  759, 26, \dodoi{10.1088/0004-637X/759/1/26}

\bibitem[{{Farrow} {et~al.}(2021){Farrow}, {S{\'a}nchez}, {Ciardullo},
  {Cooper}, {Davis}, {Fabricius}, {Gawiser}, {Grasshorn Gebhardt}, {Gebhardt},
  {Hill}, {Jeong}, {Komatsu}, {Landriau}, {Liu}, {Saito}, {Snigula}, \&
  {Wold}}]{farrow21}
{Farrow}, D.~J., {S{\'a}nchez}, A.~G., {Ciardullo}, R., {et~al.} 2021, \mnras,
  507, 3187, \dodoi{10.1093/mnras/stab1986}

\bibitem[{{Feltre} {et~al.}(2020){Feltre}, {Maseda}, {Bacon}, {Pradeep},
  {Leclercq}, {Kusakabe}, {Wisotzki}, {Hashimoto}, {Schmidt}, {Blaizot},
  {Brinchmann}, {Boogaard}, {Cantalupo}, {Carton}, {Inami}, {Kollatschny},
  {Marino}, {Matthee}, {Nanayakkara}, {Richard}, {Schaye}, {Tresse}, {Urrutia},
  {Verhamme}, \& {Weilbacher}}]{feltre20}
{Feltre}, A., {Maseda}, M.~V., {Bacon}, R., {et~al.} 2020, \aap, 641, A118,
  \dodoi{10.1051/0004-6361/202038133}

\bibitem[{{Feroz} \& {Skilling}(2013)}]{multinest}
{Feroz}, F., \& {Skilling}, J. 2013, in American Institute of Physics
  Conference Series, Vol. 1553, Bayesian Inference and Maximum Entropy Methods
  in Science and Engineering: 32nd International Workshop on Bayesian Inference
  and Maximum Entropy Methods in Science and Engineering, ed. U.~{von
  Toussaint}, 106--113, \dodoi{10.1063/1.4819989}

\bibitem[{{Finkelstein} {et~al.}(2015){Finkelstein}, {Finkelstein}, {Tilvi},
  {Malhotra}, {Rhoads}, {Grogin}, {Pirzkal}, {Dey}, {Jannuzi}, {Mobasher},
  {Pakzad}, {Salmon}, \& {Wang}}]{keely15}
{Finkelstein}, K.~D., {Finkelstein}, S.~L., {Tilvi}, V., {et~al.} 2015, \apj,
  813, 78, \dodoi{10.1088/0004-637X/813/1/78}

\bibitem[{{Finkelstein} {et~al.}(2010){Finkelstein}, {Papovich}, {Giavalisco},
  {Reddy}, {Ferguson}, {Koekemoer}, \& {Dickinson}}]{fink10}
{Finkelstein}, S.~L., {Papovich}, C., {Giavalisco}, M., {et~al.} 2010, \apj,
  719, 1250, \dodoi{10.1088/0004-637X/719/2/1250}

\bibitem[{{Finkelstein} {et~al.}(2009){Finkelstein}, {Rhoads}, {Malhotra}, \&
  {Grogin}}]{fink09}
{Finkelstein}, S.~L., {Rhoads}, J.~E., {Malhotra}, S., \& {Grogin}, N. 2009,
  \apj, 691, 465, \dodoi{10.1088/0004-637X/691/1/465}

\bibitem[{{Finkelstein} {et~al.}(2012){Finkelstein}, {Papovich}, {Salmon},
  {Finlator}, {Dickinson}, {Ferguson}, {Giavalisco}, {Koekemoer}, {Reddy},
  {Bassett}, {Conselice}, {Dunlop}, {Faber}, {Grogin}, {Hathi}, {Kocevski},
  {Lai}, {Lee}, {McLure}, {Mobasher}, \& {Newman}}]{fink12a}
{Finkelstein}, S.~L., {Papovich}, C., {Salmon}, B., {et~al.} 2012, \apj, 756,
  164, \dodoi{10.1088/0004-637X/756/2/164}

\bibitem[{{Finkelstein} {et~al.}(2019){Finkelstein}, {D'Aloisio},
  {Paardekooper}, {Ryan}, {Behroozi}, {Finlator}, {Livermore}, {Upton
  Sanderbeck}, {Dalla Vecchia}, \& {Khochfar}}]{fink19}
{Finkelstein}, S.~L., {D'Aloisio}, A., {Paardekooper}, J.-P., {et~al.} 2019,
  \apj, 879, 36, \dodoi{10.3847/1538-4357/ab1ea8}

\bibitem[{{Finkelstein} {et~al.}(2021){Finkelstein}, {Bagley}, {Song},
  {Larson}, {Papovich}, {Dickinson}, {Finkelstein}, {Koekemoer}, {Pirzkal},
  {Somerville}, {Yung}, {Behroozi}, {Ferguson}, {Giavalisco}, {Grogin},
  {Hathi}, {Hutchison}, {Jung}, {Kocevski}, {Kawinwanichakij}, {Rojas-Ruiz},
  {Ryan}, {Snyder}, \& {Tacchella}}]{fink21}
{Finkelstein}, S.~L., {Bagley}, M., {Song}, M., {et~al.} 2021, arXiv e-prints,
  arXiv:2106.13813.
\newblock \doarXiv{2106.13813}

\bibitem[{Foreman-Mackey(2016)}]{corner}
Foreman-Mackey, D. 2016, The Journal of Open Source Software, 1, 24,
  \dodoi{10.21105/joss.00024}

\bibitem[{{Foreman-Mackey} {et~al.}(2013){Foreman-Mackey}, {Hogg}, {Lang}, \&
  {Goodman}}]{emcee}
{Foreman-Mackey}, D., {Hogg}, D.~W., {Lang}, D., \& {Goodman}, J. 2013, \pasp,
  125, 306, \dodoi{10.1086/670067}

\bibitem[{{Gaia Collaboration} {et~al.}(2018){Gaia Collaboration}, {Brown},
  {Vallenari}, {Prusti}, {de Bruijne}, {Babusiaux}, {Bailer-Jones}, {Biermann},
  {Evans}, {Eyer}, {Jansen}, {Jordi}, {Klioner}, {Lammers}, {Lindegren},
  {Luri}, {Mignard}, {Panem}, {Pourbaix}, {Randich}, {Sartoretti}, {Siddiqui},
  {Soubiran}, {van Leeuwen}, {Walton}, {Arenou}, {Bastian}, {Cropper},
  {Drimmel}, {Katz}, {Lattanzi}, {Bakker}, {Cacciari}, {Casta{\~n}eda},
  {Chaoul}, {Cheek}, {De Angeli}, {Fabricius}, {Guerra}, {Holl}, {Masana},
  {Messineo}, {Mowlavi}, {Nienartowicz}, {Panuzzo}, {Portell}, {Riello},
  {Seabroke}, {Tanga}, {Th{\'e}venin}, {Gracia-Abril}, {Comoretto},
  {Garcia-Reinaldos}, {Teyssier}, {Altmann}, {Andrae}, {Audard},
  {Bellas-Velidis}, {Benson}, {Berthier}, {Blomme}, {Burgess}, {Busso},
  {Carry}, {Cellino}, {Clementini}, {Clotet}, {Creevey}, {Davidson}, {De
  Ridder}, {Delchambre}, {Dell'Oro}, {Ducourant},
  {Fern{\'a}ndez-Hern{\'a}ndez}, {Fouesneau}, {Fr{\'e}mat}, {Galluccio},
  {Garc{\'\i}a-Torres}, {Gonz{\'a}lez-N{\'u}{\~n}ez}, {Gonz{\'a}lez-Vidal},
  {Gosset}, {Guy}, {Halbwachs}, {Hambly}, {Harrison}, {Hern{\'a}ndez},
  {Hestroffer}, {Hodgkin}, {Hutton}, {Jasniewicz}, {Jean-Antoine-Piccolo},
  {Jordan}, {Korn}, {Krone-Martins}, {Lanzafame}, {Lebzelter}, {L{\"o}ffler},
  {Manteiga}, {Marrese}, {Mart{\'\i}n-Fleitas}, {Moitinho}, {Mora}, {Muinonen},
  {Osinde}, {Pancino}, {Pauwels}, {Petit}, {Recio-Blanco}, {Richards},
  {Rimoldini}, {Robin}, {Sarro}, {Siopis}, {Smith}, {Sozzetti}, {S{\"u}veges},
  {Torra}, {van Reeven}, {Abbas}, {Abreu Aramburu}, {Accart}, {Aerts},
  {Altavilla}, {{\'A}lvarez}, {Alvarez}, {Alves}, {Anderson}, {Andrei},
  {Anglada Varela}, {Antiche}, {Antoja}, {Arcay}, {Astraatmadja}, {Bach},
  {Baker}, {Balaguer-N{\'u}{\~n}ez}, {Balm}, {Barache}, {Barata}, {Barbato},
  {Barblan}, {Barklem}, {Barrado}, {Barros}, {Barstow}, {Bartholom{\'e}
  Mu{\~n}oz}, {Bassilana}, {Becciani}, {Bellazzini}, {Berihuete}, {Bertone},
  {Bianchi}, {Bienaym{\'e}}, {Blanco-Cuaresma}, {Boch}, {Boeche}, {Bombrun},
  {Borrachero}, {Bossini}, {Bouquillon}, {Bourda}, {Bragaglia}, {Bramante},
  {Breddels}, {Bressan}, {Brouillet}, {Br{\"u}semeister}, {Brugaletta},
  {Bucciarelli}, {Burlacu}, {Busonero}, {Butkevich}, {Buzzi}, {Caffau},
  {Cancelliere}, {Cannizzaro}, {Cantat-Gaudin}, {Carballo}, {Carlucci},
  {Carrasco}, {Casamiquela}, {Castellani}, {Castro-Ginard}, {Charlot},
  {Chemin}, {Chiavassa}, {Cocozza}, {Costigan}, {Cowell}, {Crifo}, {Crosta},
  {Crowley}, {Cuypers}, {Dafonte}, {Damerdji}, {Dapergolas}, {David}, {David},
  {de Laverny}, {De Luise}, {De March}, {de Martino}, {de Souza}, {de Torres},
  {Debosscher}, {del Pozo}, {Delbo}, {Delgado}, {Delgado}, {Di Matteo},
  {Diakite}, {Diener}, {Distefano}, {Dolding}, {Drazinos}, {Dur{\'a}n},
  {Edvardsson}, {Enke}, {Eriksson}, {Esquej}, {Eynard Bontemps}, {Fabre},
  {Fabrizio}, {Faigler}, {Falc{\~a}o}, {Farr{\`a}s Casas}, {Federici},
  {Fedorets}, {Fernique}, {Figueras}, {Filippi}, {Findeisen}, {Fonti},
  {Fraile}, {Fraser}, {Fr{\'e}zouls}, {Gai}, {Galleti}, {Garabato},
  {Garc{\'\i}a-Sedano}, {Garofalo}, {Garralda}, {Gavel}, {Gavras}, {Gerssen},
  {Geyer}, {Giacobbe}, {Gilmore}, {Girona}, {Giuffrida}, {Glass}, {Gomes},
  {Granvik}, {Gueguen}, {Guerrier}, {Guiraud}, {Guti{\'e}rrez-S{\'a}nchez},
  {Haigron}, {Hatzidimitriou}, {Hauser}, {Haywood}, {Heiter}, {Helmi}, {Heu},
  {Hilger}, {Hobbs}, {Hofmann}, {Holland}, {Huckle}, {Hypki}, {Icardi},
  {Jan{\ss}en}, {Jevardat de Fombelle}, {Jonker}, {Juh{\'a}sz}, {Julbe},
  {Karampelas}, {Kewley}, {Klar}, {Kochoska}, {Kohley}, {Kolenberg},
  {Kontizas}, {Kontizas}, {Koposov}, {Kordopatis}, {Kostrzewa-Rutkowska},
  {Koubsky}, {Lambert}, {Lanza}, {Lasne}, {Lavigne}, {Le Fustec}, {Le
  Poncin-Lafitte}, {Lebreton}, {Leccia}, {Leclerc}, {Lecoeur-Taibi},
  {Lenhardt}, {Leroux}, {Liao}, {Licata}, {Lindstr{\o}m}, {Lister}, {Livanou},
  {Lobel}, {L{\'o}pez}, {Managau}, {Mann}, {Mantelet}, {Marchal}, {Marchant},
  {Marconi}, {Marinoni}, {Marschalk{\'o}}, {Marshall}, {Martino}, {Marton},
  {Mary}, {Massari}, {Matijevi{\v{c}}}, {Mazeh}, {McMillan}, {Messina},
  {Michalik}, {Millar}, {Molina}, {Molinaro}, {Moln{\'a}r}, {Montegriffo},
  {Mor}, {Morbidelli}, {Morel}, {Morris}, {Mulone}, {Muraveva}, {Musella},
  {Nelemans}, {Nicastro}, {Noval}, {O'Mullane}, {Ord{\'e}novic},
  {Ord{\'o}{\~n}ez-Blanco}, {Osborne}, {Pagani}, {Pagano}, {Pailler},
  {Palacin}, {Palaversa}, {Panahi}, {Pawlak}, {Piersimoni}, {Pineau}, {Plachy},
  {Plum}, {Poggio}, {Poujoulet}, {Pr{\v{s}}a}, {Pulone}, {Racero}, {Ragaini},
  {Rambaux}, {Ramos-Lerate}, {Regibo}, {Reyl{\'e}}, {Riclet}, {Ripepi}, {Riva},
  {Rivard}, {Rixon}, {Roegiers}, {Roelens}, {Romero-G{\'o}mez}, {Rowell},
  {Royer}, {Ruiz-Dern}, {Sadowski}, {Sagrist{\`a} Sell{\'e}s}, {Sahlmann},
  {Salgado}, {Salguero}, {Sanna}, {Santana-Ros}, {Sarasso}, {Savietto},
  {Schultheis}, {Sciacca}, {Segol}, {Segovia}, {S{\'e}gransan}, {Shih},
  {Siltala}, {Silva}, {Smart}, {Smith}, {Solano}, {Solitro}, {Sordo}, {Soria
  Nieto}, {Souchay}, {Spagna}, {Spoto}, {Stampa}, {Steele},
  {Steidelm{\"u}ller}, {Stephenson}, {Stoev}, {Suess}, {Surdej}, {Szabados},
  {Szegedi-Elek}, {Tapiador}, {Taris}, {Tauran}, {Taylor}, {Teixeira},
  {Terrett}, {Teyssandier}, {Thuillot}, {Titarenko}, {Torra Clotet}, {Turon},
  {Ulla}, {Utrilla}, {Uzzi}, {Vaillant}, {Valentini}, {Valette}, {van Elteren},
  {Van Hemelryck}, {van Leeuwen}, {Vaschetto}, {Vecchiato}, {Veljanoski},
  {Viala}, {Vicente}, {Vogt}, {von Essen}, {Voss}, {Votruba}, {Voutsinas},
  {Walmsley}, {Weiler}, {Wertz}, {Wevers}, {Wyrzykowski}, {Yoldas},
  {{\v{Z}}erjal}, {Ziaeepour}, {Zorec}, {Zschocke}, {Zucker}, {Zurbach}, \&
  {Zwitter}}]{gaia18}
{Gaia Collaboration}, {Brown}, A.~G.~A., {Vallenari}, A., {et~al.} 2018, \aap,
  616, A1, \dodoi{10.1051/0004-6361/201833051}

\bibitem[{{Gawiser} {et~al.}(2007){Gawiser}, {Francke}, {Lai}, {Schawinski},
  {Gronwall}, {Ciardullo}, {Quadri}, {Orsi}, {Barrientos}, {Blanc}, {Fazio},
  {Feldmeier}, {Huang}, {Infante}, {Lira}, {Padilla}, {Taylor}, {Treister},
  {Urry}, {van Dokkum}, \& {Virani}}]{gawiser07}
{Gawiser}, E., {Francke}, H., {Lai}, K., {et~al.} 2007, \apj, 671, 278,
  \dodoi{10.1086/522955}

\bibitem[{{Gebhardt} {et~al.}(2021){Gebhardt}, {Mentuch Cooper}, {Ciardullo},
  {Acquaviva}, {Bender}, {Bowman}, {Castanheira}, {Dalton}, {Davis}, {de Jong},
  {DePoy}, {Devarakonda}, {Dongsheng}, {Drory}, {Fabricius}, {Farrow},
  {Feldmeier}, {Finkelstein}, {Froning}, {Gawiser}, {Gronwall}, {Herold},
  {Hill}, {Hopp}, {House}, {Janowiecki}, {Jarvis}, {Jeong}, {Jogee}, {Kakuma},
  {Kelz}, {Kollatschny}, {Komatsu}, {Krumpe}, {Landriau}, {Liu}, {Lujan
  Niemeyer}, {MacQueen}, {Marshall}, {Mawatari}, {McLinden}, {Mukae},
  {Nagaraj}, {Ono}, {Ouchi}, {Papovich}, {Sakai}, {Saito}, {Schneider},
  {Schulze}, {Shanmugasundararaj}, {Shetrone}, {Sneden}, {Snigula},
  {Steinmetz}, {Thomas}, {Thomas}, {Tuttle}, {Urrutia}, {Wisotzki}, {Wold},
  {Zeimann}, \& {Zhang}}]{gebhardt21}
{Gebhardt}, K., {Mentuch Cooper}, E., {Ciardullo}, R., {et~al.} 2021, arXiv
  e-prints, arXiv:2110.04298.
\newblock \doarXiv{2110.04298}

\bibitem[{{Giavalisco} {et~al.}(2004){Giavalisco}, {Ferguson}, {Koekemoer},
  {Dickinson}, {Alexander}, {Bauer}, {Bergeron}, {Biagetti}, {Brandt},
  {Casertano}, {Cesarsky}, {Chatzichristou}, {Conselice}, {Cristiani}, {Da
  Costa}, {Dahlen}, {de Mello}, {Eisenhardt}, {Erben}, {Fall}, {Fassnacht},
  {Fosbury}, {Fruchter}, {Gardner}, {Grogin}, {Hook}, {Hornschemeier}, {Idzi},
  {Jogee}, {Kretchmer}, {Laidler}, {Lee}, {Livio}, {Lucas}, {Madau},
  {Mobasher}, {Moustakas}, {Nonino}, {Padovani}, {Papovich}, {Park},
  {Ravindranath}, {Renzini}, {Richardson}, {Riess}, {Rosati}, {Schirmer},
  {Schreier}, {Somerville}, {Spinrad}, {Stern}, {Stiavelli}, {Strolger},
  {Urry}, {Vandame}, {Williams}, \& {Wolf}}]{giavalisco04}
{Giavalisco}, M., {Ferguson}, H.~C., {Koekemoer}, A.~M., {et~al.} 2004, \apjl,
  600, L93, \dodoi{10.1086/379232}

\bibitem[{{Grogin} {et~al.}(2011){Grogin}, {Kocevski}, {Faber}, {Ferguson},
  {Koekemoer}, {Riess}, {Acquaviva}, {Alexander}, {Almaini}, {Ashby}, {Barden},
  {Bell}, {Bournaud}, {Brown}, {Caputi}, {Casertano}, {Cassata}, {Castellano},
  {Challis}, {Chary}, {Cheung}, {Cirasuolo}, {Conselice}, {Roshan Cooray},
  {Croton}, {Daddi}, {Dahlen}, {Dav{\'e}}, {de Mello}, {Dekel}, {Dickinson},
  {Dolch}, {Donley}, {Dunlop}, {Dutton}, {Elbaz}, {Fazio}, {Filippenko},
  {Finkelstein}, {Fontana}, {Gardner}, {Garnavich}, {Gawiser}, {Giavalisco},
  {Grazian}, {Guo}, {Hathi}, {H{\"a}ussler}, {Hopkins}, {Huang}, {Huang},
  {Jha}, {Kartaltepe}, {Kirshner}, {Koo}, {Lai}, {Lee}, {Li}, {Lotz}, {Lucas},
  {Madau}, {McCarthy}, {McGrath}, {McIntosh}, {McLure}, {Mobasher},
  {Moustakas}, {Mozena}, {Nandra}, {Newman}, {Niemi}, {Noeske}, {Papovich},
  {Pentericci}, {Pope}, {Primack}, {Rajan}, {Ravindranath}, {Reddy}, {Renzini},
  {Rix}, {Robaina}, {Rodney}, {Rosario}, {Rosati}, {Salimbeni}, {Scarlata},
  {Siana}, {Simard}, {Smidt}, {Somerville}, {Spinrad}, {Straughn}, {Strolger},
  {Telford}, {Teplitz}, {Trump}, {van der Wel}, {Villforth}, {Wechsler},
  {Weiner}, {Wiklind}, {Wild}, {Wilson}, {Wuyts}, {Yan}, \& {Yun}}]{grogin11}
{Grogin}, N.~A., {Kocevski}, D.~D., {Faber}, S.~M., {et~al.} 2011, \apjs, 197,
  35, \dodoi{10.1088/0067-0049/197/2/35}

\bibitem[{{Gronke} {et~al.}(2016){Gronke}, {Dijkstra}, {McCourt}, \&
  {Oh}}]{gronke16}
{Gronke}, M., {Dijkstra}, M., {McCourt}, M., \& {Oh}, S.~P. 2016, \apjl, 833,
  L26, \dodoi{10.3847/2041-8213/833/2/L26}

\bibitem[{{Gronwall} {et~al.}(2007){Gronwall}, {Ciardullo}, {Hickey},
  {Gawiser}, {Feldmeier}, {van Dokkum}, {Urry}, {Herrera}, {Lehmer}, {Infante},
  {Orsi}, {Marchesini}, {Blanc}, {Francke}, {Lira}, \& {Treister}}]{gronwall07}
{Gronwall}, C., {Ciardullo}, R., {Hickey}, T., {et~al.} 2007, \apj, 667, 79,
  \dodoi{10.1086/520324}

\bibitem[{{Guaita} {et~al.}(2010){Guaita}, {Gawiser}, {Padilla}, {Francke},
  {Bond}, {Gronwall}, {Ciardullo}, {Feldmeier}, {Sinawa}, {Blanc}, \&
  {Virani}}]{guaita10}
{Guaita}, L., {Gawiser}, E., {Padilla}, N., {et~al.} 2010, \apj, 714, 255,
  \dodoi{10.1088/0004-637X/714/1/255}

\bibitem[{{Guaita} {et~al.}(2011){Guaita}, {Acquaviva}, {Padilla}, {Gawiser},
  {Bond}, {Ciardullo}, {Treister}, {Kurczynski}, {Gronwall}, {Lira}, \&
  {Schawinski}}]{guaita11}
{Guaita}, L., {Acquaviva}, V., {Padilla}, N., {et~al.} 2011, \apj, 733, 114,
  \dodoi{10.1088/0004-637X/733/2/114}

\bibitem[{{Hagen} {et~al.}(2016){Hagen}, {Zeimann}, {Behrens}, {Ciardullo},
  {Grasshorn Gebhardt}, {Gronwall}, {Bridge}, {Fox}, {Schneider}, {Trump},
  {Blanc}, {Chiang}, {Chonis}, {Finkelstein}, {Hill}, {Jogee}, \&
  {Gawiser}}]{hagen16}
{Hagen}, A., {Zeimann}, G.~R., {Behrens}, C., {et~al.} 2016, \apj, 817, 79,
  \dodoi{10.3847/0004-637X/817/1/79}

\bibitem[{Harris {et~al.}(2020)Harris, Millman, van~der Walt, Gommers,
  Virtanen, Cournapeau, Wieser, Taylor, Berg, Smith, Kern, Picus, Hoyer, van
  Kerkwijk, Brett, Haldane, del R{\'{i}}o, Wiebe, Peterson,
  G{\'{e}}rard-Marchant, Sheppard, Reddy, Weckesser, Abbasi, Gohlke, \&
  Oliphant}]{numpy}
Harris, C.~R., Millman, K.~J., van~der Walt, S.~J., {et~al.} 2020, Nature, 585,
  357, \dodoi{10.1038/s41586-020-2649-2}

\bibitem[{{Hathi} {et~al.}(2016){Hathi}, {Le F{\`e}vre}, {Ilbert}, {Cassata},
  {Tasca}, {Lemaux}, {Garilli}, {Le Brun}, {Maccagni}, {Pentericci}, {Thomas},
  {Vanzella}, {Zamorani}, {Zucca}, {Amor{\'\i}n}, {Bardelli}, {Cassar{\`a}},
  {Castellano}, {Cimatti}, {Cucciati}, {Durkalec}, {Fontana}, {Giavalisco},
  {Grazian}, {Guaita}, {Koekemoer}, {Paltani}, {Pforr}, {Ribeiro}, {Schaerer},
  {Scodeggio}, {Sommariva}, {Talia}, {Tresse}, {Vergani}, {Capak}, {Charlot},
  {Contini}, {Cuby}, {de la Torre}, {Dunlop}, {Fotopoulou},
  {L{\'o}pez-Sanjuan}, {Mellier}, {Salvato}, {Scoville}, {Taniguchi}, \&
  {Wang}}]{hathi16}
{Hathi}, N.~P., {Le F{\`e}vre}, O., {Ilbert}, O., {et~al.} 2016, \aap, 588,
  A26, \dodoi{10.1051/0004-6361/201526012}

\bibitem[{{Hayes}(2015)}]{hayes15}
{Hayes}, M. 2015, \pasa, 32, e027, \dodoi{10.1017/pasa.2015.25}

\bibitem[{{Hayes} {et~al.}(2014){Hayes}, {{\"O}stlin}, {Duval}, {Sand berg},
  {Guaita}, {Melinder}, {Adamo}, {Schaerer}, {Verhamme}, {Orlitov{\'a}},
  {Mas-Hesse}, {Cannon}, {Atek}, {Kunth}, {Laursen}, {Ot{\'\i}-Floranes},
  {Pardy}, {Rivera-Thorsen}, \& {Herenz}}]{hayes14}
{Hayes}, M., {{\"O}stlin}, G., {Duval}, F., {et~al.} 2014, \apj, 782, 6,
  \dodoi{10.1088/0004-637X/782/1/6}

\bibitem[{{Hill} {et~al.}(2008){Hill}, {Gebhardt}, {Komatsu}, {Drory},
  {MacQueen}, {Adams}, {Blanc}, {Koehler}, {Rafal}, {Roth}, {Kelz}, {Gronwall},
  {Ciardullo}, \& {Schneider}}]{hill08}
{Hill}, G.~J., {Gebhardt}, K., {Komatsu}, E., {et~al.} 2008, Astronomical
  Society of the Pacific Conference Series, Vol. 399, {The Hobby-Eberly
  Telescope Dark Energy Experiment (HETDEX): Description and Early Pilot Survey
  Results}, ed. T.~{Kodama}, T.~{Yamada}, \& K.~{Aoki}, 115

\bibitem[{{Hill} {et~al.}(2021){Hill}, {Lee}, {MacQueen}, {Kelz}, {Drory},
  {Vattiat}, {Good}, {Ramsey}, {Kriel}, {Peterson}, {DePoy}, {Gebhardt},
  {Marshall}, {Tuttle}, {Bauer}, {Chonis}, {Fabricius}, {Froning}, {Haeuser},
  {Indahl}, {Jahn}, {Landriau}, {Leck}, {Montesano}, {Prochaska}, {Snigula},
  {Zeimann}, {Bryant}, {Damm}, {Fowler}, {Janowiecki}, {Martin}, {Mrozinski},
  {Odewahn}, {Rostopchin}, {Shetrone}, {Spencer}, {Mentuch Cooper},
  {Armandroff}, {Bender}, {Dalton}, {Hopp}, {Komatsu}, {Lambert}, {Nicklas},
  {Ramsey}, {Roth}, {Schneider}, {Sneden}, \& {Steinmetz}}]{hill21}
{Hill}, G.~J., {Lee}, H., {MacQueen}, P.~J., {et~al.} 2021, arXiv e-prints,
  arXiv:2110.03843.
\newblock \doarXiv{2110.03843}

\bibitem[{{Horne}(1986)}]{horne86}
{Horne}, K. 1986, \pasp, 98, 609, \dodoi{10.1086/131801}

\bibitem[{{Huang} {et~al.}(2021){Huang}, {Lee}, {Shi}, {Malavasi}, {Xue}, \&
  {Dey}}]{huang21}
{Huang}, Y., {Lee}, K.-S., {Shi}, K., {et~al.} 2021, arXiv e-prints,
  arXiv:2104.11354.
\newblock \doarXiv{2104.11354}

\bibitem[{{Jung} {et~al.}(2018){Jung}, {Finkelstein}, {Livermore}, {Dickinson},
  {Larson}, {Papovich}, {Song}, {Tilvi}, \& {Wold}}]{jung18}
{Jung}, I., {Finkelstein}, S.~L., {Livermore}, R.~C., {et~al.} 2018, \apj, 864,
  103, \dodoi{10.3847/1538-4357/aad686}

\bibitem[{{Jung} {et~al.}(2020){Jung}, {Finkelstein}, {Dickinson}, {Hutchison},
  {Larson}, {Papovich}, {Pentericci}, {Straughn}, {Guo}, {Malhotra}, {Rhoads},
  {Song}, {Tilvi}, \& {Wold}}]{jung20}
{Jung}, I., {Finkelstein}, S.~L., {Dickinson}, M., {et~al.} 2020, \apj, 904,
  144, \dodoi{10.3847/1538-4357/abbd44}

\bibitem[{{Kelz} {et~al.}(2014){Kelz}, {Jahn}, {Haynes}, {Hill}, {Lee},
  {Murphy}, {Neumann}, {Nicklas}, {Rutowska}, {Sandin}, {Streicher}, {Tuttle},
  {Fabricius}, {Bauer}, {Vattiat}, {Anwand}, \& {Savage}}]{kelz14}
{Kelz}, A., {Jahn}, T., {Haynes}, D., {et~al.} 2014, Society of Photo-Optical
  Instrumentation Engineers (SPIE) Conference Series, Vol. 9147, {VIRUS:
  assembly, testing and performance of 33,000 fibres for HETDEX}, 914775,
  \dodoi{10.1117/12.2056384}

\bibitem[{{Khostovan} {et~al.}(2021){Khostovan}, {Malhotra}, {Rhoads},
  {Harish}, {Jiang}, {Wang}, {Wold}, {Zheng}, {Barrientos}, {Coughlin}, {Hu},
  {Infante}, {Perez}, {Pharo}, {Valdes}, \& {Walker}}]{khostovan21}
{Khostovan}, A.~A., {Malhotra}, S., {Rhoads}, J.~E., {et~al.} 2021, \mnras,
  503, 5115, \dodoi{10.1093/mnras/stab778}

\bibitem[{{Koekemoer} {et~al.}(2011){Koekemoer}, {Faber}, {Ferguson}, {Grogin},
  {Kocevski}, {Koo}, {Lai}, {Lotz}, {Lucas}, {McGrath}, {Ogaz}, {Rajan},
  {Riess}, {Rodney}, {Strolger}, {Casertano}, {Castellano}, {Dahlen},
  {Dickinson}, {Dolch}, {Fontana}, {Giavalisco}, {Grazian}, {Guo}, {Hathi},
  {Huang}, {van der Wel}, {Yan}, {Acquaviva}, {Alexander}, {Almaini}, {Ashby},
  {Barden}, {Bell}, {Bournaud}, {Brown}, {Caputi}, {Cassata}, {Challis},
  {Chary}, {Cheung}, {Cirasuolo}, {Conselice}, {Roshan Cooray}, {Croton},
  {Daddi}, {Dav{\'e}}, {de Mello}, {de Ravel}, {Dekel}, {Donley}, {Dunlop},
  {Dutton}, {Elbaz}, {Fazio}, {Filippenko}, {Finkelstein}, {Frazer}, {Gardner},
  {Garnavich}, {Gawiser}, {Gruetzbauch}, {Hartley}, {H{\"a}ussler},
  {Herrington}, {Hopkins}, {Huang}, {Jha}, {Johnson}, {Kartaltepe},
  {Khostovan}, {Kirshner}, {Lani}, {Lee}, {Li}, {Madau}, {McCarthy},
  {McIntosh}, {McLure}, {McPartland}, {Mobasher}, {Moreira}, {Mortlock},
  {Moustakas}, {Mozena}, {Nandra}, {Newman}, {Nielsen}, {Niemi}, {Noeske},
  {Papovich}, {Pentericci}, {Pope}, {Primack}, {Ravindranath}, {Reddy},
  {Renzini}, {Rix}, {Robaina}, {Rosario}, {Rosati}, {Salimbeni}, {Scarlata},
  {Siana}, {Simard}, {Smidt}, {Snyder}, {Somerville}, {Spinrad}, {Straughn},
  {Telford}, {Teplitz}, {Trump}, {Vargas}, {Villforth}, {Wagner}, {Wandro},
  {Wechsler}, {Weiner}, {Wiklind}, {Wild}, {Wilson}, {Wuyts}, \&
  {Yun}}]{koekemoer11}
{Koekemoer}, A.~M., {Faber}, S.~M., {Ferguson}, H.~C., {et~al.} 2011, \apjs,
  197, 36, \dodoi{10.1088/0067-0049/197/2/36}

\bibitem[{{Kornei} {et~al.}(2010){Kornei}, {Shapley}, {Erb}, {Steidel},
  {Reddy}, {Pettini}, \& {Bogosavljevi{\'c}}}]{kornei10}
{Kornei}, K.~A., {Shapley}, A.~E., {Erb}, D.~K., {et~al.} 2010, \apj, 711, 693,
  \dodoi{10.1088/0004-637X/711/2/693}

\bibitem[{{Kusakabe} {et~al.}(2018){Kusakabe}, {Shimasaku}, {Ouchi},
  {Nakajima}, {Goto}, {Hashimoto}, {Konno}, {Harikane}, {Silverman}, \&
  {Capak}}]{kusakabe18}
{Kusakabe}, H., {Shimasaku}, K., {Ouchi}, M., {et~al.} 2018, \pasj, 70, 4,
  \dodoi{10.1093/pasj/psx148}

\bibitem[{{Lang} {et~al.}(2016){Lang}, {Hogg}, \& {Mykytyn}}]{lang16}
{Lang}, D., {Hogg}, D.~W., \& {Mykytyn}, D. 2016, {The Tractor: Probabilistic
  astronomical source detection and measurement}.
\newblock \doeprint{1604.008}

\bibitem[{{Lee} {et~al.}(2010){Lee}, {Ferguson}, {Somerville}, {Wiklind}, \&
  {Giavalisco}}]{lee10}
{Lee}, S.-K., {Ferguson}, H.~C., {Somerville}, R.~S., {Wiklind}, T., \&
  {Giavalisco}, M. 2010, \apj, 725, 1644, \dodoi{10.1088/0004-637X/725/2/1644}

\bibitem[{{Leja} {et~al.}(2017){Leja}, {Johnson}, {Conroy}, {van Dokkum}, \&
  {Byler}}]{leja17}
{Leja}, J., {Johnson}, B.~D., {Conroy}, C., {van Dokkum}, P.~G., \& {Byler}, N.
  2017, \apj, 837, 170, \dodoi{10.3847/1538-4357/aa5ffe}

\bibitem[{{Leung} {et~al.}(2017){Leung}, {Acquaviva}, {Gawiser}, {Ciardullo},
  {Komatsu}, {Malz}, {Zeimann}, {Bridge}, {Drory}, {Feldmeier}, {Finkelstein},
  {Gebhardt}, {Gronwall}, {Hagen}, {Hill}, \& {Schneider}}]{leung17}
{Leung}, A.~S., {Acquaviva}, V., {Gawiser}, E., {et~al.} 2017, \apj, 843, 130,
  \dodoi{10.3847/1538-4357/aa71af}

\bibitem[{{Madau} \& {Dickinson}(2014)}]{madau14}
{Madau}, P., \& {Dickinson}, M. 2014, \araa, 52, 415,
  \dodoi{10.1146/annurev-astro-081811-125615}

\bibitem[{{Malhotra} \& {Rhoads}(2004)}]{malhotra04}
{Malhotra}, S., \& {Rhoads}, J.~E. 2004, \apjl, 617, L5, \dodoi{10.1086/427182}

\bibitem[{{Marchi} {et~al.}(2019){Marchi}, {Pentericci}, {Guaita}, {Talia},
  {Castellano}, {Hathi}, {Schaerer}, {Amorin}, {Bolzonella}, {Carnall},
  {Charlot}, {Chevallard}, {Cullen}, {Finkelstein}, {Fontana}, {Fontanot},
  {Garilli}, {Hibon}, {Koekemoer}, {Maccagni}, {McLure}, {Papovich},
  {Pozzetti}, \& {Saxena}}]{marchi19}
{Marchi}, F., {Pentericci}, L., {Guaita}, L., {et~al.} 2019, \aap, 631, A19,
  \dodoi{10.1051/0004-6361/201935495}

\bibitem[{{Martin} {et~al.}(2015){Martin}, {Dijkstra}, {Henry}, {Soto},
  {Danforth}, \& {Wong}}]{martin15}
{Martin}, C.~L., {Dijkstra}, M., {Henry}, A., {et~al.} 2015, \apj, 803, 6,
  \dodoi{10.1088/0004-637X/803/1/6}

\bibitem[{{Matthee} {et~al.}(2016){Matthee}, {Sobral}, {Oteo}, {Best}, {Smail},
  {R{\"o}ttgering}, \& {Paulino-Afonso}}]{matthee16}
{Matthee}, J., {Sobral}, D., {Oteo}, I., {et~al.} 2016, \mnras, 458, 449,
  \dodoi{10.1093/mnras/stw322}

\bibitem[{{Matthee} {et~al.}(2021){Matthee}, {Sobral}, {Hayes}, {Pezzulli},
  {Gronke}, {Schaerer}, {Naidu}, {R{\"o}ttgering}, {Calhau}, {Paulino-Afonso},
  {Santos}, \& {Amor{\'\i}n}}]{matthee21}
{Matthee}, J., {Sobral}, D., {Hayes}, M., {et~al.} 2021, \mnras, 505, 1382,
  \dodoi{10.1093/mnras/stab1304}

\bibitem[{{Miralda-Escud{\'e}}(1998)}]{miralda-escude98}
{Miralda-Escud{\'e}}, J. 1998, \apj, 501, 15, \dodoi{10.1086/305799}

\bibitem[{{Neufeld}(1991)}]{neufeld91}
{Neufeld}, D.~A. 1991, \apjl, 370, L85, \dodoi{10.1086/185983}

\bibitem[{{Ono} {et~al.}(2021){Ono}, {Itoh}, {Shibuya}, {Ouchi}, {Harikane},
  {Yamanaka}, {Inoue}, {Amagasa}, {Miura}, {Okura}, {Shimasaku}, {Iwata},
  {Taniguchi}, {Fujimoto}, {Iye}, {Jaelani}, {Kashikawa}, {Kikuchihara},
  {Kikuta}, {Kobayashi}, {Kusakabe}, {Lee}, {Liang}, {Matsuoka}, {Momose},
  {Nagao}, {Nakajima}, \& {Tadaki}}]{ono2021}
{Ono}, Y., {Itoh}, R., {Shibuya}, T., {et~al.} 2021, \apj, 911, 78,
  \dodoi{10.3847/1538-4357/abea15}

\bibitem[{{Ouchi} {et~al.}(2020){Ouchi}, {Ono}, \& {Shibuya}}]{ouchi20}
{Ouchi}, M., {Ono}, Y., \& {Shibuya}, T. 2020, \araa, 58, 617,
  \dodoi{10.1146/annurev-astro-032620-021859}

\bibitem[{{Oyarz{\'u}n} {et~al.}(2017){Oyarz{\'u}n}, {Blanc}, {Gonz{\'a}lez},
  {Mateo}, \& {Bailey}}]{oyarzun17}
{Oyarz{\'u}n}, G.~A., {Blanc}, G.~A., {Gonz{\'a}lez}, V., {Mateo}, M., \&
  {Bailey}, John~I., I. 2017, \apj, 843, 133, \dodoi{10.3847/1538-4357/aa7552}

\bibitem[{{Papovich} {et~al.}(2001){Papovich}, {Dickinson}, \&
  {Ferguson}}]{papovich01}
{Papovich}, C., {Dickinson}, M., \& {Ferguson}, H.~C. 2001, \apj, 559, 620,
  \dodoi{10.1086/322412}

\bibitem[{{Papovich} {et~al.}(2011){Papovich}, {Finkelstein}, {Ferguson},
  {Lotz}, \& {Giavalisco}}]{papovich11}
{Papovich}, C., {Finkelstein}, S.~L., {Ferguson}, H.~C., {Lotz}, J.~M., \&
  {Giavalisco}, M. 2011, \mnras, 412, 1123,
  \dodoi{10.1111/j.1365-2966.2010.17965.x}

\bibitem[{{Papovich} {et~al.}(2016){Papovich}, {Shipley}, {Mehrtens}, {Lanham},
  {Lacy}, {Ciardullo}, {Finkelstein}, {Bassett}, {Behroozi}, {Blanc}, {de
  Jong}, {DePoy}, {Drory}, {Gawiser}, {Gebhardt}, {Gronwall}, {Hill}, {Hopp},
  {Jogee}, {Kawinwanichakij}, {Marshall}, {McLinden}, {Mentuch Cooper},
  {Somerville}, {Steinmetz}, {Tran}, {Tuttle}, {Viero}, {Wechsler}, \&
  {Zeimann}}]{papovich16}
{Papovich}, C., {Shipley}, H.~V., {Mehrtens}, N., {et~al.} 2016, \apjs, 224,
  28, \dodoi{10.3847/0067-0049/224/2/28}

\bibitem[{{Partridge} \& {Peebles}(1967)}]{partridge67}
{Partridge}, R.~B., \& {Peebles}, P.~J.~E. 1967, \apj, 147, 868,
  \dodoi{10.1086/149079}

\bibitem[{{Pentericci} {et~al.}(2009){Pentericci}, {Grazian}, {Fontana},
  {Castellano}, {Giallongo}, {Salimbeni}, \& {Santini}}]{pentericci09}
{Pentericci}, L., {Grazian}, A., {Fontana}, A., {et~al.} 2009, \aap, 494, 553,
  \dodoi{10.1051/0004-6361:200810722}

\bibitem[{{Pentericci} {et~al.}(2010){Pentericci}, {Grazian}, {Scarlata},
  {Fontana}, {Castellano}, {Giallongo}, \& {Vanzella}}]{pentericci10}
{Pentericci}, L., {Grazian}, A., {Scarlata}, C., {et~al.} 2010, \aap, 514, A64,
  \dodoi{10.1051/0004-6361/200913425}

\bibitem[{{Ramsey} {et~al.}(1994){Ramsey}, {Sebring}, \& {Sneden}}]{rams94}
{Ramsey}, L.~W., {Sebring}, T.~A., \& {Sneden}, C.~A. 1994, Society of
  Photo-Optical Instrumentation Engineers (SPIE) Conference Series, Vol. 2199,
  {Spectroscopic survey telescope project}, ed. L.~M. {Stepp}, 31--40,
  \dodoi{10.1117/12.176221}

\bibitem[{{Reddy} {et~al.}(2021){Reddy}, {Topping}, {Shapley}, {Steidel},
  {Sanders}, {Du}, {Coil}, {Mobasher}, \& {Price}}]{reddy21}
{Reddy}, N.~A., {Topping}, M.~W., {Shapley}, A.~E., {et~al.} 2021, arXiv
  e-prints, arXiv:2108.05363.
\newblock \doarXiv{2108.05363}

\bibitem[{{Rhoads} {et~al.}(2000){Rhoads}, {Malhotra}, {Dey}, {Stern},
  {Spinrad}, \& {Jannuzi}}]{rhoads00}
{Rhoads}, J.~E., {Malhotra}, S., {Dey}, A., {et~al.} 2000, \apjl, 545, L85,
  \dodoi{10.1086/317874}

\bibitem[{{Robertson} {et~al.}(2015){Robertson}, {Ellis}, {Furlanetto}, \&
  {Dunlop}}]{robertson15}
{Robertson}, B.~E., {Ellis}, R.~S., {Furlanetto}, S.~R., \& {Dunlop}, J.~S.
  2015, \apjl, 802, L19, \dodoi{10.1088/2041-8205/802/2/L19}

\bibitem[{{Sanders} {et~al.}(2018){Sanders}, {Shapley}, {Kriek}, {Freeman},
  {Reddy}, {Siana}, {Coil}, {Mobasher}, {Dav{\'e}}, {Shivaei}, {Azadi},
  {Price}, {Leung}, {Fetherolf}, {de Groot}, {Zick}, {Fornasini}, \&
  {Barro}}]{sanders18}
{Sanders}, R.~L., {Shapley}, A.~E., {Kriek}, M., {et~al.} 2018, \apj, 858, 99,
  \dodoi{10.3847/1538-4357/aabcbd}

\bibitem[{{Santos} {et~al.}(2020){Santos}, {Sobral}, {Matthee}, {Calhau}, {da
  Cunha}, {Ribeiro}, {Paulino-Afonso}, {Arrabal Haro}, \&
  {Butterworth}}]{santos20}
{Santos}, S., {Sobral}, D., {Matthee}, J., {et~al.} 2020, \mnras, 493, 141,
  \dodoi{10.1093/mnras/staa093}

\bibitem[{{Scarlata} {et~al.}(2009){Scarlata}, {Colbert}, {Teplitz}, {Panagia},
  {Hayes}, {Siana}, {Rau}, {Francis}, {Caon}, {Pizzella}, \&
  {Bridge}}]{scarlata09}
{Scarlata}, C., {Colbert}, J., {Teplitz}, H.~I., {et~al.} 2009, \apjl, 704,
  L98, \dodoi{10.1088/0004-637X/704/2/L98}

\bibitem[{{Shapley} {et~al.}(2003){Shapley}, {Steidel}, {Pettini}, \&
  {Adelberger}}]{shapley03}
{Shapley}, A.~E., {Steidel}, C.~C., {Pettini}, M., \& {Adelberger}, K.~L. 2003,
  \apj, 588, 65, \dodoi{10.1086/373922}

\bibitem[{{Shimakawa} {et~al.}(2017){Shimakawa}, {Kodama}, {Shibuya},
  {Kashikawa}, {Tanaka}, {Matsuda}, {Tadaki}, {Koyama}, {Hayashi}, {Suzuki}, \&
  {Yamamoto}}]{shimakawa17}
{Shimakawa}, R., {Kodama}, T., {Shibuya}, T., {et~al.} 2017, \mnras, 468, 1123,
  \dodoi{10.1093/mnras/stx091}

\bibitem[{{Sobral} {et~al.}(2018){Sobral}, {Santos}, {Matthee},
  {Paulino-Afonso}, {Ribeiro}, {Calhau}, \& {Khostovan}}]{sobral18}
{Sobral}, D., {Santos}, S., {Matthee}, J., {et~al.} 2018, \mnras, 476, 4725,
  \dodoi{10.1093/mnras/sty378}

\bibitem[{{Steidel} {et~al.}(2010){Steidel}, {Erb}, {Shapley}, {Pettini},
  {Reddy}, {Bogosavljevi{\'c}}, {Rudie}, \& {Rakic}}]{steidel10}
{Steidel}, C.~C., {Erb}, D.~K., {Shapley}, A.~E., {et~al.} 2010, \apj, 717,
  289, \dodoi{10.1088/0004-637X/717/1/289}

\bibitem[{{Trainor} {et~al.}(2019){Trainor}, {Strom}, {Steidel}, {Rudie},
  {Chen}, \& {Theios}}]{trainor19}
{Trainor}, R.~F., {Strom}, A.~L., {Steidel}, C.~C., {et~al.} 2019, \apj, 887,
  85, \dodoi{10.3847/1538-4357/ab4993}

\bibitem[{{Vargas} {et~al.}(2014){Vargas}, {Bish}, {Acquaviva}, {Gawiser},
  {Finkelstein}, {Ciardullo}, {Ashby}, {Feldmeier}, {Ferguson}, {Gronwall},
  {Guaita}, {Hagen}, {Koekemoer}, {Kurczynski}, {Newman}, \&
  {Padilla}}]{vargas14}
{Vargas}, C.~J., {Bish}, H., {Acquaviva}, V., {et~al.} 2014, \apj, 783, 26,
  \dodoi{10.1088/0004-637X/783/1/26}

\bibitem[{{Weiss} {et~al.}(2021){Weiss}, {Bowman}, {Ciardullo}, {Zeimann},
  {Gronwall}, {Mentuch Cooper}, {Gebhardt}, {Hill}, {Blanc}, {Farrow},
  {Finkelstein}, {Gawiser}, {Janowiecki}, {Jogee}, {Schneider}, \&
  {Wisotzki}}]{weiss21}
{Weiss}, L.~H., {Bowman}, W.~P., {Ciardullo}, R., {et~al.} 2021, \apj, 912,
  100, \dodoi{10.3847/1538-4357/abedb9}

\bibitem[{{Wold} {et~al.}(2014){Wold}, {Barger}, \& {Cowie}}]{wold14}
{Wold}, I. G.~B., {Barger}, A.~J., \& {Cowie}, L.~L. 2014, \apj, 783, 119,
  \dodoi{10.1088/0004-637X/783/2/119}

\bibitem[{{Xue} {et~al.}(2016){Xue}, {Luo}, {Brandt}, {Alexander}, {Bauer},
  {Lehmer}, \& {Yang}}]{xue16}
{Xue}, Y.~Q., {Luo}, B., {Brandt}, W.~N., {et~al.} 2016, \apjs, 224, 15,
  \dodoi{10.3847/0067-0049/224/2/15}

\bibitem[{{York} {et~al.}(2000){York}, {Adelman}, {Anderson}, {Anderson},
  {Annis}, {Bahcall}, {Bakken}, {Barkhouser}, {Bastian}, {Berman}, {Boroski},
  {Bracker}, {Briegel}, {Briggs}, {Brinkmann}, {Brunner}, {Burles}, {Carey},
  {Carr}, {Castander}, {Chen}, {Colestock}, {Connolly}, {Crocker}, {Csabai},
  {Czarapata}, {Davis}, {Doi}, {Dombeck}, {Eisenstein}, {Ellman}, {Elms},
  {Evans}, {Fan}, {Federwitz}, {Fiscelli}, {Friedman}, {Frieman}, {Fukugita},
  {Gillespie}, {Gunn}, {Gurbani}, {de Haas}, {Haldeman}, {Harris}, {Hayes},
  {Heckman}, {Hennessy}, {Hindsley}, {Holm}, {Holmgren}, {Huang}, {Hull},
  {Husby}, {Ichikawa}, {Ichikawa}, {Ivezi{\'c}}, {Kent}, {Kim}, {Kinney},
  {Klaene}, {Kleinman}, {Kleinman}, {Knapp}, {Korienek}, {Kron}, {Kunszt},
  {Lamb}, {Lee}, {Leger}, {Limmongkol}, {Lindenmeyer}, {Long}, {Loomis},
  {Loveday}, {Lucinio}, {Lupton}, {MacKinnon}, {Mannery}, {Mantsch}, {Margon},
  {McGehee}, {McKay}, {Meiksin}, {Merelli}, {Monet}, {Munn}, {Narayanan},
  {Nash}, {Neilsen}, {Neswold}, {Newberg}, {Nichol}, {Nicinski}, {Nonino},
  {Okada}, {Okamura}, {Ostriker}, {Owen}, {Pauls}, {Peoples}, {Peterson},
  {Petravick}, {Pier}, {Pope}, {Pordes}, {Prosapio}, {Rechenmacher}, {Quinn},
  {Richards}, {Richmond}, {Rivetta}, {Rockosi}, {Ruthmansdorfer}, {Sandford},
  {Schlegel}, {Schneider}, {Sekiguchi}, {Sergey}, {Shimasaku}, {Siegmund},
  {Smee}, {Smith}, {Snedden}, {Stone}, {Stoughton}, {Strauss}, {Stubbs},
  {SubbaRao}, {Szalay}, {Szapudi}, {Szokoly}, {Thakar}, {Tremonti}, {Tucker},
  {Uomoto}, {Vanden Berk}, {Vogeley}, {Waddell}, {Wang}, {Watanabe},
  {Weinberg}, {Yanny}, {Yasuda}, \& {SDSS Collaboration}}]{york00}
{York}, D.~G., {Adelman}, J., {Anderson}, John~E., J., {et~al.} 2000, \aj, 120,
  1579, \dodoi{10.1086/301513}

\end{thebibliography}

\clearpage

\appendix

\section{Model-Dependence of Measured Galaxy Properties}
\label{sec:appendixa}

Bayesian approaches to SED fitting, like the one implemented in \bagpipes\, provide robust constraints on the parameter uncertainties and their interdependence, but the model chosen for comparison to the data (as well as the chosen priors) determines the accuracy of those estimates. In other words, an inaccurate model yields inaccurate measurements of galaxy properties. Many galaxy SED fitting studies have shown that model choices, such as the SFH, systematically impact the measured galaxy properties (see \citealt{conroy13} for review). 

To test the robustness of our results to different modeling choices, we performed an additional analysis of our entire sample using an alternate model. We did not seek to find a more (or less) accurate model; we simply wanted a different model to determine if the median properties or correlations between \lya\ emission and galaxy properties changed. To this end, we adopted a constant SFH parametrization as well as the dust absorption model of \citealt{CharlotFall}. The constant SFH required two parameters: the time when star formation began and the constant star formation rate. For dust attentuation, we adopted the recipe given in \citet{CharlotFall} by using an absorption curve proportional to $\lambda^{-0.7}$, and a factor of three reduction in the dust extinction normalization for stellar populations older than $10^7$ years to account for the dispersal of stellar birth clouds. The authors found this recipe to match the absorption of stellar continuum and nebular emission for nearby starburst galaxies very well, and the differential extinction toward young stars differs markedly from the treatment by \citet{calzetti94} used in our ``fiducial'' model presented above. 

\begin{figure}[h]
    \centering
    \includegraphics[width=0.8\textwidth]{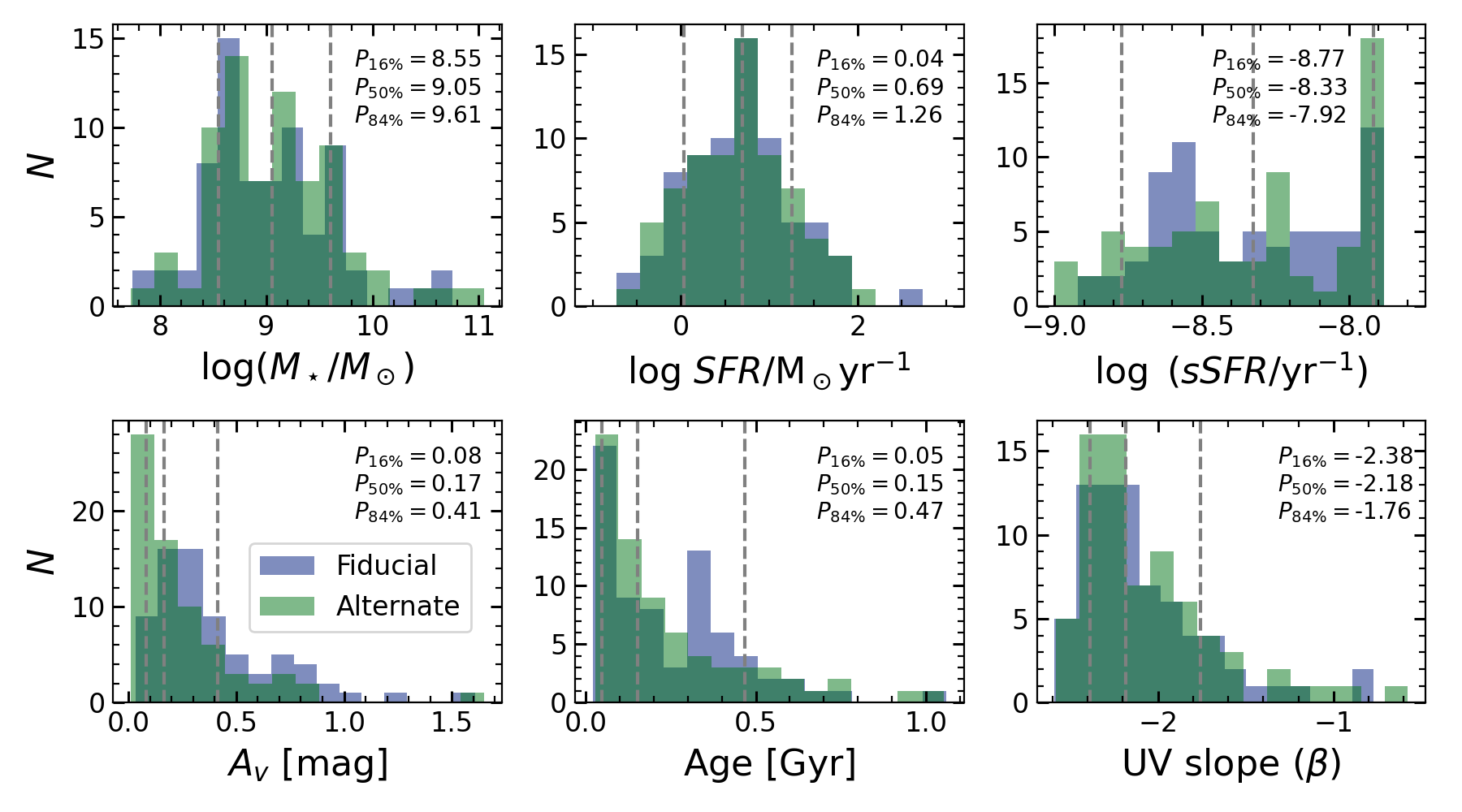}
    \caption{Comparison of galaxy properties as measured using our ``fiducial'' model (light blue) versus our ``alternate'' model (sea green). The 16$^{th}$, 50$^{th}$, and 84$^{th}$ percentiles calculated using the alternate model are indicated by vertical dashed grey lines, and their values are indicated with text. The distributions are consistent, save for dust extinction ($\mathrm{A_V}$), which has lower values by a factor of $\sim 2$ for the alternate model. }
    \label{fig:modcomphist}
\end{figure}

\begin{figure}
    \centering
    \includegraphics[width=0.8\textwidth]{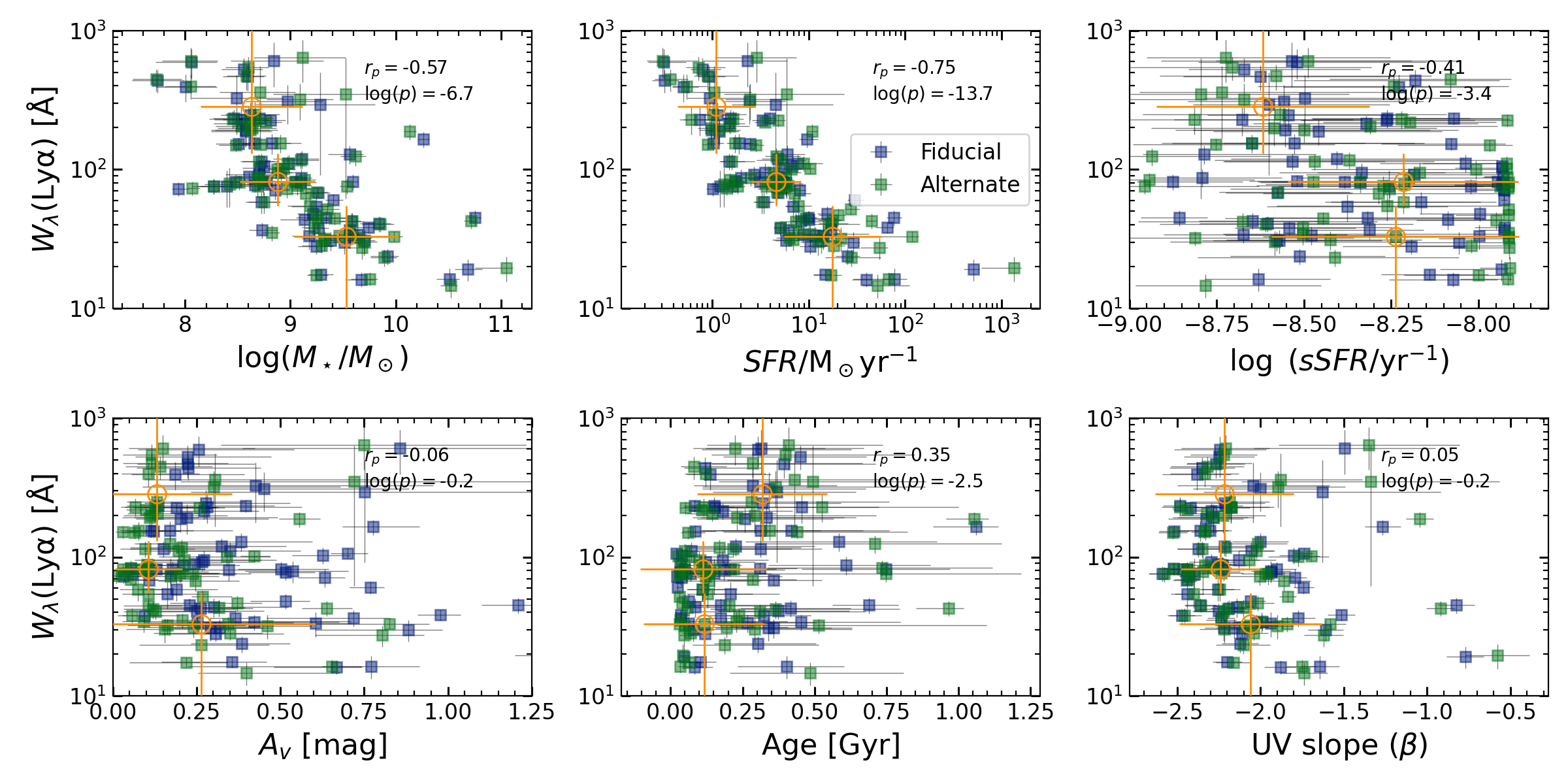}
    \caption{Comparison of correlations between $\ewlya$ and galaxy properties as measured using our ``fiducial'' model (light blue) versus our ``alternate'' model (sea green). Binned values from the alternate model are indicated as open gold circles. The correlations presented in \S \ref{ssec:corrs} appear robust when different models are adopted.}
    \label{fig:modcompscat}
\end{figure}

Figure~\ref{fig:modcomphist} shows the distribution of \lae\ properties measured using the alternate model compared with the fiducial model. The sample median stellar mass increased by 0.1 dex, as did the median SFR. These two changes do not affect our results or interpretation significantly. The median dust dropped from $A_V = 0.30$ to 0.17, a fairly substantial change, but not unusual given the common factors of $\sim$ a few discrepancies between different models and SED-fitting codes (see \citealt{leja17}). Nonetheless, the correlations between galaxy properties and $\ewlya$ remained unaffected by the model modifications, as shown in Figure~\ref{fig:modcompscat}. Stellar mass and SFR correlated strongly and negatively with \lya\ emission strength, while other parameters, like dust extinction, continued to show no significant correlations.

\section{Imaging, Emission lines, and SED fits for LAEs in this study}
\label{sec:appendixb}

In this section, for all \nsamp\ \laes\ in our sample, we present \hst\ imaging cutouts in Figure~\ref{fig:hst_stamps} showing the sources and any neighbors, the \hd\ \lya\ emission line detections in Figure~\ref{fig:all_lines}, and the SED fits with \bagpipes\ \citep{carnall18} used to measure physical properties in Figure~\ref{fig:all_sed_2}.  

\begin{figure}
    \centering
    \includegraphics[width=0.9\textwidth]{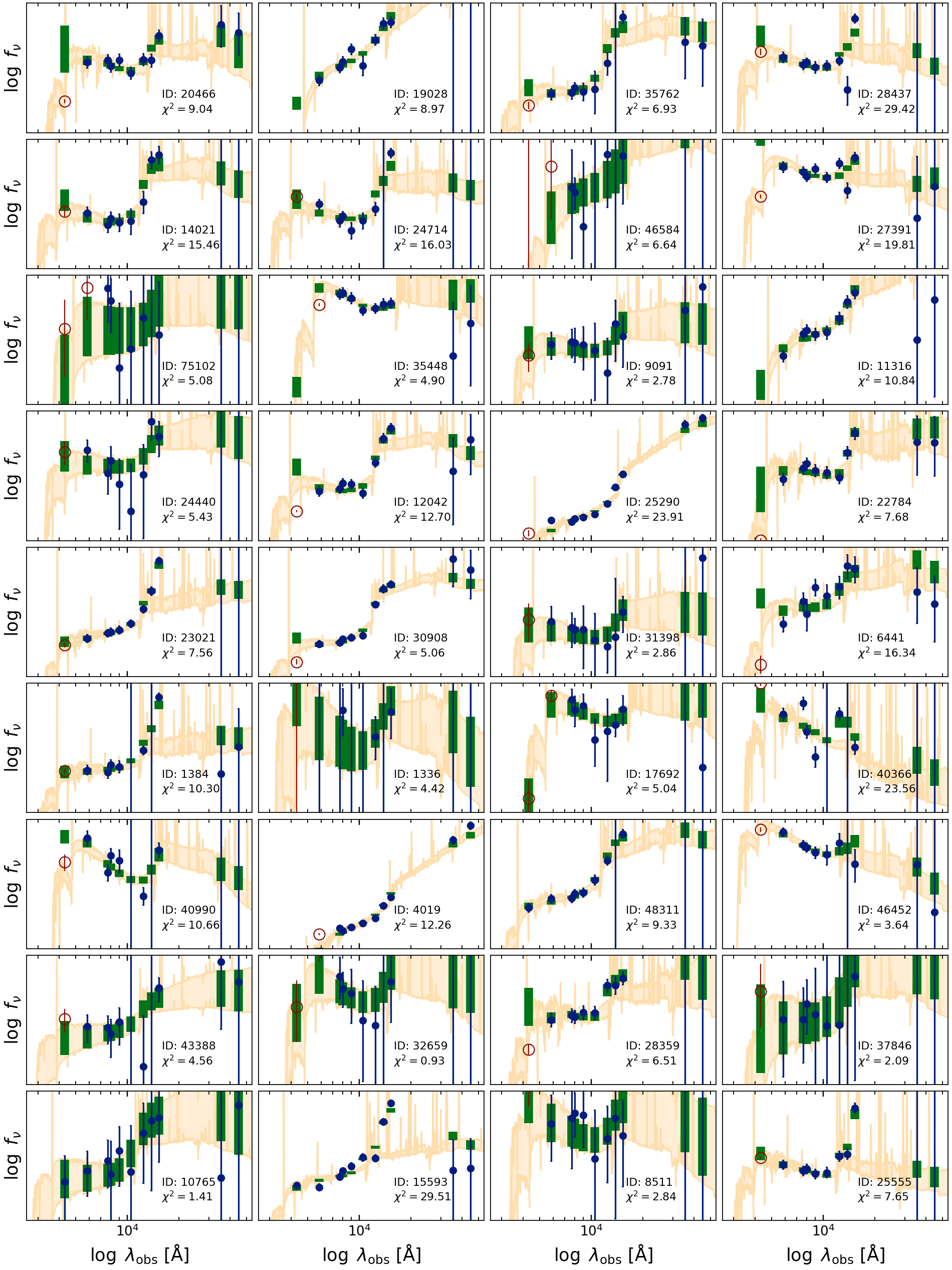}
    \label{fig:all_sed_1}
\end{figure}

\begin{figure}
    \centering
    \includegraphics[width=0.9\textwidth]{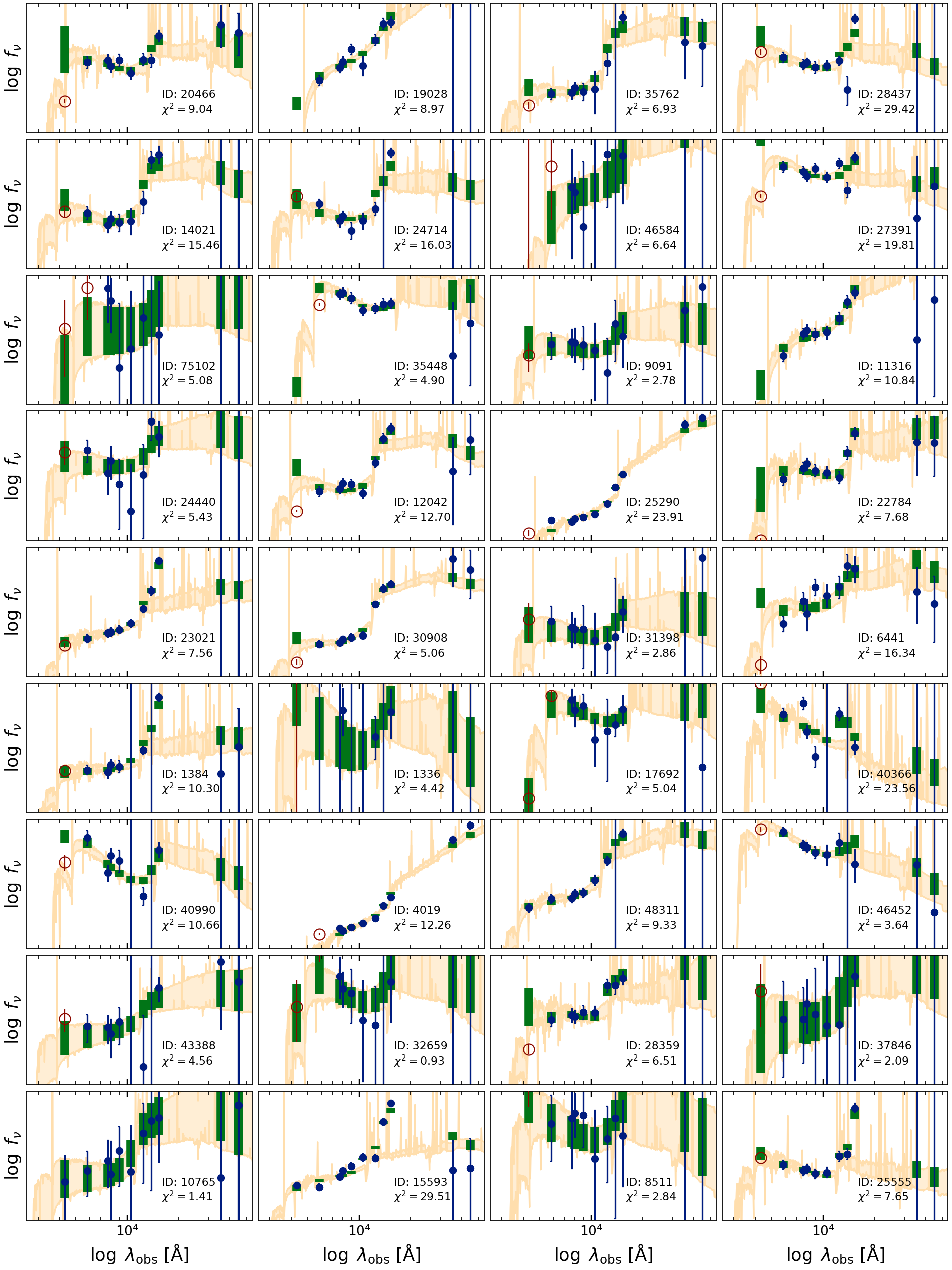}
    \caption{All SED fits for LAEs in the sample (see Figure~\ref{fig:sedfit} for a description of the plots). The $\chi^2$ value for each fit is also given with text.}
    \label{fig:all_sed_2}
\end{figure}

\begin{figure}
    \centering
    \includegraphics[width=0.9\textwidth]{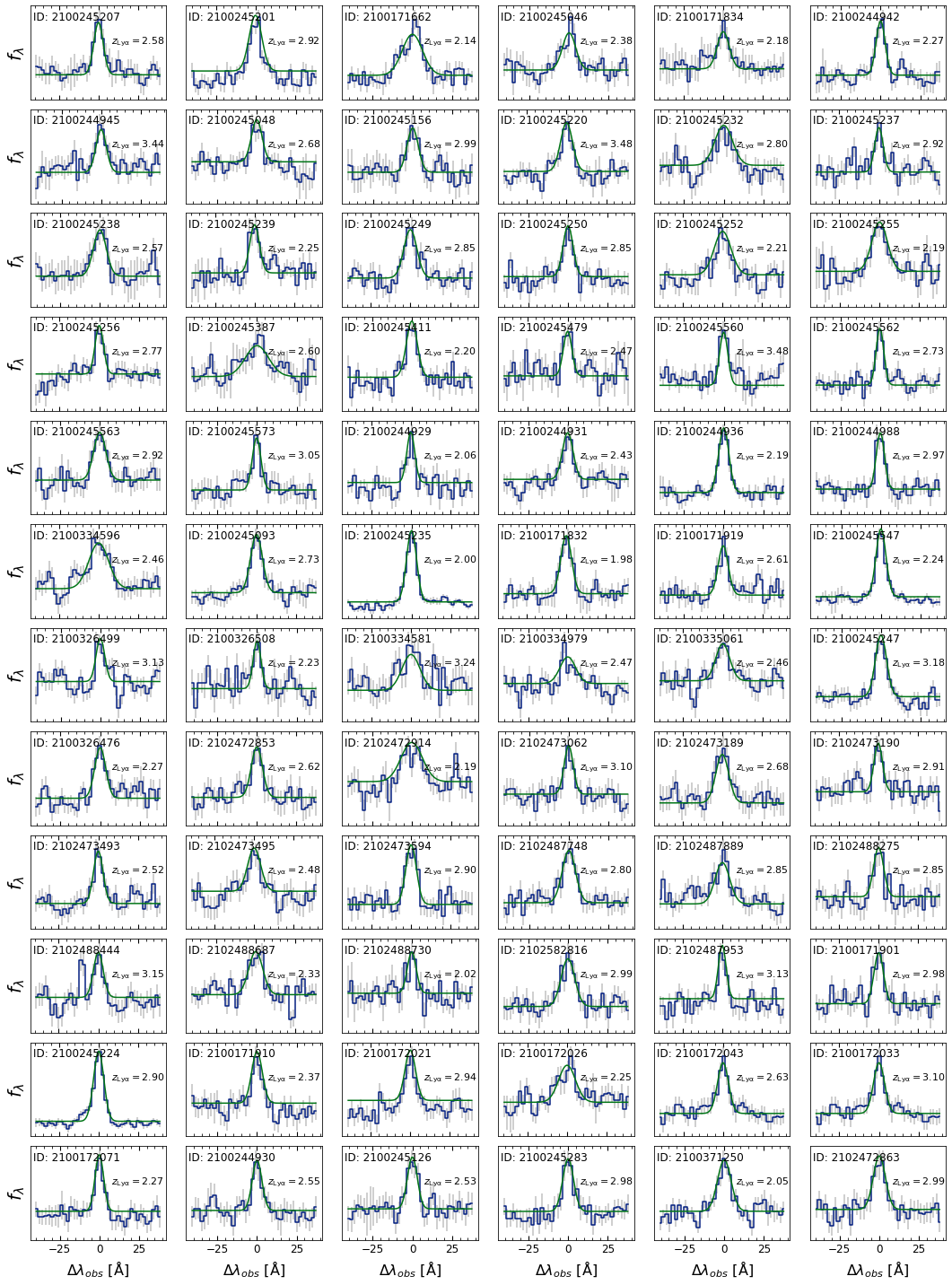}
    \caption{All emission line detections from the \hd\ Survey for LAEs in the sample. The observed data are indicated by the blue lines with grey error bars. A Gaussian model fit to the data is shown in green. The x-axis is scaled in Angstroms relative to the line center. Detection IDs and \lya\ line redshifts are indicatd with text.}
    \label{fig:all_lines}
\end{figure}

\begin{figure}
    \centering
    \includegraphics[width=0.9\textwidth]{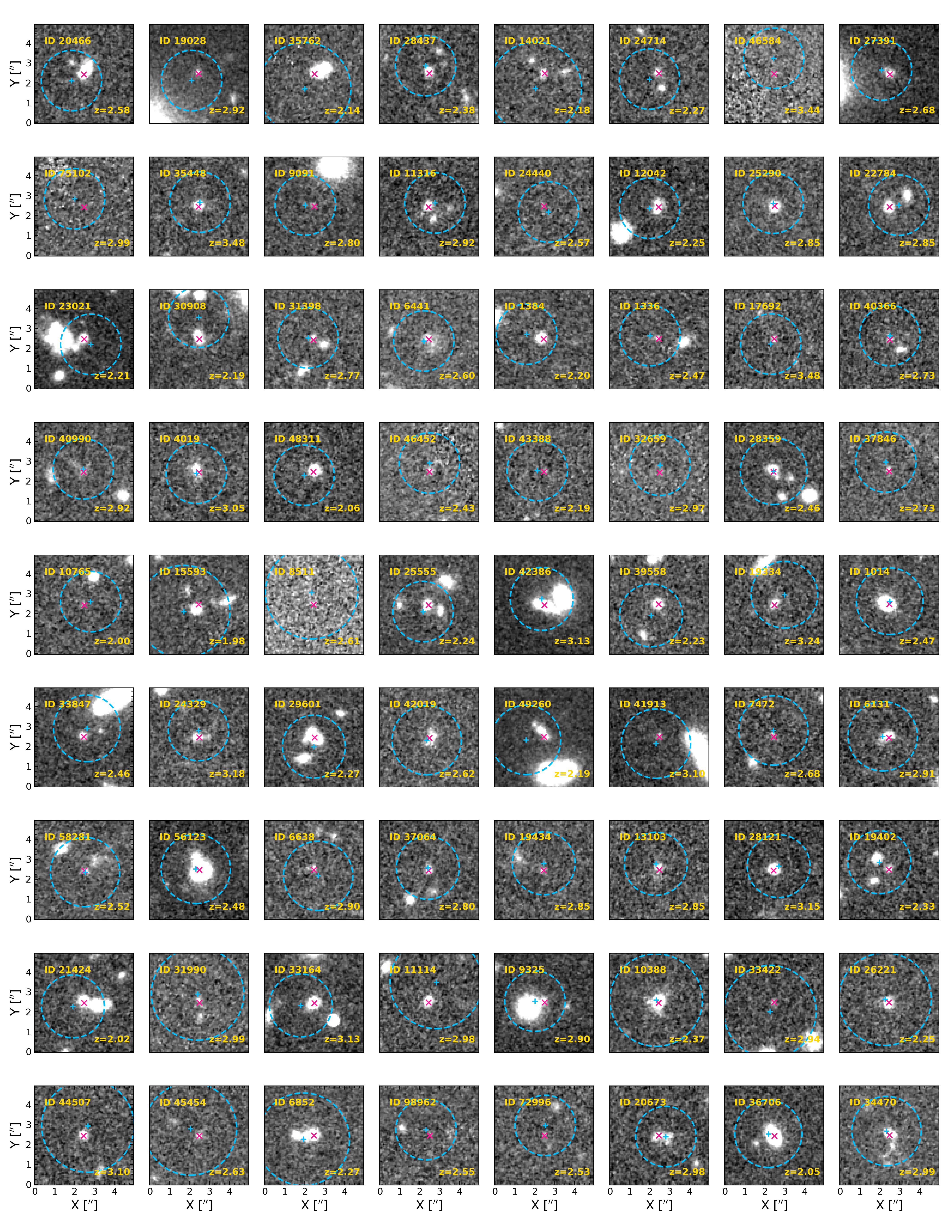}
    \caption{All \hst\ F160W ($H$-band) images of LAEs in the sample. Each cutout shows a $5 \arcsec \times 5 \arcsec$ image centered on each galaxy in our sample. The galaxy centroid is indicated with a pink diamond, and the \hd\ detection position and PSF FWHM are indicated by a light blue cross and dashed circle, respectively. We also include object IDs and redshifts with text.}
    \label{fig:hst_stamps}
\end{figure}

\end{document}